\documentclass[fleqn,usenatbib,useAMS]{mnras}
\usepackage{newtxtext,newtxmath}
\usepackage[T1]{fontenc}
\usepackage{subfloat}

\usepackage{graphicx}	
\usepackage{amsmath}	
\usepackage{lipsum}
\usepackage{caption}
\usepackage{textgreek} 
\usepackage{deluxetable}
\usepackage{comment}
\usepackage{xcolor}  
\newcommand{\rev}[1]{{\color{black} #1}} 
\usepackage[intoc, english]{nomencl}

\title[Cosmic quenching for SMBH \& Host Evolution]{Cosmic quenching and scaling laws for the evolution of supermassive black holes and host galaxies}

\author[Z. Xu]{Zhijie (Jay) Xu,$^{1}$\thanks{E-mail: \href{mailto:zhijie.xu@pnnl.gov}{zhijie.xu@pnnl.gov}; \href{mailto:zhijiexu@hotmail.com}{zhijiexu@hotmail.com}}
\\
$^{1}$Physical and Computational Sciences Directorate, Pacific Northwest National Laboratory; Richland, WA 99354, USA\\
}

\date{Accepted XXX. Received YYY; in original form ZZZ}

\pubyear{2024}

\makenomenclature

\begin{document}
\label{firstpage}
\pagerange{\pageref{firstpage}--\pageref{lastpage}}
\maketitle

\begin{abstract}
Observations suggest a strong coevolution of supermassive black holes (SMBHs) and host galaxies. In this paper, we consider the mass and energy flow in a near-equilibrium bulge suffused by gases of varying temperatures. By assuming the rate of energy flow independent of the distance $r$ from the bulge center and the local virial equilibrium for permeated gases on scale $r$, a key parameter $\varepsilon_b$ (unit: m$^2$/s$^3$) was identified that quantifies the rate of mass and energy flow in gases and the efficiency of gas cooling (or the "specific" cooling rate per unit mass), and thus regulates the coevolution of both SMBHs and hosts. With the help of Illustris simulations and observations, we determined the redshift variation $\varepsilon_b\propto (1+z)^{5/2}$. A higher $\varepsilon_b$ in the early Universe means a higher specific cooling rate that allows rapid evolution of SMBHs and hosts. This simple theory, characterized by a single parameter $\varepsilon_b$, provides the dominant mean cosmic evolution of SMBHs and hosts. All other transient phenomena may only contribute to the dispersion around this mean evolution. Based on this theory and relevant assumptions, scaling laws involving $\varepsilon_b$ were identified for the evolution of SMBHs and hosts. For host galaxies, the mass-size relation $M_b\propto \varepsilon_b^{2/3}r_b^{5/3}G^{-1}$, the dispersion-size relation $\sigma_b^2\propto(\varepsilon_b r_b)^{2/3}\propto (1+z)$, or the mass-dispersion relation $M_b\propto \varepsilon_b^{-1}G^{-1}\sigma_b^5$ were identified, where $r_b\propto (1+z)^{-1}$ is the bulge size. For SMBHs, three evolution phases were found involving an initial rapid growth stage with a rising luminosity $L_B\propto (\varepsilon_b M_{BH})^{4/5}G^{-1/5}c$, a transition stage with a declining $L_B\propto \varepsilon_b^2 M_{BH} \propto (1+z)^5$, and a dormant stage with $L_B\propto (\varepsilon_b M_{BH})^{4/3}G^{1/3}c^{-5/3}$. Our results suggest a rapid initial super-Eddington growth in a short period with a new redshift-dependent luminosity limit $L_X\propto\varepsilon_b^{4/5}M_{BH}^{4/5}G^{-1/5}c$, in contrast to the Eddington limit. Analytical solutions were formulated for the BH mass function $\Phi_{BH}$, AGN mass function $\Phi_{AGN}$, and duty cycle $U$ that predict $\Phi_L\propto L^{-1/5}$ for the faint-end luminosity function, $\Phi_{AGN}\propto M^{-1/5}$ for small-mass-end AGN mass function $\Phi_L$, and $U\propto M^{-1/5}$ at high redshift. 
\end{abstract}


\begingroup
\let\clearpage\relax
\tableofcontents
\endgroup


\section{Introduction}
\label{sec:1}
Supermassive black holes (SMBHs) are ubiquitously associated with the center of massive galaxies that contain bulges \citep{Kormendy:2013-Coevolution-Supermassive-Black-Holes}. Numerous observations suggest that SMBHs and their host galaxies are "co-evolving". The first evidence is the strong and tight correlations between SMBHs and host galaxies. Early black hole demography reveals the correlation between the BH mass $M_{BH}$ and the bulge luminosity $L_b$ \citep{Magorrian:1998-The-Demography-of-Massive-Dark-Objects,Marconi:2003-The-relation-between-blac,Graham:2013-Relation-at-High-and-Low-Masses,McConnell:2013-Revisiting-the-Scaling-Re}. Since bulge luminosity $L_b$ is related to bulge mass $M_b$ and velocity dispersion $\sigma_b$ \citep{Faber:1976-Velocity-Dispersions-and-Mass-,Curtis:1981-Dissipation-and-the-global-properties}, the $M_{BH}$-$L_b$ correlation strongly hints direct correlations between $M_{BH}$ and $M_b$ or $\sigma_b$. The mass of the bulge $M_b$ can be directly related to the velocity dispersion $\sigma_b$ of the bulge by the virial theorem (that is, $M_b \propto r_b \sigma_b^2$, where $r_b$ is the size of the bulge). The correlation between SMBH mass and bulge mass is generally consistent with a linear relation $M_{BH} \propto M_b$ \citep{Magorrian:1998-The-Demography-of-Massive-Dark-Objects, Marconi:2003-The-relation-between-blac, Haring:2004-On-the-black-hole-mass-bu}. A tighter correlation with a smaller intrinsic scatter was discovered between the BH mass and the bulge velocity dispersion ($M_{BH}$-$\sigma_b$). This tighter correlation provides strong evidence for a fundamental relationship between SMBHs and their host galaxies \citep{Ferrarese:2000-A-Fundamental-Relation-between-Supermassive-Black-Holes, Merritt:2001-Black-hole-demographics-from-the,Hopkins:2007-An-Observed-Fundamental-Plane-Relation, Hu:2008-The-black-hole-mass-stell,Gultekin:2009-THE-M-sigma-AND-M-L-RELAT,McConnell:2013-Revisiting-the-Scaling-Re}. Many studies suggest a power law $M_{BH} \propto \sigma_b^\alpha$ with $\alpha \approx 5$. Examples are $\alpha=4.8$ \citep{Ferrarese:2000-A-Fundamental-Relation-between-Supermassive-Black-Holes}, $\alpha=4.86$ \citep{Ferrarese:2005-Supermassive-Black-Holes-in-Galactic-Nuclei}, $\alpha=5.4$ \citep{Marsden:2020-The-Case-for-the-Fundamen}, $\alpha=5$ \citep{Woo:2015-THE-BLACK-HOLE-MASS-STELLAR-VELOCITY-DISPERSION}, $\alpha=4.24$ \citep{Gultekin:2009-THE-M-sigma-AND-M-L-RELAT}, and $\alpha=4$ or 4.5 \citep{Hu:2008-The-black-hole-mass-stell}. 

The physical mechanisms responsible for these tight correlations are not fully understood. This is partially because of the vast disparity in scales from black holes ($\sim$$10^{-4}$pc) to their host galaxies ($\sim 10^4$pc) and the complex physics involved on the black hole and bulge scales. Various mechanisms have been proposed to interpret these tight correlations. A possible mechanism involves the SMBH feedback during its active galactic nucleus (AGN) phase, where a significant amount of energy/momentum is injected into the surrounding gas. The energy or momentum released unbinds the surrounding gas, prevents star formation, and shapes the evolution of the host galaxy \citep{Silk:1998-Quasars-and-galaxy-formation, King:2003-Black-Holes-Galaxy-Formation-and-the}. An alternative mechanism proposes that coevolution is established through the same source of gas supply provided for black hole mass accretion and star formation \citep{Menci:2016-Relative-growth-of-black-holes-and-the-stellar}. A statistical interpretation is also presented such that the tight correlation is just a consequence of statistical convergence during the hierarchical formation of the galaxy structure \citep{Peng:2007-How-Mergers-May-Affect-the-Mass-Scaling-Relation}. 

\begin{figure}
\includegraphics*[width=\columnwidth]{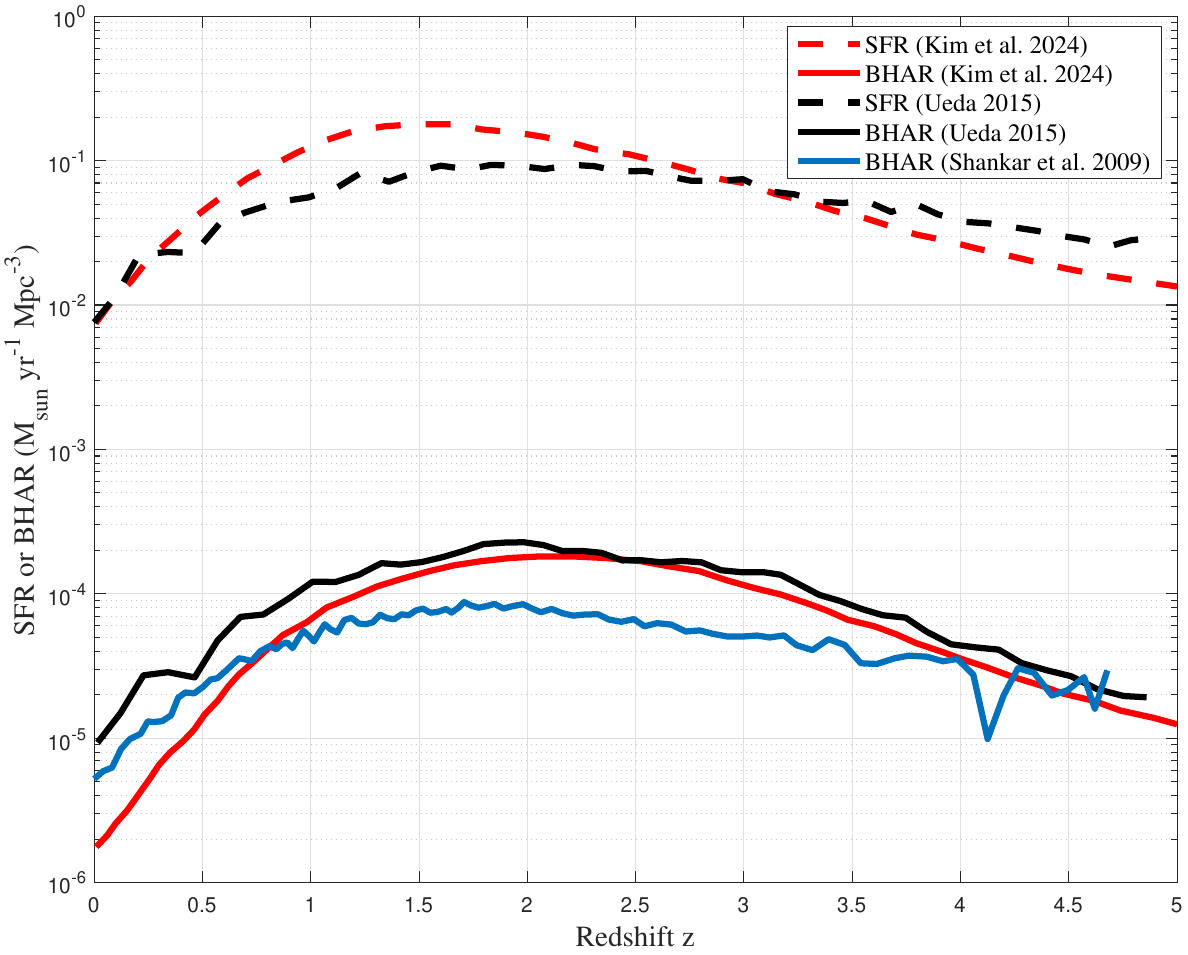}
\caption{The cosmic evolution of the stellar mass and the mass of supermassive black holes from Kim et al. \citep{Kim:2024-Cosmic-star-formation-history-and-black-hole-accretion-history}, Ueda \citep{UEDA:2015-Cosmological-evolution-of-supermassive-black-holes}, and Shankar et al. \citep{Shankar:2009-Self-Consistent-Models-of-the-AGN-and-Black-Hole}. The black hole accretion rate (BHAR) is compared to the star-formation rate (SFR) as a function of redshift $z$. A striking similarity between the SFR and BHAR can be found over cosmic time, suggesting a synchronized evolution of SMBHs and their hosts.} 
\label{fig:S37}
\end{figure}

The second evidence for the "co-evolution" between SMBHs and host galaxies follows from the striking similarity between the star formation rate (SFR) and the BH accretion rate (BHAR) over cosmic time. Figure \ref{fig:S37} presents the cosmic evolution of SFR and BHAR obtained from Kim et al. \citep{Kim:2024-Cosmic-star-formation-history-and-black-hole-accretion-history}, Ueda \citep{UEDA:2015-Cosmological-evolution-of-supermassive-black-holes} and Shankar et al. \citep{Shankar:2009-Self-Consistent-Models-of-the-AGN-and-Black-Hole}. The evolution of both rates exhibits a steep increase from redshift $z$ = 0 to 1, followed by a maximum around $z\approx 2$, and then a steep decline at higher redshifts. The ratio between the two growth rates remained roughly constant on the order of $10^3$. In the volume-averaged sense, the evolution of galaxies and SMBHs has somehow been synchronized \citep{Heckman:2014-The-Coevolution-of-Galaxies-and-Supermassive-Black-Holes}. This strongly suggests the need for a holistic view of the evolution of SMBHs and their hosts and potential common mechanisms that regulate and synchronize their evolution.  

Such mechanisms can be made possible if one or a group of common parameters simultaneously regulate the evolution of both SMBH and star formation. In this way, both evolutions are automatically synchronized during the cosmic evolution of these parameters. To find these parameters, we recall the observed SMBH-host correlations. A very interesting finding is: combining the $M_{BH}$-$\sigma_b$ correlation $M_{BH}=A(z)\sigma_b^5$, the linear mass correlation $M_{BH}=B(z)M_b$, and the virial theorem $GM_b=r_b \sigma_b^2/\gamma_r$, leads to a simple parameter
\begin{equation} 
\label{eq:2} 
\varepsilon_b(z)=\frac{\sigma_b^3}{r_b}=\frac{\sigma_b^2}{r_b/\sigma_b}=\frac{B(z)}{A(z)}\frac{1}{\gamma_rG},
\end{equation}
where the parameter $\varepsilon_b(z)$ (unit: m$^2$/$s^3$) is a new physical quantity and the focus of this work. Here, A(z), B(z), and $\gamma_r$ are proportional coefficients that may depend on the redshift $z$. Inversely, if parameter $\varepsilon_b(z)$ is known, it also enables us to derive the observed correlations. For example, with $M_{BH}\propto M_b$, the virial theorem $M_b \propto r_b \sigma_b^2$, and Eq. \eqref{eq:2}, we can recover the $M_{BH}-\sigma_b$ correlation as
\begin{equation} 
\label{eq:2-2} 
M_{BH} = A(z)\sigma_b^5 = \frac{B(z)}{\varepsilon_b(z)}\frac{\sigma_b^5}{\gamma_rG}.
\end{equation}
The redshift variation of $M_{BH}-\sigma_b$ correlation can be fully determined if the redshift variations of $B(z)$ and $\varepsilon_b(z)$ are known. 

For the local Universe, the value of $\varepsilon_b$ can be estimated from the well-established SMBH-host correlations and the virial theorem
\begin{equation} 
\label{eq:1-2} 
\begin{split}
&\frac{M_{BH}}{10^8M_{\odot}} = 1.66\left(\frac{\sigma_b}{200km/s}\right)^{4.86} \quad \textrm{\citep{Ferrarese:2005-Supermassive-Black-Holes-in-Galactic-Nuclei}}, \\
&M_{BH} \approx 0.002M_b \quad \textrm{\citep{Marconi:2003-The-relation-between-blac}}, \\
&M_b \approx 3r_b \sigma_b^2 /G \quad \textrm{\citep{Marconi:2003-The-relation-between-blac}}, 
\end{split}
\end{equation}
which leads to $\varepsilon_{b0}\equiv\varepsilon_b(z=0)\approx 10^{-4}m^2/s^3$ as an average value in the local Universe. Here, the proportional coefficient $\gamma_r=1/3$.

\begin{figure}
\includegraphics*[width=\columnwidth]{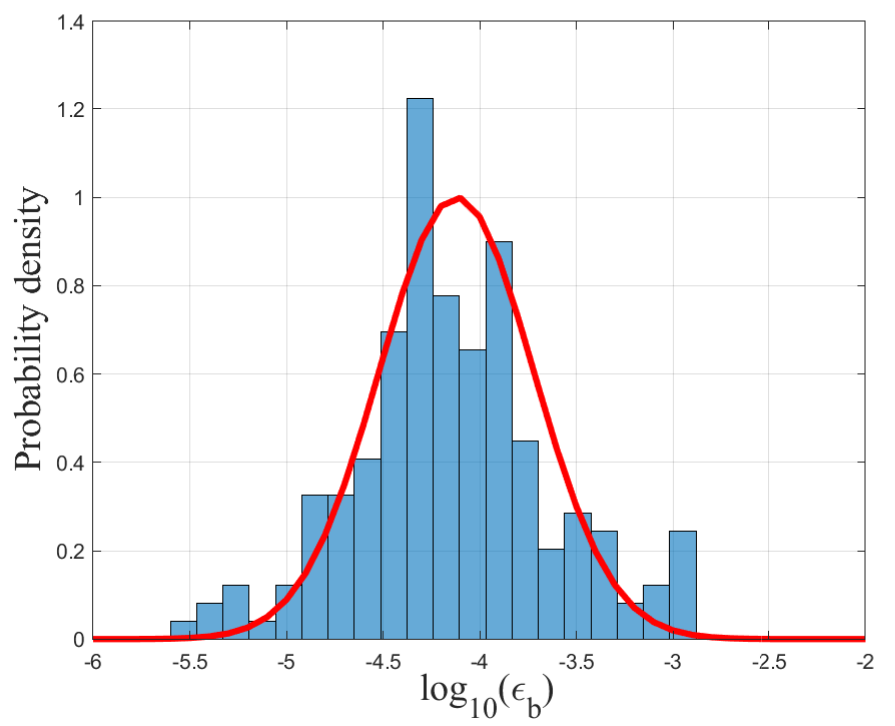}
\caption{The probability distribution of parameter $\varepsilon_b$ for 180 local galaxies listed in Table \ref{tab:A1}. The distribution is approximately log-normal with a mean rate $\left<\varepsilon_b\right>\approx 10^{-4}m^2/s^3$. This confirms the value obtained from the SMBH-host correlations in Eq. \eqref{eq:1-2}. The red solid line plots the best Gaussian fit in Eq. \eqref{eq:12-1}. The dispersion in $\varepsilon_b$ reflects the spatial intermittency (fluctuation) of $\varepsilon_b$ for different galaxies.} 
\label{fig:1-1}
\end{figure}

Alternatively, the parameter $\varepsilon_b=\sigma_b^3/r_b$ (Eq. \eqref{eq:2}) can also be calculated directly for each galaxy if the velocity dispersion $\sigma_b^2$ and size $r_b$ are known. Generally speaking, $\varepsilon_b$ can be different for individual galaxies. We compiled existing data, including the velocity dispersion ($\sigma_b^2$) and the size ($r_b$) for about 180 galaxies (Table \ref{tab:A1}). Figure \ref{fig:1-1} presents the distribution of $\varepsilon_b$ for all 180 galaxies with a mean value of $\left<\varepsilon_{b0}\right>\approx 10^{-4}m^2/s^3$. This confirms the results obtained from the SMBH-host correlations (Eq. \eqref{eq:1-2}). The solid red line in Fig. \ref{fig:1-1} plots the best log-normal fit that reads 
\begin{equation} 
\label{eq:12-1} 
f(\log_{10}(\varepsilon_{b0}))=\exp\left[-\frac{1}{2}\left(\frac{\log_{10}(\varepsilon_{b0})+4.12}{0.4}\right)\right].
\end{equation} 
In particular, the same value can also be obtained from cosmological N-body simulations (Fig. \ref{fig:S6} from Illustris simulations). 

In this paper, we propose that the parameter $\varepsilon_b$ is a key parameter of a cosmic quenching mechanism that regulates the evolution of the SMBH and the host (or "key-$\varepsilon$" theory for the sake of brevity). We focus on the evolution of $\varepsilon_b$ on the global cosmic scale rather than for each galaxy. Such a cosmic scale $\varepsilon_b$ in Eq. \eqref{eq:2} is not a coincidence. We can find support from many observations of SMBHs and host galaxies. However, there are still important questions that we naturally ask: What is the physical meaning of $\varepsilon_b$? What is the value and redshift evolution of $\varepsilon_b$? How does this key parameter impact the evolution of SMBHs and host galaxies? These are the key questions we will focus on. 

The remainder of this paper is organized as follows. Section \ref{sec:1-1-1} presents the basic concepts of the cosmic quenching mechanism and the physical meaning of the key parameter $\varepsilon_b$. Section \ref{sec:2-1-1} demonstrates the parameter $\varepsilon_b$ and its redshift evolution from Illustris simulations, followed by observational data in Section \ref{sec:2-1-2} for the scaling laws involving $\varepsilon_b$ that govern the evolution of host galaxies. Section \ref{sec:4} identifies the relevant length scales in the SMBH-bulge system based on these scaling laws. This provides a holistic view of the coevolution of the SMBH and host, which is then applied in Sections \ref{sec:4-2} to \ref{sec:7} to derive the upper and lower limits for the SMBH distribution and the three-phase evolution of SMBHs. To validate the evolution model, Sections \ref{sec:7-1} to \ref{sec:7-2} apply that model to analytically derive the BH mass function, the AGN mass function, the AGN duty cycle, and the Eddington ratio distributions and compare these analytical solutions with numerical solutions and observations. Section \ref{sec:9-2} introduces a new luminosity limit allowing for a super-Eddington growth at the early stage of SMBHs and compares it with the standard Eddington limit. Finally, Section \ref{sec:8} in the Appendix applies the evolution model to several observed high-redshift SMBHs to predict their complete redshift evolution, which can be potentially compared with high-resolution simulations.  

\section{The basic concepts of cosmic quenching}
\label{sec:1-1-1}
Since the key parameter $\varepsilon_b$ from SMBH-host correlations (Eq. \eqref{eq:2}) is the central quantity of cosmic quenching, in this section, we start from the physical meaning of $\varepsilon_b$ and introduce the basic physical picture and relevant assumptions. 
First, the radial gas flow in the bulge is generally required to continuously supply fresh gas to feed and maintain star formation \citep{Teodoro:2021-Radial-Motions-and-Radial-Gas-Flows,Trapp:Gas-infall-and-radial-transport-in-cosmological}. The mass flow rate is on the order of 1$M_{\odot}/yr$, which is comparable to the star formation rate. Along with the radial mass flow, there should also exist a continuous flow of energy carried by the random motion of gases. We postulate that parameter $\varepsilon_b$ quantifies the mass and energy flow in the radial direction of the bulge. 

\begin{figure}
\includegraphics*[width=\columnwidth]{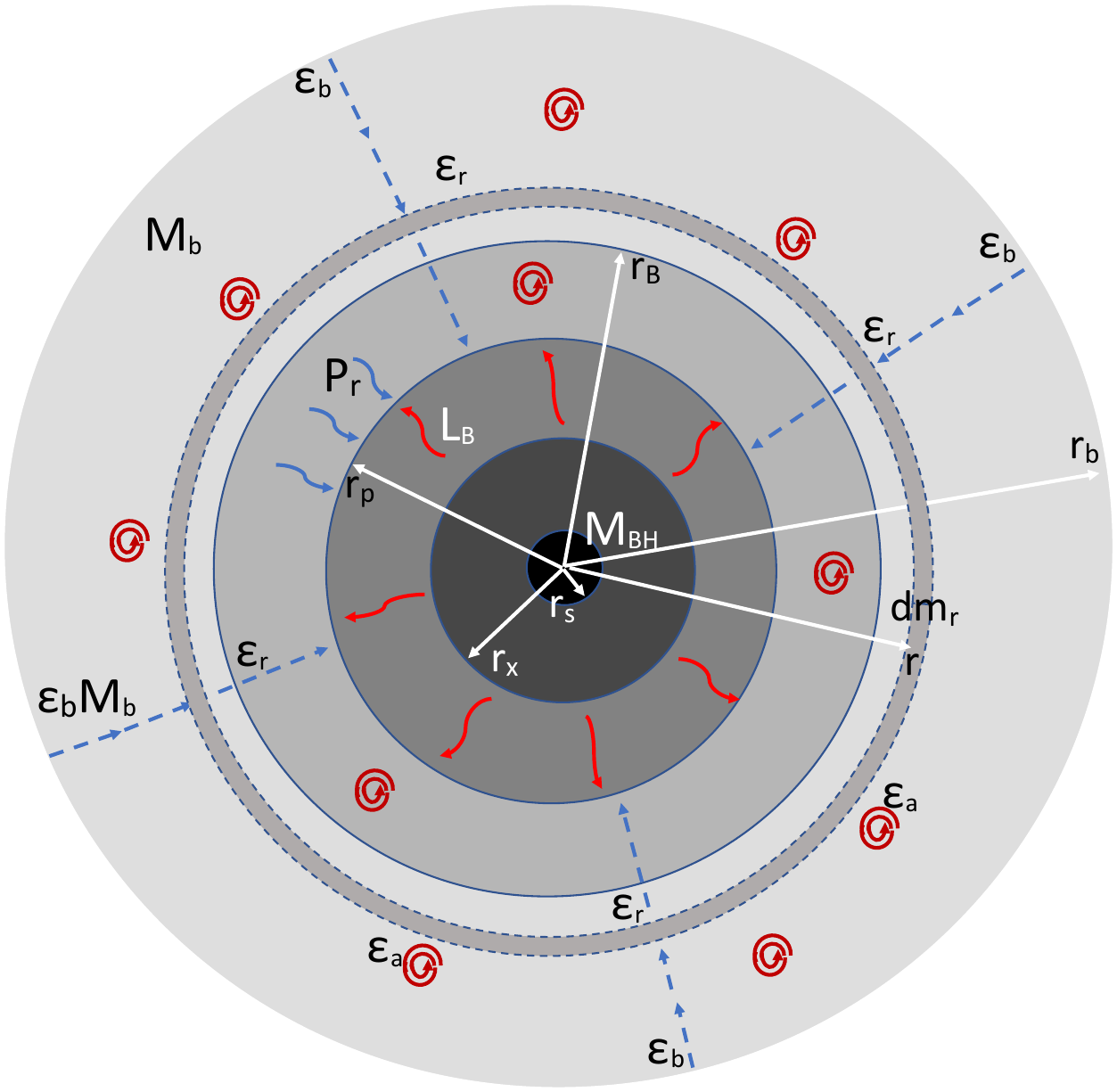}
\caption{Schematic plot of a SMBH-bulge system with a bulge mass $M_b$, BH mass $M_{BH}$, and BH bolometric luminosity $L_B$. On scale $r$, associated with the mass flow of $\dot{m}_r$, the energy flux is on the order of $\varepsilon_rm_r$ (Eq. \eqref{eq:2-2-5}). The mass and energy flow down to small scales that are controlled by the parameter $\varepsilon_r$ (Eq. \eqref{eq:2-2-5}). The red spirals represent the local baryonic energy dissipation at a rate of $\varepsilon_a$ that impacts the gas cooling and star formation. The energy balance in an infinitesimal mass $dm_r$ requires $\varepsilon_r=\varepsilon_a$ (Eq. \eqref{eq:3-1-2}). Five relevant length scales are also shown, that is, the bulge scale $r_b$ (Eq. \eqref{eq:15}), the BH sphere of influence $r_B$ (Eq. \eqref{eq:15}), the radiation scale $r_p$ (Eq. \eqref{eq:16}), the dissipation scale $r_x$ (Eq. \eqref{eq:17}), and the Schwarzschild radius $r_s$ (Eq. \eqref{eq:15}). With these key scales, the upper and lower limits of the distribution and evolution of SMBH can be developed (Figs. \ref{fig:3} and \ref{fig:4}). The BH radiation pressure due to $L_B$ must be balanced by the dynamic gas pressure $P_r$ due to the mass and energy flow (Eqs. \eqref{eq:10} and \eqref{eq:54a-2}) such that super-Eddington accretion is required at the early stage of SMBH evolution (Section \ref{sec:9-2}).} 
\label{fig:1}
\end{figure}

To understand this, as shown in Fig. \ref{fig:1}, we consider a bulge suffused by gases of highly varying thermal states ranging from cold gas at small scales $r$ to warm and hot gases at large scales $r$. The temperature of the gas can be characterized by a random velocity dispersion $\sigma_r^2$ at scale $r$ (or specific kinetic energy). An example of the variation of gas temperature with scale $r$ is shown in Fig. \ref{fig:S4}, which is obtained from the Illustris simulation. 

Next, we focus on the mass and energy flow in the radial direction and the energy dissipation in gases, which govern the energy budget of gases suffused in the bulge. We introduce a parameter $\varepsilon_r$, similarly to the parameter $\varepsilon_b$ in Eq. \eqref{eq:2} but is defined on a given scale $r$,
\begin{equation} 
\label{eq:2-2-2} 
\varepsilon_r=\frac{\sigma_r^3}{r}=\frac{\sigma_r^2}{r/\sigma_r}=\frac{\sigma_r^2}{t_r},
\end{equation}
where $\sigma_r^2$ is the velocity dispersion of gas on scale $r$, while $t_r=r/\sigma_r$ is the typical time for gas particles to travel from $r$ to the center of bulge. The velocity $\sigma_r$ vanishes at the center of the bulge with $r\to 0$. With changes in the specific energy from $\sigma_r^2$ to zero during time $t_r$, the parameter $\varepsilon_r$ represents the average rate of change in the specific energy of the gas particles. Obviously, the key parameter $\varepsilon_b\equiv \varepsilon_r(r=r_b)$ is the value of $\varepsilon_r$ on the bulge scale $r_b$. 

To facilitate the discussion, we need to introduce two important hypotheses: i) the parameter $\varepsilon_r$ is independent of the scale $r$ such that $\varepsilon_b\equiv \varepsilon_r$, i.e., $\varepsilon_r$ is scale-independent; ii) the gas cooling time is less than the Hubble time but greater than the free-fall time. The system is in a quasi-static equilibrium such that gases in the bulge are always in a near virial equilibrium at any scale $r$ and at any moment $t$. Of course, these hypotheses must be tested and validated by both simulations and observations, as we demonstrate in Sections \ref{sec:2-1-1} and \ref{sec:2-1-2} (Figs. \ref{fig:S4} and \ref{fig:S5}). At this point, we simply take these hypotheses.

From the first hypothesis, the parameter $\varepsilon_r$ is independent of the radial scale $r$ such that
\begin{equation} 
\label{eq:2-2-3} 
\varepsilon_r=\frac{\sigma_r^3}{r}=\frac{\sigma_r^2}{t_r}=\varepsilon_r\frac{dt_r}{dt_r}=\frac{d(\varepsilon_r t_r)}{dt_r}=\frac{d\sigma_r^2}{dt_r}.
\end{equation}
From the second hypothesis, the virial equilibrium of gases requires $\sigma_r^2\propto G m_r/r$, where $m_r$ is the total mass of all cosmic components contained in all scales below $r$. Combining the virial equilibrium with Eq. \eqref{eq:2-2-3} leads to relations
\begin{equation} 
\label{eq:2-2-4} 
m_r \propto G^{-1}\varepsilon_r^{-1}\sigma_r^5 \quad \textrm{and} \quad \frac{d\sigma_r^2}{dm_r}=\frac{2}{5}\frac{\sigma_r^2}{m_r}.
\end{equation}
To illustrate the meaning of $\varepsilon_r$, we write (from Eqs. \eqref{eq:2-2-3} and \eqref{eq:2-2-4})
\begin{equation} 
\label{eq:2-2-5} 
\varepsilon_r=\frac{d\sigma_r^2}{dt_r}=\frac{2}{5}\frac{\sigma_r^2}{m_r}\frac{dm_r}{dt_r} \quad \textrm{and} \quad \varepsilon_rm_r=\frac{2}{5}\frac{dm_r}{dt_r}\sigma_r^2=\frac{2}{5}\dot{m}_r\sigma_r^2,
\end{equation}
where $\dot{m}_r=dm_r/dt_r$ is the flux of mass flow at scale $r$, i.e. during the infinitesimal time $dt_r$, an infinitesimal mass of $dm_r$ is passed from scales above $r$ to scales below $r$. Simultaneously, an energy flow is associated with the mass flow because the kinetic energy $\sigma_r^2$ is carried by the infinitesimal mass $dm_r$. Therefore, from Eq. \eqref{eq:2-2-5}, the product $\varepsilon_rm_r$ describes the energy flux on scale $r$ (Fig. \ref{fig:1}), that is, the infinitesimal energy of ($dm_r\cdot\sigma_r^2$) is transferred across the scale $r$ during infinitesimal time $dt_r$. The mass flux $\dot{m}_r$ and the energy flux $\varepsilon_rm_r$ are clearly related to the parameter $\varepsilon_r$. From Eq. \eqref{eq:2-2-5}, $\varepsilon_r$ also describes the flux of the specific energy across the scale $r$, i.e., the specific energy on the scale $r$ changes by an infinitesimal amount of $\sigma_r^2dm_r/m_r$ over time $dt_r$. 

Since the rate of energy flow into the sphere of size $r$ can be written as the product $\varepsilon_r m_r$ (unit: J/s) (Eq. \eqref{eq:2-2-5}), for a spherical shell of a differential mass $dm_r$ in Fig. \ref{fig:1}, the energy conservation requires
\begin{equation} 
\label{eq:3-1-2}
d(\varepsilon_r m_r)=\varepsilon_a {dm_r},
\end{equation}
where $m_r$ is the total mass enclosed within scale $r$. Term $d(\varepsilon_r m_r)$ represents the net energy flux due to the difference between the flux into and the flux out of the mass shell $dm_r$. The energy is dissipated at a rate of $\varepsilon_a$ (unit: m$^2$/s$^3$) in the same shell $dm_r$. In principle, the net accumulation of the energy due to the energy flux in and out must balance the energy dissipated ($\varepsilon_a {dm_r}$ due to baryonic dissipation) 
to respect the energy conservation in the mass shell $dm_r$.
Since $\varepsilon_r$ is independent of scale $r$ (and therefore of the enclosed mass $m_r$), Eq. \eqref{eq:3-1-2} implies that $\varepsilon_b=\varepsilon_r=\varepsilon_a$, that is, the rate of energy flow should be equal to the rate of energy dissipation. Therefore, the gas suffused in the bulge is self-regulated in a way that the net energy accumulated on any scale $r$ due to the energy flow always balances the energy dissipated on the same scale. Both rates $\varepsilon_b$ and $\varepsilon_a$ are independent of the scale $r$. Since $\varepsilon_r=\varepsilon_a$, the time $t_r$ in Eq. \eqref{eq:2-2-3} can be written as $t_r=\sigma_r^2/\varepsilon_a$ that represents the cooling time for a parcel of gases on scale $r$ to lose its thermal energy $\sigma_r^2$ at a rate of $\varepsilon_a$. 

The rate of energy dissipation $\varepsilon_a$ reflects the cooling rate per unit mass (or specific cooling rate) that is directly related to the gas cooling efficiency. The primary gas cooling processes are the two-body radiative processes, where gas loses energy through the emission of photons as a result of two-body interactions. By assuming the radiative cooling as the dominant cooling process responsible for the energy dissipation, we arrive at the third hypothesis: iii) the rate of energy flow $\varepsilon_b$ balances the rate of energy dissipation $\varepsilon_a$. Both rates can be related to the gas cooling function $\Gamma$ (in the average sense),
\begin{equation} 
\label{eq:3-1-5}
\varepsilon_b \equiv \varepsilon_r =\varepsilon_a = {\Gamma(T,Z) n_H}/{m_p},
\end{equation}
where $n_H$ is the number density of gas particles, $m_p$ is the mass of hydrogen atom, $T$ is the temperature of gas, and $Z$ is the metallicity of gas. The value of $\varepsilon_b=10^{-4}m^2/s^3$ at redshift $z=0$ is equivalent to an average gas cooling function on the order of $10^{-24} $erg s$^{-1}$cm$^3$ for a number density $n_H$ of one or two particles per cubic centimeter. Therefore, the key parameter $\varepsilon_b$ also describes how quickly gas can cool and condense and impacts the rate of star formation.

Finally, in this section, we focus on the mass and energy flow in a near-equilibrium bulge suffused by gases of different temperatures on different scales $r$. Based on three assumptions: i) the rate of energy flow is independent of the scale $r$; ii) the permeated gas is in virial equilibrium; and iii) the energy flow balances the energy dissipation in gas, we found that the key parameter $\varepsilon_b$ quantifies the mass and energy flow in gases and also reflects the efficiency of gas cooling and the supply of cold gas. The same parameter impacts the evolution of SMBHs through the energy and mass flow in the bulge and the star formation through the energy dissipation in gases. This is the parameter that synchronizes the evolution of both SMBHs and their hosts (Fig. \ref{fig:S37}). Since parameter $\varepsilon_b\propto (1+z)^{5/2}$ (Eq. \eqref{ZEqnNum9863112994}) that decreases rapidly with time, a larger $\varepsilon_b$ (and $\varepsilon_a$) in the early Universe means more efficient gas cooling and a richer supply of cold gas for faster structure evolution. At lower redshifts, a smaller $\varepsilon_b$ means less efficient gas cooling, less cold gas supply, and slower star formation and SMBH growth. Therefore, the rapid decrease in $\varepsilon_b$ represents a global quenching process on the cosmic scale that slows down the structure formation and evolution (i.e., a cosmic quenching).  

This relatively simple theory, characterized by a single parameter $\varepsilon_b$, neglects all the transient phenomena, such as the massive mass accretion and merging, the merging-induced disruptions, the bar formation/evolution, and any transient AGN jets and winds. When properly calibrated by simulations and observations, this simple theory provides the dominant mean cosmic evolution of SMBHs and host galaxies. At the same time, all transient phenomena only contribute to the dispersion around these mean cosmic evolutions. With this in mind, the scaling laws involving $\varepsilon_b$ will be developed for the mean evolution of SMBHs and host galaxies. Based on these scaling laws, the evolution of the BH mass function, the AGN mass function, and the AGN duty cycle can all be derived analytically. In the remainder of this paper, these results are presented and compared with simulations and observations. It should be noted that the same concept can also be applied to dark matter haloes, where the energy flow in haloes dominates the halo internal structures and may reveal the relevant dark matter properties \citep{Xu:2023-Universal-scaling-laws-and-density-slope,Xu:2023-Dark-matter-halo-mass-functions-and, Xu:2021-Inverse-mass-cascade-mass-function, Xu:2022-Postulating-dark-matter-partic}.

\section{Bulge dynamics from Illustris simulations}
\label{sec:2-1-1}
In this section, we use large-scale cosmological simulations to illustrate the energy flow in the bulge and quantify the value of $\varepsilon_b$ that is related to the rate of energy flow, the efficiency of gas cooling, and the supply of cold gas. Illustris is a suite of large-volume DM-only cosmological simulations (Illustris-1-Dark) and hydrodynamical simulations (Illustris-1) \citep{NELSON:2015-The-illustris-simulation}. The selected Illustris-1-Dark is a suite of DM-only simulations of a 106.5Mpc$^3$ cosmological volume with 1820$^3$ DM particles. Each DM particle has a mass around $7.6\times 10^6 M_{\odot}$. The gravitational softening length is around 1.4 kpc. The simulation has cosmological parameters of a total matter density $\Omega_m=0.2726$, a dark energy density $\Omega_{DE}=0.7274$ at $z=0$, and a Hubble constant $h=0.704$. 

The selected Illustris-1 is a suite of hydrodynamic simulations of the same volume, dark energy density, and Hubble constant. For standard $\Lambda$CDM cosmology, Illustris-1 includes dark matter, stars, gas, black holes, and dark energy with a baryonic matter density of $\Omega_b=0.0456$. Each gas particle has a mass around $1.3\times 10^6 M_{\odot}$, and the DM particle has a mass of $6.3\times 10^6 M_{\odot}$. The gravitational softening length for baryons is around 0.7 kpc. More details on baryonic and BH physics models can be found in \cite{NELSON:2015-The-illustris-simulation}. Dark matter haloes were identified using a standard Friends of Friends (FoF) algorithm with the link length parameter $b=0.2$. The center of the halo is placed at a minimum of the gravitational potential of the entire halo. Illustris-1 is used to study the dynamics in bulges that involve complex baryonic physics, while the Illustris-1-Dark (DM-only) simulation provides a reference for comparison. The halo mass $m_h^*$ defines a characteristic mass scale in halo mass functions. Haloes greater than $m_h^*$ are rare. The evolution of $m_h^*$ follows an approximate scaling $\propto a^{3/2}$ in the matter-dominant era. More details on the evolution of $m_h^*$ can be found in \citep{Xu:2023-Dark-matter-halo-mass-functions-and,Xu:2022-Postulating-dark-matter-partic}. In this work, we focus mainly on the bulge dynamics in haloes of a characteristic mass $m_h^*(z)$ to provide representative evolution dynamics.

To study the bulge's dynamics, we introduce the cumulative function of the mass of different cosmic components (dark matter, gas, stars, BHs, etc.). The cumulative mass function $\Lambda^i_m(m_h,r,z)$ represents the total mass of a given component enclosed in a sphere of size $r$ that centers around the center of the halo. This mass is averaged for all haloes of the same mass $m_h$ such that
\begin{equation} 
\label{ZEqnNum98631129} 
\Lambda^i_m(m_h,r,z) = \int _{0}^{r} \rho^i_h \left(m_h,r',z) \right)4\pi r'^2 dr'.  
\end{equation} 
where $\rho^i_h$ is the mean mass density of a given component for all haloes of the same mass $m_h$ ($i$=d for dark matter, g for gas, s for stars, and BH for black holes, respectively). 

\begin{figure}
\includegraphics*[width=\columnwidth]{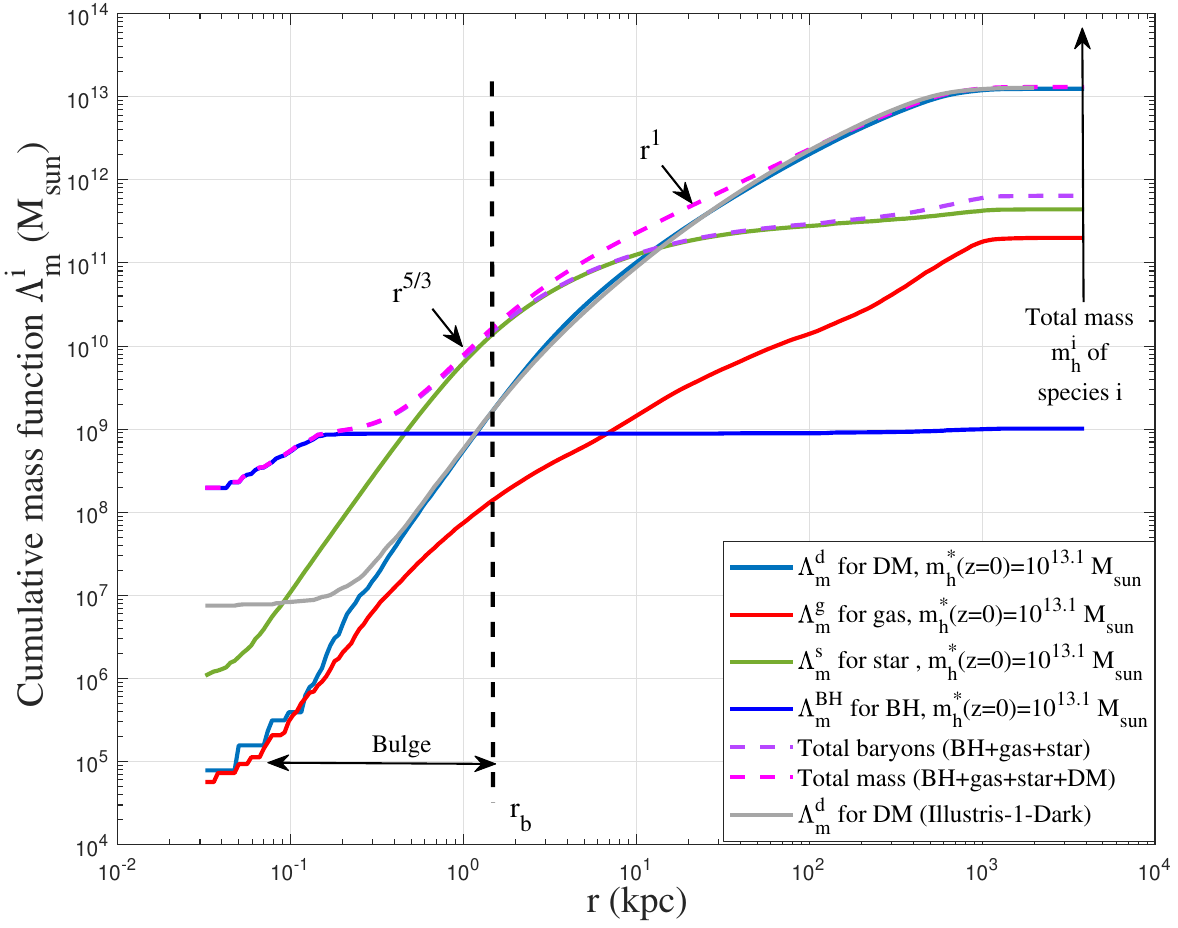}
\caption{The variation of cumulative mass function $\Lambda^i_m(m_h,r,z)$ with the length scale $r$ for all haloes of the characteristic mass $m_h^*$=$10^{13.1}M_{\odot}$ at redshift $z=0$. The dark matter mass dominates baryonic matter on scales $r$>10kpc. In the bulge, the baryonic matter is dominant, with star mass much greater than the mass of gas in this particular size of haloes. The total mass of all comic components follows a 5/3 law ($\propto r^{5/3}$) in the bulge that is consistent with the prediction in Eq. \eqref{ZEqnNum98631129992}. The total mass is $\propto r$ outside the bulge, consistent with a flat velocity in the outer region (Fig. \ref{fig:S4}). The dashed line denotes the approximate size of the bulge $r_b$.} 
\label{fig:S2}
\end{figure}

\begin{figure}
\includegraphics*[width=\columnwidth]{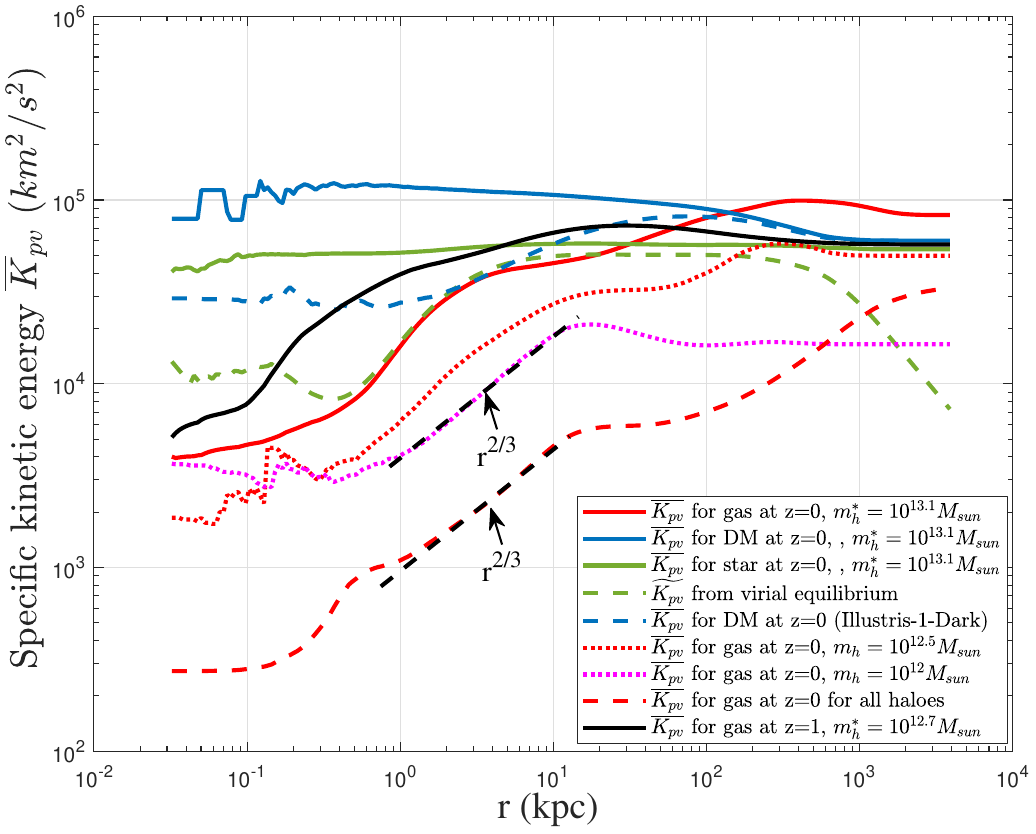}
\caption{The mean specific kinetic energy $\overline K_{pv}$ of different cosmic components varying with scale $r$ for haloes of mass $m_h^*$. Because of their collisionless nature, dark matter, and stars are hotter in the bulge and have higher specific kinetic energy. The gas in the bulge is colder, with lower kinetic energy due to energy dissipation. Three components have comparable kinetic energy outside the bulge with an almost flat velocity consistent with the mass scaling $\propto r$ in Fig. \ref{fig:S2}. Compared to the kinetic energy calculated from the virial theorem (green dashed line from Eq. \eqref{ZEqnNum9863112999}), only gas velocity satisfies the virial theorem in the bulge. The 2/3 scaling for gas temperature or specific kinetic energy $T_g\propto \overline K_{pv} \propto r^{2/3}$ is consistent with the mass scaling $\propto r^{5/3}$ in Fig. \ref{fig:S2} (Eq. \eqref{ZEqnNum98631129992}). This 2/3 scaling exists for all haloes of different sizes. An important parameter $\varepsilon$ can be introduced based on the 2/3 scaling (Eq. \eqref{ZEqnNum98631129991}) and presented in Fig. \ref{fig:S5}. For comparison, the same quantity $\overline K_{pv}$ from the DM-only simulation is also plotted (blue dashed line). Due to the gravitational interaction with baryons (baryonic feedback), dark matter is hotter (heated) with a higher kinetic energy in the bulge than in DM-only simulations.} 
\label{fig:S4}
\end{figure}

Figure \ref{fig:S2} plots the variation of the cumulative mass function $\Lambda^i_m(m_h,r,z=0)$ with scale $r$ for all haloes of the same mass between $m_h^*=10^{13.1\pm0.1}M_{\odot}$, where $m_h^*$ is the characteristic halo mass. 
The total mass of the component $i$ ($m_h^i$) in haloes of mass $m_h$ can be obtained by setting $r \rightarrow \infty$ in Eq. \eqref{ZEqnNum98631129}, i.e. $m_h^i(m_h,z)=\Lambda_m^i(m_h,r\rightarrow \infty,z)$. For haloes of characteristic mass $m_h^*$, the dark matter mass dominates baryonic matter on scales $r$>10kpc. The total mass of baryonic components (BH + gas + stars) dominates the dark matter mass in the bulge. The total mass of all components $\propto r^{5/3}$ in the bulge that can be predicted by the scaling laws in Eq. \eqref{ZEqnNum98631129992} that involves $\varepsilon_b$. While on scales >10kpc, the total mass is $\propto r$, which leads to a flat velocity in Fig. \ref{fig:S4} according to the virial theorem. In bulges, the mass of stars dominates over the mass of gas and dark matter for this size of haloes. 

To better describe the bulge dynamics, we decompose the kinetic energy into two parts of a different nature. In N-body simulations, every particle of cosmic component $i$ has a velocity vector $\boldsymbol{\mathrm{v}}_{\boldsymbol{\mathrm{p}}}$. The velocity vector $\boldsymbol{\mathrm{v}}_{p}$ can be decomposed as \citep{Xu:2023-Maximum-entropy-distributions-of-dark-matter}
\begin{equation} 
\label{ZEqnNum502045} 
\boldsymbol{\mathrm{v}}_{p} =\boldsymbol{\mathrm{v}}_{h} +\boldsymbol{\mathrm{v}}_p',           
\end{equation} 
namely, the halo mean velocity, $\boldsymbol{\mathrm{v}}_{h}=\langle \boldsymbol{\mathrm{v}}_{p} \rangle_h$, and the velocity fluctuation, $\boldsymbol{\mathrm{v}}_{p}^{'} $. Here, $\boldsymbol{\mathrm{v}}_{h}$ represents the velocity of that halo, that is, the average velocity of all components. In the simulation, it is calculated as the sum of the mass-weighted velocities of all particles of different components in the same halo. Consequently, a given particle's total kinetic energy $K_p$ can be divided into $K_p = K_{ph}+K_{pv}$. Here $K_{ph}=\boldsymbol{\mathrm{v}}_{h}^2/2$ (halo kinetic energy) is the contribution from the motion of entire haloes due to the inter-halo interaction of that particle with all other particles outside that halo \citep{Xu:2022-Postulating-dark-matter-partic}. This part of the kinetic energy is related to interactions on large scales in the linear regime. The other part, $K_{pv}={\boldsymbol{\mathrm{v}}_p'}^2/2$ (the virial kinetic energy), is the contribution of the velocity fluctuation $\boldsymbol{\mathrm{v}}_p'$ due to the intra-halo interaction of that particle with all other particles in the same halo. This part of the kinetic energy is due to interactions on a shorter distance and smaller scales in the non-linear regime \citep{Xu:2022-Postulating-dark-matter-partic}. Since only the velocity fluctuation is relevant to the bulge dynamics, we focus on the kinetic energy $K_{pv}$. Similarly, we introduce a cumulative function $\Lambda_{pv}^i$ for $K_{pv}$

\begin{equation} 
\label{ZEqnNum986311299} 
\begin{split}
&\Lambda^i_{pv}(m_h,r,z) = \int _{0}^{r} K_{pv} \rho^i_h \left(m_h,r',z) \right)4\pi r'^2 dr',  \\
& \overline {K_{pv}^i}(m_h,r,z) = \frac{\Lambda^i_{pv}}{\Lambda^i_{m}}=\frac{\int _{0}^{r} K_{pv} \rho^i_h \left(m_h,r',z) \right)4\pi r'^2 dr'}{\int _{0}^{r} \rho^i_h \left(m_h,r',z) \right)4\pi r'^2 dr'}.
\end{split}
\end{equation} 
where $K_{pv} = {{v_p'}^2}/2$ is the specific kinetic energy of any cosmic component $i$ due to velocity fluctuation $v_p'$. The cumulative function $\Lambda^i_{pv}$ represents the total kinetic energy contained in the scale $r$. While $\overline {K_{pv}^i}$ is the specific energy (energy per unit mass) contained in the sphere of size $r$, an important quantity to determine $\varepsilon_b$ in Eq. \eqref{ZEqnNum98631129991}.  

Figure \ref{fig:S4} plots the variation of the specific kinetic energy $\overline {K_{pv}}$ with the scale $r$ for different cosmic components in haloes of different sizes. This is an important figure with much information:\\
\newline
\noindent i) In bulge, dark matter and stars are hotter with a higher specific kinetic energy than gas due to their collisionless nature. When compared to the dark matter-only simulation (Illustris-1-Dark), the gravitational interaction between dark matter and baryons seems to contribute to a hotter dark matter in the bulge. \\
\newline
\noindent ii) In the bulge, gas is colder, with decreasing kinetic energy over time due to energy dissipation mainly through radiative cooling. The gas velocity satisfies the virial theorem on all scales $r\le r_b$. This can be demonstrated by comparing $\overline {K_{pv}}$ for gas (solid red line) with the kinetic energy expected from the virial theorem (green dashed line), 
\begin{equation} 
\label{ZEqnNum9863112999} 
{\widetilde{K_{pv}}} = \frac{G\Sigma (\Lambda^i_m)}{2r} \approx \frac{Gm_r}{2r},
\end{equation} 
where $\Sigma(\Lambda^i_m)$ is the total mass of all cosmic components within scale $r$ (see Fig. \ref{fig:S2}). The velocity of other components (solid blue for dark matter and solid green for stars) does not satisfy the virial theorem in bugle due to their collisionless nature. In the bulge, the total mass is dominated by baryons such that $m_r\approx \Sigma\Lambda^i_m$, where $m_r$ is the baryonic mass contained in the scale $r$. This validates the Hypothesis ii) in Section \ref{sec:1-1-1}).
\newline

\begin{figure}
\includegraphics*[width=\columnwidth]{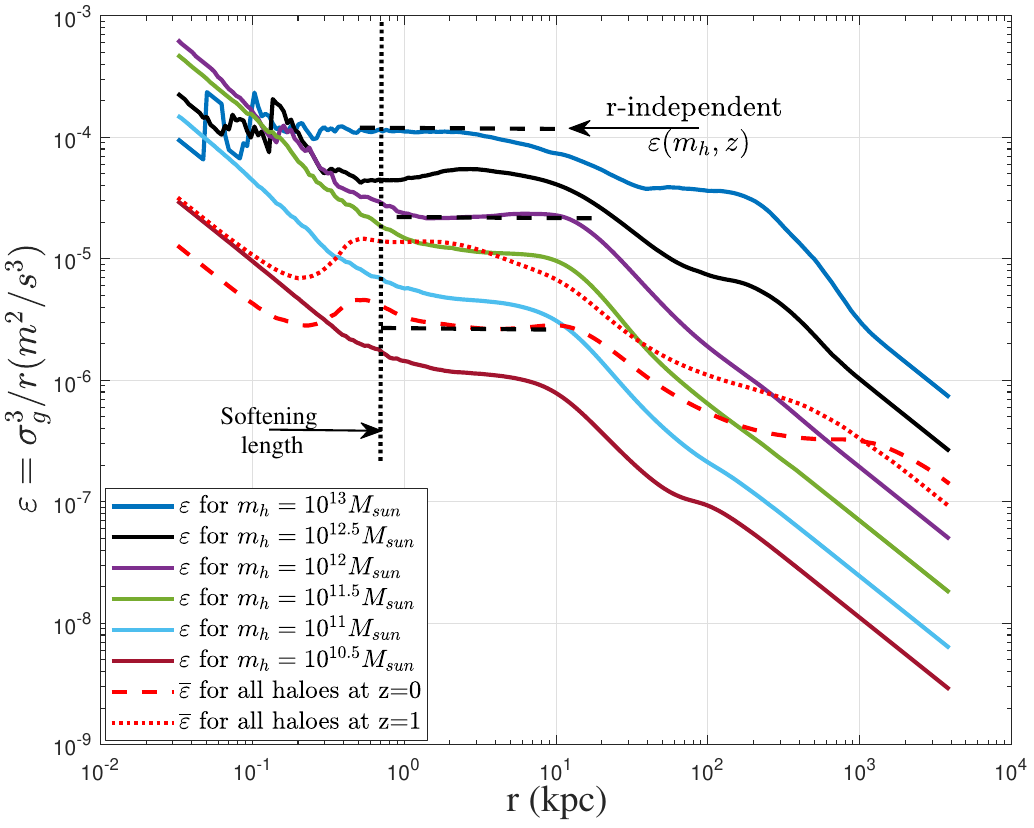}
\caption{The variation of key parameter $\varepsilon$ with the scale $r$ for haloes of mass $m_h$ at redshift $z=0$ using the data from Fig. \ref{fig:S4} and Eq. \eqref{ZEqnNum98631129991}. In the bulge, $\varepsilon$ is independent of the scale $r$ and increases with the mass of the halo. The variation of $\varepsilon$ with the halo mass is presented in Fig. \ref{fig:S6}. The red dashed lines ($z=0$) and the dotted lines ($z=1$) show the variation of $\overline \varepsilon$ obtained using the kinetic energy calculated for the composite halo built from all haloes at the same redshift (red dashed line in Fig. \ref{fig:S4}). Clearly, $\varepsilon$ is larger at a higher redshift. The variation of $\varepsilon$ with $z$ is shown in Fig. \ref{fig:S7}.}
\label{fig:S5}
\end{figure}

\noindent iii) More importantly, the 2/3 scaling ($\overline {K_{pv}} \propto r^{2/3}$) exists for gas in the bulge of haloes of different masses $m_h$. All haloes of different masses can also be stacked together to form a composite halo. The specific kinetic energy $\overline {K_{pv}}$ can be calculated similarly for that composite halo, which also exhibits the same scaling (red dashed line). This allows us to introduce a parameter $\varepsilon$, 
\begin{equation} 
\label{ZEqnNum98631129991} 
\varepsilon(m_h,z) = \frac{\sigma_g^3}{r}=\frac{\sigma_g^2}{r/\sigma_g}=\frac{\sigma_b^3}{r_b}, \quad \overline {K_{pv}} = \frac{1}{2}\sigma_g^2=\frac{1}{2} (\varepsilon r)^{2/3}.
\end{equation} 
Here, $\sigma_g^2(r)$ is the velocity dispersion of gas on the scale $r$, while $\sigma_b^2=\sigma_g^2(r\equiv r_b)$ is the velocity dispersion on the bulge scale $r_b$. This equation is the same as Eq. \eqref{eq:2} obtained independently from the SMBH-host correlations. The parameter $\varepsilon$ is independent of the scale $r$, which is relevant to the rate of energy flow from large to small $r$ and the rate of energy dissipation in gas (Section \ref{sec:1-1-1}). Since the 2/3 scaling in Eq. \eqref{ZEqnNum98631129991} is valid on all scales below the bulge size $r_b$, the parameter $\varepsilon$ describes an r-independent rate of energy flow in the radial direction that is associated with the mass flow. The cosmic evolution of this rate $\varepsilon$ reflects the supply of cold gas and regulates the evolution of host galaxies and SMBHs (Section \ref{sec:1-1-1}). 

Figure \ref{fig:S5} plots the variation of the parameter $\varepsilon(m_h,z)$ using Eq. \eqref{ZEqnNum98631129991} and the kinetic energy $\overline {K_{pv}}$ in Fig. \ref{fig:S4}. This figure shows a r-independent parameter $\varepsilon$ below the bulge size $r_b$. The key parameter $\varepsilon$ increases with the halo mass $m_h$ and the redshift $z$. To estimate the values of $\varepsilon(m_h,z)$, we require a sufficient number of DM, stars, and gas particles in bulges and haloes and a sufficient number of haloes at a given mass $m_h$ for reliable statistics. This figure confirms the hypothesis that the rate of energy flow $\varepsilon_b$ is independent of the scale $r$, i.e., the Hypothesis i) in Section \ref{sec:1-1-1}. The r-independent $\varepsilon_b$ can be a crucial feature when the mass and energy flow establishes a statistically steady state. If this is not the case, there would be a net accumulation of energy on some intermediate scale $r$. This should be excluded if the statistical structures in the flow are self-similar and scale-free on scales $r\le r_b$. When such a statistically steady state is established, the flow fields on these scales are statistically similar. The fast motion on small scales does not feel the slow motion on large scales directly, except through $\varepsilon$. 
\newline

\begin{figure}
\includegraphics*[width=\columnwidth]{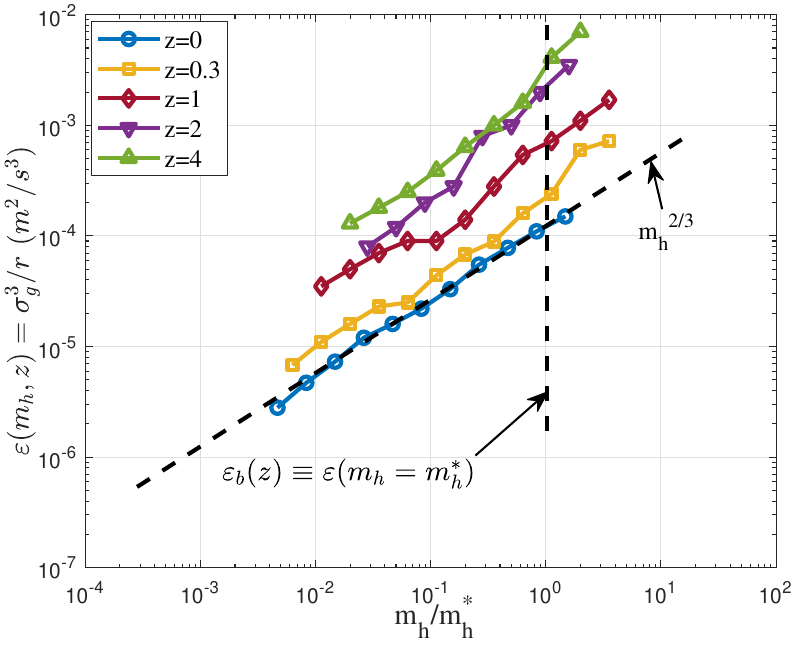}
\caption{The variation of $\varepsilon$ with halo mass $m_h$ at different redshifts $z$ using data from Fig. \ref{fig:S5}. The figure demonstrates that parameter $\varepsilon$ increases with halo mass as $\varepsilon \propto m_h^{2/3}$ and increases with redshift $z$ such that $\varepsilon$ is much greater in the early Universe. The black dashed line points to the parameter $\varepsilon_b(z)\equiv \varepsilon(m_h=m_h^*,z)$ for haloes with a characteristic mass $m_h^*(z)$ in Fig. \ref{fig:S2}. The redshift variation of $\varepsilon_b$ is presented in Fig. \ref{fig:S7}.} 
\label{fig:S6}
\end{figure}

\noindent iv) In Fig. \ref{fig:S4}, comparing the kinetic energy $\overline {K_{pv}}$ for bulges in haloes of mass $m_h^*(z)$ at two different redshifts $z=0$ (solid red line) and $z=1$ (solid black line), the bulge size $r_b$ roughly increases with time, while the velocity dispersion $\sigma_b^2$ decreases with time. The bulges expand over time with decreasing specific kinetic energy (Eq. \eqref{eq:13}).

Using the data in Fig. \ref{fig:S5}, Figure \ref{fig:S6} plots the variation of $\varepsilon(m_h,z)$ with the halo mass $m_h$ and the redshift $z$. The figure shows that $\varepsilon \propto m_h^{2/3}$ and increases with redshift $z$. We can write 
\begin{equation} 
\label{ZEqnNum9863112993} 
\varepsilon(m_h,z) = \varepsilon_b(z) \nu = \varepsilon_b(z) \left({m_h}/{m_h^*}\right)^{2/3}, 
\end{equation} 
where $\varepsilon_b(z)\equiv \varepsilon(m_h^*,z)$ is the rate of energy flow in haloes of mass $m_h^*$. The parameter $\nu$ is defined as $\nu = ({m_h}/{m_h^*})^{2/3}$ \citep{Xu:2023-Dark-matter-halo-mass-functions-and}. Since $\varepsilon$ is relevant to the rate of energy dissipation in gas, dwarf galaxies generally have small $\varepsilon$ and low star formation efficiency. 

\begin{figure}
\includegraphics*[width=\columnwidth]{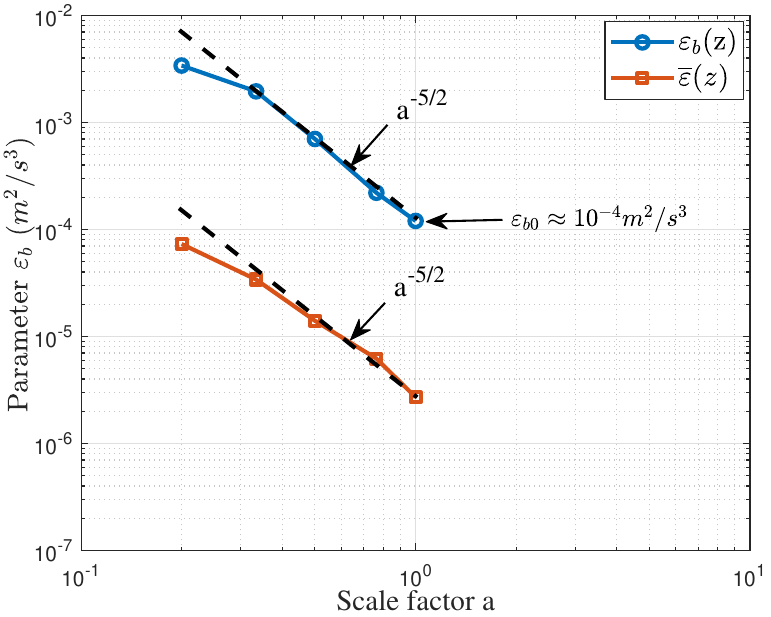}
\caption{The variation of $\varepsilon_b$ with scale factor $a$ from the Illustris-1 simulation using data from Fig. \ref{fig:S6}. An important relation $\varepsilon_b \propto a^{-5/2}$ can be found for haloes with a characteristic mass $m_h^*$. The figure demonstrates that $\varepsilon_b$ is much larger in the early Universe, which means a faster energy flow, more efficient gas cooling, and a richer supply of cold gas. The fast decreasing $\varepsilon_b$ leads to a rapid cosmic quenching over time. For comparison, the variation of mean rate $\overline \varepsilon$ (see Fig. \ref{fig:S5}) obtained for composite haloes (including all haloes at any given $z$) is also presented with the same scaling.} 
\label{fig:S7}
\end{figure}

We are interested in the dynamics of the bulge in haloes with a characteristic mass $m_h^*$ that is representative. Therefore, we will focus on the evolution of $\varepsilon_b(z)\equiv \varepsilon(m_h^*,z)$. Using the data in Fig. \ref{fig:S6}, Figure \ref{fig:S7} shows the variation of $\varepsilon_b(z)$
\begin{equation} 
\label{ZEqnNum9863112994} 
\varepsilon_b(z) = \varepsilon_{b0} a^{-5/2},
\end{equation} 
where $\varepsilon_{b0}\equiv \varepsilon_b(z=0) \approx 10^{-4}m^2/s^3$. Again, this value obtained from the Illustris simulation is the same as the value we obtained from the SMBH-host correlations in the local Universe (Eq. \eqref{eq:1-2}) and local galaxies (Fig. \ref{fig:1-1}). This is not a surprise if both simulations and correlations are grounded in the same underlying physical principles. The deviation at high redshift can be due to the limited mass resolution in hydrodynamic simulations. The variation of the average rate $\overline \varepsilon$ (see Fig. \ref{fig:S5}) obtained for composite haloes (including all dark matter haloes of all masses at a given $z$) is also presented. The same scaling is also found for $\overline \varepsilon$. 
\newline

\noindent v) Finally, since the gas satisfies the virial equilibrium (see Eq. \eqref{ZEqnNum9863112999} and Fig. \ref{fig:S4}), a 5/3 scaling can be obtained for the mass-size relation (combining Eqs. \eqref{ZEqnNum9863112999} and \eqref{ZEqnNum98631129991}). For haloes of mass $m_h^*$, these key relations are (2/3 law for kinetic energy and 5/3 law for mass) 
\begin{equation} 
\label{ZEqnNum98631129992} 
\begin{split}
&\sigma_g(r)^2 \propto \varepsilon_b^{2/3} r^{2/3},\quad \textrm{and} \quad \sigma_b^2 \propto \varepsilon_b^{2/3} r_b^{2/3},\\
&m_r \propto \varepsilon_b^{2/3} G^{-1} r^{5/3}, \quad \textrm{and} \quad M_b=\alpha_r\varepsilon_b^{2/3} G^{-1} r_b^{5/3}, \\
&\sigma_g(r)^5 \propto \varepsilon_b G m_r, \quad \textrm{and} \quad \sigma_b^5\propto \varepsilon_b G M_b,\\
&\varepsilon_bm_r \propto \sigma_g(r)^5 G^{-1}, \quad \textrm{and} \quad \varepsilon_b M_b \propto \sigma_b^5 G^{-1},
\end{split}
\end{equation} 
where $M_b=m_r(r\equiv r_b)$ is the mass of entire bulge and $\alpha_r$ is a numerical factor. This 5/3 scaling ($M_b\propto r_b^{5/3}$) is also consistent with the Illustris simulation in Fig. \ref{fig:S2}. Here, the scaling of the velocity dispersion $\sigma_b^5\propto M_b$ can be directly tested by observations. These scaling laws involving $\varepsilon_b$ will also facilitate the analytical derivation of the mass functions and the duty cycle in Sections \ref{sec:7-1} and \ref{sec:7-2-1}. For haloes of other masses, we use $\varepsilon$ to replace $\varepsilon_b$ in these scaling laws. It should be noted that similar scaling laws were also observed for dark matter haloes \citep{Xu:2023-Universal-scaling-laws-and-density-slope, Xu:2023-Dark-matter-halo-mass-functions-and, Xu:2021-Inverse-mass-cascade-mass-function, Xu:2022-Postulating-dark-matter-partic}. 

Plugging the redshift dependence (Eq. \eqref{ZEqnNum9863112994}) into the scaling laws (Eq. \eqref{ZEqnNum98631129992}), for a fixed bulge mass $M_b$, the time evolution of relevant quantities on the bulge scale $r_b$ is
\begin{equation} 
\label{eq:13} 
\begin{split}
&r_b \propto a,\quad \sigma_b^2 \propto a^{-1}, \quad \rho_b \propto a^{-3},
\end{split}
\end{equation}
where $M_b$ and $\rho_b$ are the mass and density of the bulge. This is consistent with the commonly accepted picture that massive galaxies roughly doubled their size from $z\sim 1$ and by 3 to 5 from $z\sim 2$ \citep{Huertas_Company:2013-The-evolution-of-the-mass}. The redshift variation of $\varepsilon_b\propto a^{-5/2}$ is also consistent with a decreasing bulge density ($\rho_b \propto a^{-3}$), i.e., the bulge density of baryons follows the evolution of background density of matter, just like the mean density of dark matter haloes. 

So far, we have presented the bulge dynamics that involve the parameter $\varepsilon_b$ from the Illustris simulations. The key findings are the 2/3 law for the kinetic energy and the 5/3 law for the mass-size relation (Eq. \eqref{ZEqnNum98631129992}). Both scaling laws involve $\varepsilon_b$, which quantifies the rate of energy flow in the bulge. The parameter $\varepsilon_b\propto a^{-5/2}$ rapidly decreases with time (Eq. \eqref{ZEqnNum9863112994}), i.e., a cosmic quenching process that slows down the SMBH evolution and star formation. In the next section, we present more comparisons with observations. 

\section{Bulge dynamics from galaxy survey}
\label{sec:2-1-2}

\begin{figure}
\includegraphics*[width=\columnwidth]{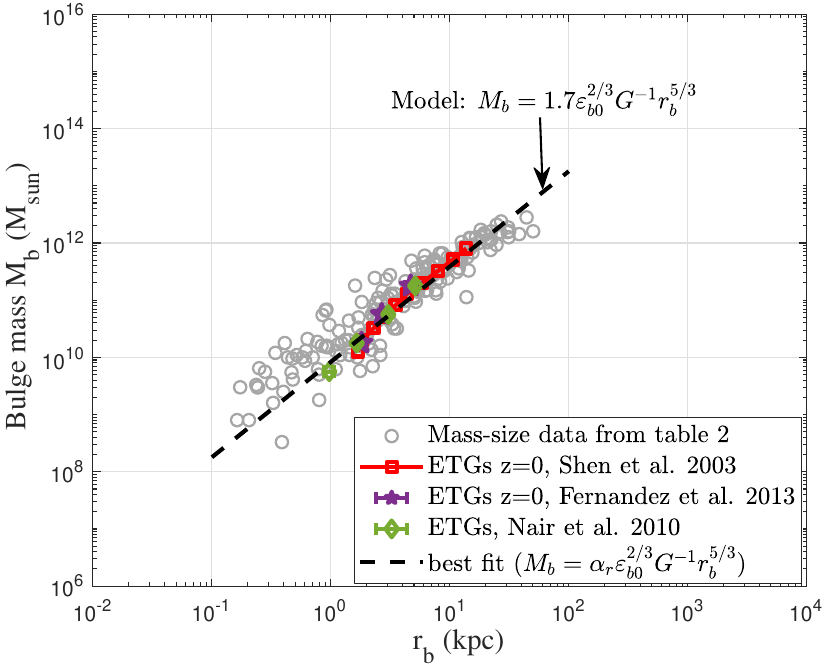}
\caption{The mass-size relation between mass $M_b$ and size $r_b$ (Eq. \eqref{ZEqnNum98631129992}). Circles are taken from Table \ref{tab:A1} for 180 galaxies with reference listed for each data. Also reported are the measurements from Shen et al. \citep{Shen:2003-The-size-distribution-of-}, Fernandez et al. \citep{Lorenzo:2013-The-stellar-mass-size-rel}, and Nair et al. \citep{Nair:2010-A-CATALOG-OF-DETAILED-VISUAL} for early-type galaxies. The dashed line indicates the predicted 5/3 scaling (see Eq. \eqref{ZEqnNum98631129992}). The good agreement confirms the 5/3 scaling and the value of $\varepsilon_{b0}=10^{-4}m^2/s^3$ in local Universe.} 
\label{fig:S8}
\end{figure}



This section focuses on observational evidence on scaling laws involving $\varepsilon_b$. Let us first check the scaling laws in Eq. \eqref{ZEqnNum98631129992}. The 5/3 law between the bulge mass-size relation ($M_b\propto r_b^{5/3}$ or $r_b\propto {M_b}^{0.6}$) is supported by many studies, especially for early-type and quiescent galaxies (ETGs). These studies show a mass-size relation $r_b \propto M_b^\alpha$ with $\alpha \approx$ [0.5 0.6] (Huertas-Company et al. \citep{Huertas_Company:2013-The-evolution-of-the-mass}), $\alpha \approx 0.6$ (Mowla et al. \citep{Mowla:2019-A-Mass-dependent-Slope-of}), $\alpha \approx$ [0.5 0.7] (Mowla et al. \citep{Mowla:2019-COSMOS-DASH-The-Evolution-of-the-Galaxy-Size-Mass}), $\alpha \approx$ [0.51 0.64] (Damjanov et al. \citep{Damjanov:2022-Quiescent-Galaxy-Size-Velocity-Dispersion}), $\alpha \approx$ [0.41 0.56] (Williams et al. \citep{Williams:2010-THE-EVOLVING-RELATIONS-BETWEEN-SIZE}), $\alpha \approx 0.55$ (Shen et al. \citep{Shen:2003-The-size-distribution-of-}). For comparison, we predict $\alpha=0.6$ from Eq. \eqref{ZEqnNum98631129992}. Figure \ref{fig:S8} presents the mass and size of 180 local galaxies (gray circles) in Table \ref{tab:A1}. Also reported are some early-type galaxies. The predicted 5/3 scaling in Eq. \eqref{ZEqnNum98631129992} is plotted as a black dashed line. Here, we use $\varepsilon_{b0}=10^{-4}m^2/s^3$ at $z=0$. The good agreement confirms the 5/3 scaling. 

\begin{figure}
\includegraphics*[width=\columnwidth]{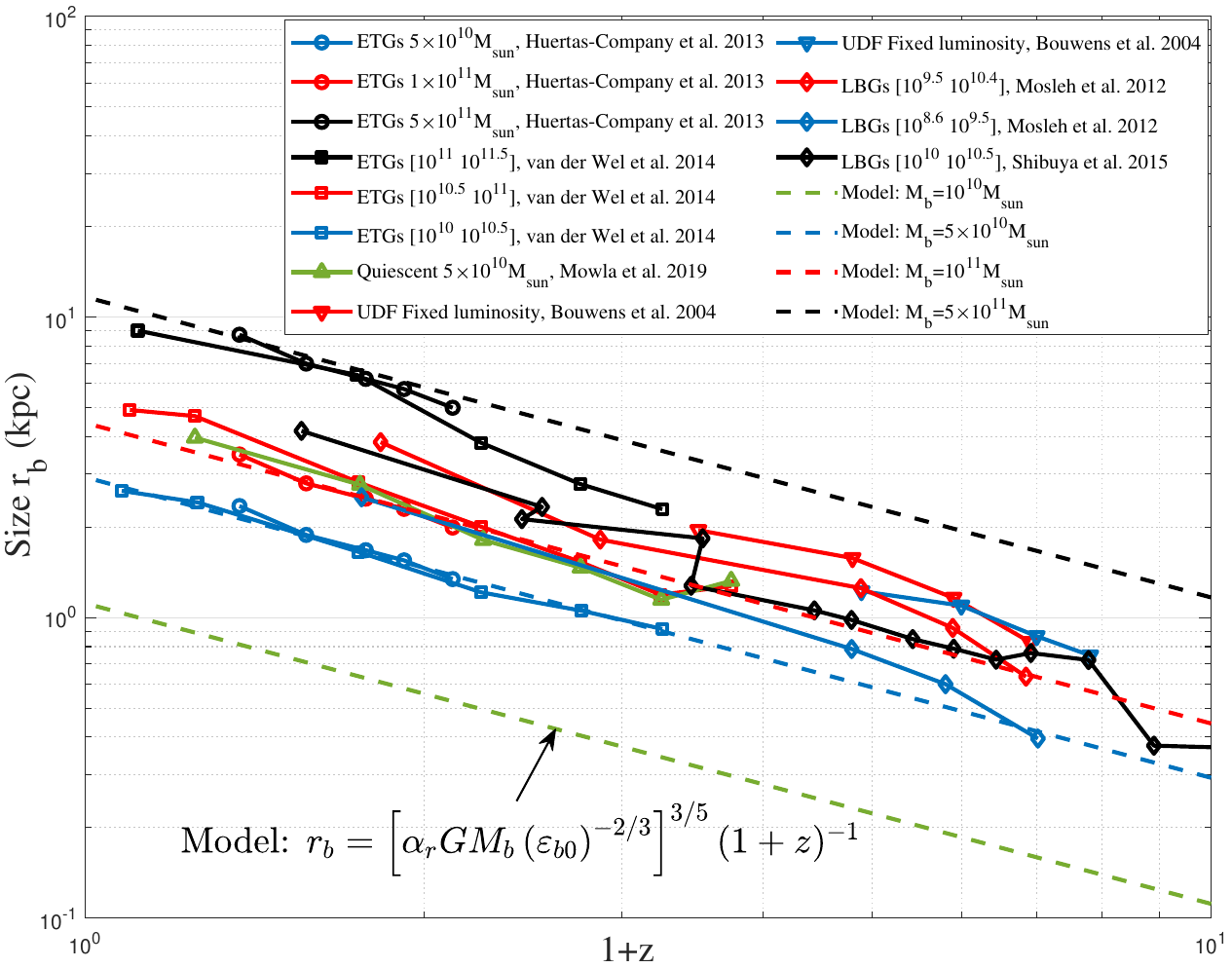}
\caption{The observed size evolution of different types of galaxies and different mass $M_b$. Early-type galaxies (ETGs) are taken from Huertas-Company et al. \citep{Huertas_Company:2013-The-evolution-of-the-mass} (circles) and van der Wel et al. \citep{van_der_Wel:2014-3D-HST-CANDELS--THE-EVOLU} (squares) for three different masses. The prediction of the model ($\alpha_r=1.7$) for the size evolution (Eq. \eqref{ZEqnNum986311299922}) is in good agreement with observations for ETGs. Quiescent galaxies (triangles) are taken from Mowla et al. \citep{Mowla:2019-COSMOS-DASH-The-Evolution-of-the-Galaxy-Size-Mass}. Lyman-break galaxies (LBGs) (diamonds) are taken from Mosleh et al. \citep{Mosleh:2012-THE-EVOLUTION-OF-MASS–SIZE-RELATION-FOR-LYMAN} and Shibuya et al. \citep{Shibuya:2015-Morphologies-of-190000-Galaxies}. The mass-size relation can be different with a different $\alpha_r$. All data agree with a size evolution of $r_b \propto a \propto (1+z)^{-1}$.} 
\label{fig:S9}
\end{figure}

Next, we check the redshift variation of the galaxy size $r_b$. A power law is usually observed with $r_b \propto (1+z)^{-\beta} \propto a^{\beta}$ with $\beta \approx 1$, especially for early-type galaxies (ETGs) and quiescent galaxies. This is supported by studies with $\beta \approx 1.01$ (Huertas-Company et al. \citep{Huertas_Company:2013-The-evolution-of-the-mass}), $\beta \approx 1.05$ (Yang et al. \citep{Yang:2020-The-evolution-of-the-size-mass-relation-at}), $\beta \approx$ [0.75 1.3] (Williams et al. \citep{Williams:2010-THE-EVOLVING-RELATIONS-BETWEEN-SIZE}) and $\beta \approx 0.95$ (Mowla et al. \citep{Mowla:2019-COSMOS-DASH-The-Evolution-of-the-Galaxy-Size-Mass}). This is consistent with our prediction in Eq. \eqref{eq:13}, where $r_b \propto a$. More specifically, from the 5/3 law (Eq. \eqref{ZEqnNum98631129992}) and the redshift dependence of $\varepsilon_b\propto a^{-5/2}$ (Eq. \eqref{ZEqnNum9863112994}), we can write the size $r_b$ as a simple function of the mass $M_b$,
\begin{equation} 
\label{ZEqnNum986311299922} 
r_b=\left[\frac{GM_b}{\alpha_r\left(\varepsilon_{b}\right)^{2/3}}\right]^{3/5}=\left[\frac{GM_b}{\alpha_r\left(\varepsilon_{b0}\right)^{2/3}}\right]^{3/5}(1+z)^{-1}.
\end{equation}

Figure \ref{fig:S9} presents the size evolution for different types of galaxies and different masses. The open circles and squares represent the early-type galaxies of Huertas-Company et al. \citep{Huertas_Company:2013-The-evolution-of-the-mass} and van der Wel et al. \citep{van_der_Wel:2014-3D-HST-CANDELS--THE-EVOLU}. The model for the size evolution (Eq. \eqref{ZEqnNum986311299922} with $\alpha_r=1.7$) agrees with different observations. Triangles represent the quiescent galaxies of Mowla et al. \citep{Mowla:2019-COSMOS-DASH-The-Evolution-of-the-Galaxy-Size-Mass}. Diamonds represent the Lyman-break galaxies of Mosleh et al. \citep{Mosleh:2012-THE-EVOLUTION-OF-MASS–SIZE-RELATION-FOR-LYMAN} and Shibuya et al. \citep{Shibuya:2015-Morphologies-of-190000-Galaxies}. For all galaxies, the mass-size relation might differ with different $\alpha_r$ or slope $\alpha$. However, all data suggest $r_b \propto a$, in agreement with the prediction (Eqs. \eqref{eq:13} and \eqref{ZEqnNum986311299922}).

\begin{figure}
\includegraphics*[width=\columnwidth]{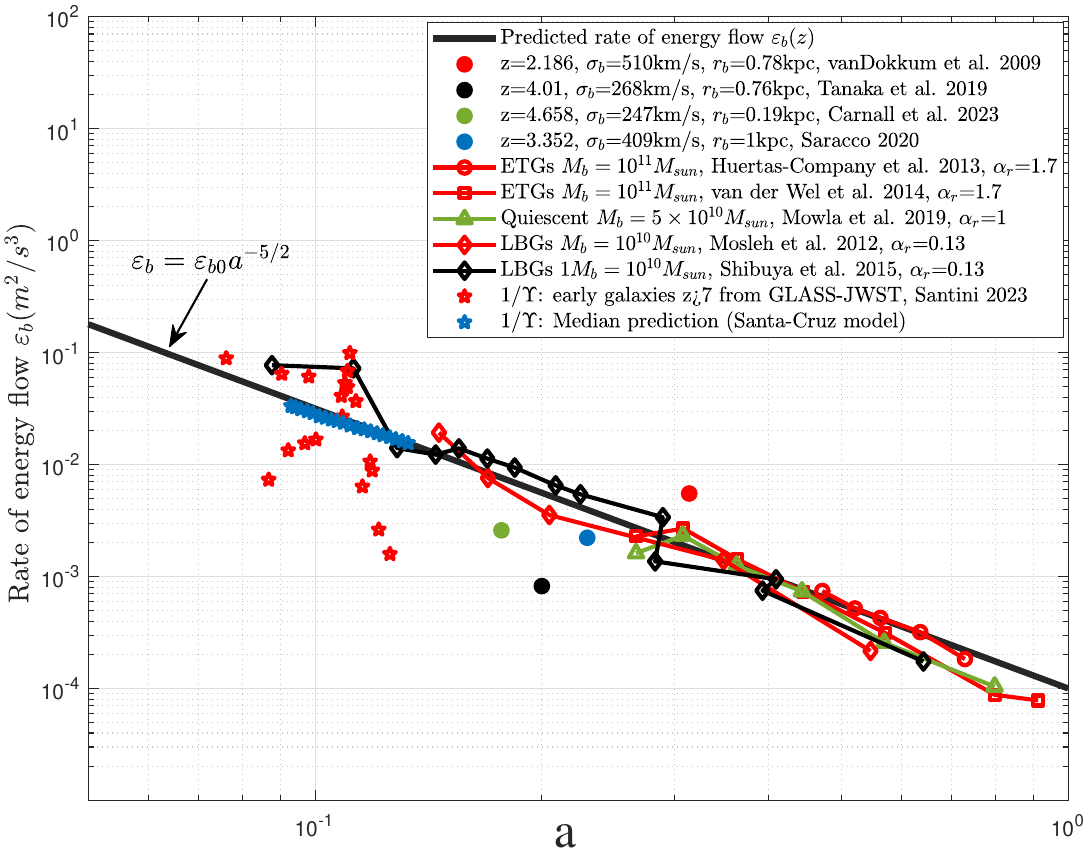}
\caption{The redshift dependence for the rate of energy flow $\varepsilon_b(z)$. The solid black line presents the variation $\varepsilon_b\propto a^{-5/2}$ with $\varepsilon_{b0}=10^{-4}m^2/s^3$ (Eq. \eqref{ZEqnNum9863112994}). The filled circles are four high-redshift galaxies with known velocity dispersion $\sigma_b^2$ and size $r_b$ to calculate $\varepsilon_b=\sigma_b^3/r_b$ (Eq. \eqref{ZEqnNum98631129991}). These include a compact massive galaxy \citep{van_Dokkum:2009-A-high-stellar-velocity-dispersion-for-a-compact}, a quenching galaxy \citep{Tanaka:2019-Stellar-Velocity-Dispersion-of-a-Massive}, a quiescent galaxy GS-9209 \citep{Carnall:2023-A-massive-quiescent-galaxy-at-redshift}, and an early type galaxy \citep{Saracco:2020-The-Rapid-Buildup-of-Massive-Early-type-Galaxies}. The size evolution in Fig. \ref{fig:S9} was also used to calculate $\varepsilon_b$ by Eq. \eqref{ZEqnNum986311299922} with known mass $M_b$ and size $r_b$ (solid lines with symbols). The figure confirms a rapidly decreasing $\varepsilon_b$ with time. In addition, blue stars plot the Santa Cruz semi-analytic model for the median prediction of the light-to-mass ratio $1/\Upsilon$ over time \citep{Santini:2023-Early-Results-from-GLASS-JWST}. Red stars plot results from GLASS-JWST for the light-to-mass ratio (1/$\Upsilon$) of galaxies $z>7$ \citep{Santini:2023-Early-Results-from-GLASS-JWST}. Good agreement also suggests strong connections between the light-to-mass ratio 1/$\Upsilon$ and parameter $\varepsilon_b$.} 
\label{fig:S10}
\end{figure}

From Illustris simulations, we find that the rate of energy flow $\varepsilon_b$ has a strong dependence on the redshift, which can be much greater in the early Universe (Fig. \ref{fig:S7}). Figure \ref{fig:S10} presents some relevant observations for the redshift dependence of $\varepsilon_b(z)$. The proposed variation $\varepsilon_b\propto a^{-5/2}$ is plotted as the solid black line with $\varepsilon_{b0}=10^{-4}m^2/s^3$ (Eq. \eqref{ZEqnNum9863112994}). The filled circles present four high-redshift galaxies with known velocity dispersion $\sigma_b^2$ and size $r_b$ \citep{van_Dokkum:2009-A-high-stellar-velocity-dispersion-for-a-compact,Tanaka:2019-Stellar-Velocity-Dispersion-of-a-Massive,Carnall:2023-A-massive-quiescent-galaxy-at-redshift,Saracco:2020-The-Rapid-Buildup-of-Massive-Early-type-Galaxies}. The parameter $\varepsilon_b=\sigma_b^3/r_b$ can be calculated for each galaxy using Eq. \eqref{ZEqnNum98631129991}. The rate of energy flow $\varepsilon_b$ is obviously higher at a higher redshift. In the same figure, the size evolution in Fig. \ref{fig:S9} was also used to calculate $\varepsilon_b$ for galaxies with known mass $M_b$ and size $r_b$ (using Eq. \eqref{ZEqnNum986311299922}). The figure confirms a rapidly decreasing $\varepsilon_b$ or a less efficient gas cooling and cold gas supply at a lower redshift. 

In addition, the parameter $\varepsilon_b(z)$ is also related to the light-to-mass ratio (1/$\Upsilon$) of galaxies. For example, the Santa Cruz semi-analytic model for the median prediction of $\Upsilon$ is represented by the blue stars that show a decrease 1/$\Upsilon$ over time \citep{Santini:2023-Early-Results-from-GLASS-JWST}. Early results from GLASS-JWST for the light-to-mass ratio (1/$\Upsilon$) of galaxies $z>7$ show a much higher 1/$\Upsilon$ in the early Universe (red stars) \citep{Santini:2023-Early-Results-from-GLASS-JWST}. Good agreement also suggests strong connections between the light-to-mass ratio 1/$\Upsilon$ and $\varepsilon_b$. At higher redshifts, the larger value of $\varepsilon_b$ leads to more efficient gas cooling, faster star formation, and a higher light-to-mass ratio $1/\Upsilon$.

In this section, we briefly discuss the mass-size relation and the size evolution of galaxies from observations. Both simulations and observations support the scaling laws involving $\varepsilon_b$ (Eq. \eqref{ZEqnNum98631129992}) and its redshift evolution (Eq. \eqref{ZEqnNum9863112994}). In the next section, we focus on the effects of $\varepsilon_b$ on the co-evolution of SMBHs and hosts.  

\section{Length scales for SMBH-bulge evolution}
\label{sec:4}

In Illustris simulations, the SMBH scale ($10^{-7}$ kpc) cannot be directly resolved due to the vast scale disparity (Fig. \ref{fig:S4}). 
However, neglecting the complexities of inner regions, it is very instructive to extend scaling laws established to small scales, where critical length scales can be identified based on these scaling laws and the dominant physics on relevant scales. In this section, we identify these critical length scales that are highly relevant to the distribution and evolution of SMBHs and their hosts. 

The scaling laws for the bugle mass and density read (Eq. \eqref{ZEqnNum98631129992}) 
\begin{equation} 
\label{eq:9} 
m_r = \alpha_r \varepsilon_b^{2/3}G^{-1}r^{5/3}, \quad  \rho_r = \frac{m_r}{4/3\pi r^3}= \beta_r \varepsilon_b^{2/3}G^{-1}r^{-4/3},
\end{equation} 
where $\alpha_r$ and $\beta_r=3\alpha_r/(4\pi)$ are two constants of order unity. Similarly, the characteristic time $t_r$, the velocity dispersion $\sigma_r^2$, the pressure $P_r$ (dynamic pressure due to the random motion in gas), and the force $F_r$ (pressure gradient) read
\begin{equation} 
\label{eq:10} 
\begin{split}
&t_r \propto \varepsilon_b^{-1/3} r^{2/3}, \quad \sigma_r^2=\gamma_rGm_r/r=\alpha_r\gamma_r(\varepsilon_br)^{2/3}, \\
&P_r=\rho_r\sigma_r^2=\alpha_r\beta_r\gamma_r\varepsilon_b^{4/3}G^{-1}r^{-2/3},\\
&F_r=4\pi r^2P_r=4\pi\alpha_r\beta_r\gamma_r\varepsilon_b^{4/3}G^{-1}r^{4/3}=3\sigma_r^4/(\gamma_rG),
\end{split}
\end{equation} 
where $\alpha_r\gamma_r$ is on the order of unity, $\gamma_r<1$ is a numerical constant with $\gamma_r \approx 1/3$ for galaxy bulge \citep{Marconi:2003-The-relation-between-blac}. In the spherical Jeans equations, the product $\rho_r\sigma_r^2$ due to the random motion of the gas has a similar effect as the pressure \citep{Mo:2010-Galaxy-formation-and-evolution}. Therefore, the dynamic pressure $P_r$ in Eq. \eqref{eq:10} comes from the random motion $\sigma_r^2$ of gases. The force $F_r\propto \sigma_r^4/G$ is associated with the dynamic pressure $P_r$ and should balance the radiation force $L_B/c$ of the SMBHs to reflect the effect of turbulent, dynamic, and random motion in gases. This radial force $F_r$ can be much greater than the static weight of the gas in Eddington's theory to enable super-Eddington accretion of SMBHs (Sections \ref{sec:9-2} and \ref{sec:8}). 

Similarly, the energy flux and the mass flux on scale $r$ read
\begin{equation} 
\label{eq:11} 
\begin{split}
&\varepsilon_b m_r\propto \dot m_r \sigma_r^2 \propto {Gm_r\dot m_r}/{r}, \\
&\dot m_r \propto \varepsilon_bG^{-1}r \quad \textrm{and} \quad \dot M_b \propto \varepsilon_bG^{-1}r_b \propto a^{-3/2} \propto t^{-1},
\end{split}
\end{equation} 
where $\varepsilon_b m_r$ is the rate of energy flow into scales below $r$, while $\dot m_r=m_r/t_r$ is the average rate of mass flow into scale $r$. Here, $\dot M_b\propto t^{-1}$ is the average mass flow rate into the entire bulge such that $M_b$ only slowly (logarithmically) increases with time. 

Similar scaling laws were also identified for dark matter haloes \citep{Xu:2023-Universal-scaling-laws-and-density-slope}. A specific example is the 2/3 law for velocity dispersion $\sigma_r^2$ on scale $r$ that can be demonstrated by the two-point statistics for collisionless dark matter \citep{Xu:2023-On-the-statistical-theory-of-self-gravitating, Xu:2024-On-the-statistical-theory-of-self-gravitating, Xu:2024-On-the-statistical-theory-of-self-gravitating-scale_redshift}. In principle, a similar analysis can also be extended to the baryonic matter in hydrodynamic simulations to confirm these scaling laws.

Here, six physical quantities can be identified for the SMBH-bulge system: the bulge mass $M_b$, the black hole mass $M_{BH}$, the black hole bolometric luminosity $L_B$, the rate of energy flow $\varepsilon_b$, plus two physical constants: the gravitational constant $G$ and the speed of light $c$. These six quantities lead to five critical length scales determined by the dominant physics on relevant scales. Figure \ref{fig:1} provides a schematic plot of these scales. The first two scales are the size of the bulge $r_b$ and the BH sphere of influence $r_B$, both of which are determined by the rate of energy flow $\varepsilon_b$ and the mass on that scale. The smallest scale is the Schwarzschild radius $r_s$. Three length scales from large to small read (from Eq. \eqref{eq:9}):
\begin{equation}
\begin{split}
&r_b=(1/\alpha_r)^{3/5} M_b^{3/5}G^{3/5}\varepsilon_b^{-2/5} \\
&r_B=(1/\alpha_r)^{3/5} M_{BH}^{3/5}G^{3/5}\varepsilon_b^{-2/5} \\
&r_s = 2GM_{BH}/c^2.
\end{split}
\label{eq:15}
\end{equation}

\begin{figure}
\includegraphics*[width=\columnwidth]{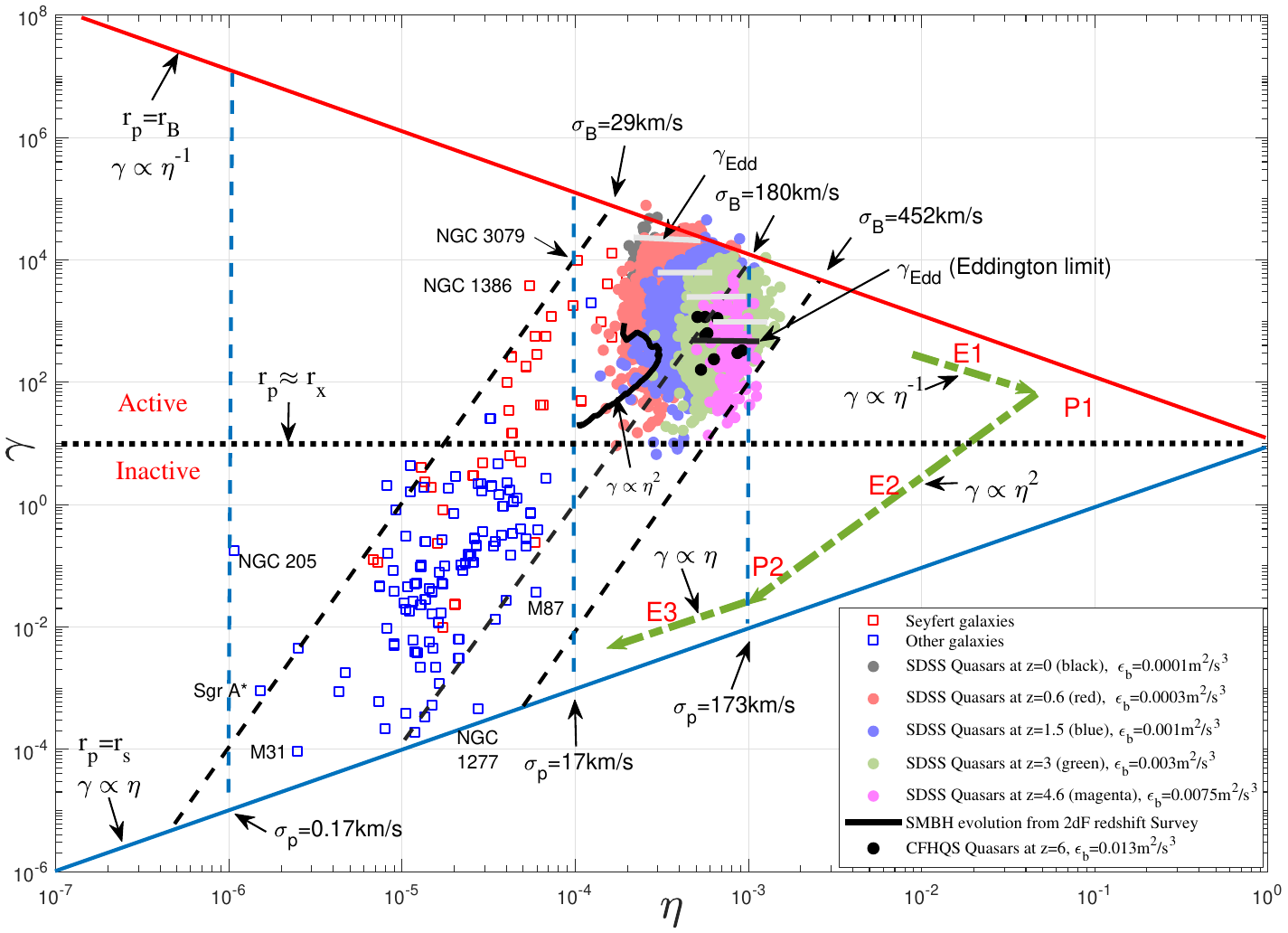}
\caption{The distribution and evolution of SMBHs in $\gamma$-$\eta$ plane. The distribution of SMBHs from Table \ref{tab:A1} is presented as square symbols in red (Seyfert) and blue (other). Color circles are the distributions of SDSS DR7 quasars at $z$ = 0, 0.6, 1.5, 3, and 4.6, respectively. The black circles are high redshift quasars from the CFHQS survey at $z$=6. Eddington limit is presented as short horizontal lines, showing that many high redshift quasars are beyond the Eddington limit. The quasars are distributed in the upper half-triangle, shifting toward smaller $\eta$ (or lower luminosity) over time. The upper limit of that distribution (solid red line), the lower limit (solid blue line), and the boundary for active and inactive SMBHs (dotted black line) are discussed in Section \ref{sec:6}. The evolution of a typical SMBH (solid black line) is mapped onto the $\gamma$-$\eta$ plane using data from the 2dF redshift survey in Fig. \ref{fig:4}. The green dashed line represents the evolution path of SMBH with three stages (E1, E2, and E3) and two turning points (P1 and P2) (see Section \ref{sec:7}).} 
\label{fig:3}
\end{figure}

\begin{figure}
\includegraphics*[width=\columnwidth]{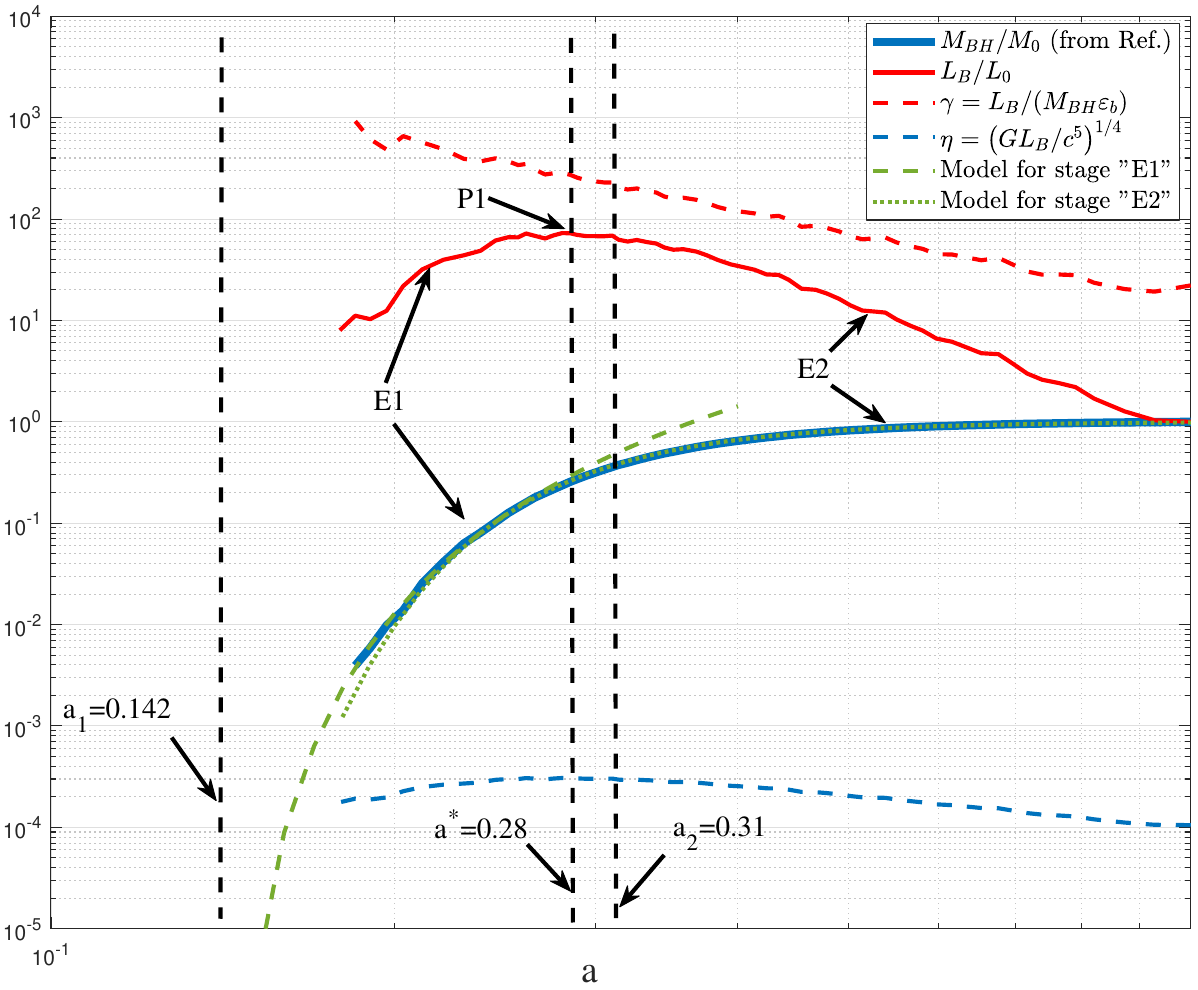}
\caption{The variation of BH mass ($M_{BH}$) and luminosity ($L_B$) of a typical SMBH with scale factor $a$. The mass evolution is obtained from the quasar luminosity function of the 2dF Redshift Survey \citep{Yu:2002-Observational-constraints}. The luminosity evolution is derived from $M_{BH}$ using Eq. \eqref{eq:22} with a maximum BH luminosity at $a^* \approx 0.28$. The mass and the luminosity are normalized by their values at $z=0$ ($M_0=10^9M_{\odot}$ and $L_0=4.45\times 10^{43}\textrm{erg/s}$). The evolution of two parameters $\gamma$ and $\eta$ is computed by Eq. \eqref{eq:18} and mapped in the $\gamma$-$\eta$ plane (solid black line in Fig. \ref{fig:3}). Luminosity $L_B$ increases with time in stage E1, followed by a decrease in stage E2. Proposed models of the evolution of SMBH in Eqs. \eqref{eq:43} and \eqref{eq:48} (green dashed and dotted) are also presented for comparison, which is in good agreement with the data. The models involve a scale factor $a_1$ for the time BH formed in stage E1 and a characteristic scale factor $a_2$ in stage E2.} 
\label{fig:4}
\end{figure}

The fourth length scale (radiation scale) $r_p$ can be determined by balancing the pressure of BH radiation with the radial pressure $P_r$ due to random motion in gas (Eq. \eqref{eq:10}). Since the radiation pressure due to the luminosity of BH is $P_{rad}=L_B/(4\pi r^2c)$, the radiation scale $r_p$ can be obtained by equating the radiation pressure with the dynamic pressure (Eq. \eqref{eq:10}), that is, $P_{rad}=P_r$. Radiation pressure dominates on scales below $r_p$. The radiation scale $r_p$ sets the smallest scale for the energy flow with a r-independent rate of $\varepsilon_b$. The rate of energy flow becomes r-dependent, and the BH accretion physics becomes important on scales smaller than $r_p$. Set $P_{rad}=P_r$ in Eq. \eqref{eq:10}, the radiation scale reads
\begin{equation}
r_p=\left(\frac{GL_B}{3\alpha_r^2\gamma_rc}\right)^{\frac{3}{4}}\varepsilon_b^{-1}. 
\label{eq:16}
\end{equation}
The radiation scale $r_p$ should initially increase with time due to increasing BH luminosity, followed by a decreasing stage when luminosity decreases. Intuitively, the radiation scale $r_p$ should not exceed the BH sphere of influence $r_B$. Beyond that limit, the black hole's gravity is insufficient to hold the gas repelled by the radiation. The radiation scale $r_p$ should also be greater than the Schwarzschild radius $r_s$. With $r_s \le r_p\le r_B$, the upper and lower limits of the SMBH distributions can be developed (Figs. \ref{fig:3} and \ref{fig:4}). 

The fifth length scale is related to the energy dissipation in the BH accretion disk. In the alpha disk model, Shakura and Sunyaev suggested that the kinematic viscosity due to the turbulence of gas in the accretion disk at a given radius $r$ can be written as $\nu = \alpha c_s(r) H(r)$ \citep{Shakura:1973-Black-holes-in-binary-systems}, where $\alpha\le 1$ is a numerical factor, $H$ is the half thickness of the disk, $c_s\approx (H/r) V_k$ is the sound speed and $V_k^2=GM_{BH}/r$ is the orbital speed. The maximum viscosity of $\nu$ should be at the Schwarzschild radius $r_s$ that reads
\begin{equation}
\nu_s\equiv\nu(r=r_s)=z_r r_s c \quad \textrm{and} \quad z_r=\frac{\alpha}{\sqrt 2}\left(\frac{H_s}{r_s}\right)^2,
\label{eq:17-1}
\end{equation}
where $z_r$ is a numerical factor and $H_s\equiv H(r=r_s)$ is the disk thickness at $r_s$. In the kinetic theory of gas, $z_r\approx 1/3$. Combining the kinematic viscosity $\nu_s$ and the rate of energy flow $\varepsilon_b$, the fifth length scale (dissipation scale) $r_x$ can be introduced,
\begin{equation}
r_x =\left(\frac{\nu_s^3}{\varepsilon_b}\right)^{\frac{1}{4}}=\left(\frac{8z_r^3G^3M_{BH}^3}{c^3\varepsilon_b}\right)^{\frac{1}{4}}. 
\label{eq:17}
\end{equation}
For constant $z_r$, the scale $r_x$ increases over time due to decreasing $\varepsilon_b$. The effect of the viscous dissipation of the accretion disk is only dominant on scales below $r_x$, where the energy is dissipated by the turbulent gas in the accretion disk. The bulge dynamics are not affected by the accretion disk on scales greater than $r_x$. The boundary between active and inactive SMBH can be obtained by comparing the two scales $r_p$ and $r_x$ (Fig. \ref{fig:3}). For the active phase (quasar), we should have $r_p>r_x$ so that the energy flow in the gas can be at a constant rate $\varepsilon_b$ down to the radiation scale $r_p$ without being affected by the BH accretion disk. However, for $r_p<r_x$, due to the viscous dissipation of the accretion disk, the rate of energy flow that reaches the radiation scale $r_p$ can be much lower than $\varepsilon_b$, leading to less mass and energy supply and the inactive phase of the SMBH.

Finally, Table \ref{tab:A1} presents all five length scales computed for every galaxy with length scales $r_s \le r_p \le r_B < r_b$, while the scale $r_x$ depends on the phase of SMBH: active or inactive. In addition, three fundamental dimensionless parameters can be obtained from these six physical quantities,
\begin{equation}
\beta=\frac{L_B}{M_b\varepsilon_b} \textrm{,} \quad \gamma=\frac{L_B}{M_{BH}\varepsilon_b}\textrm{,} \quad \textrm{and} \quad \eta=\left(\frac{GL_B}{c^5}\right)^{\frac{1}{4}},
\label{eq:18}
\end{equation}
where $\beta$ is the ratio of the rate of energy dissipated in BH luminosity to the rate of energy injected on the bulge scale $r_b$. Since the rate of energy flow $\varepsilon_b$ also represents the rate of energy dissipation in gas ($\varepsilon_b=\varepsilon_a$ in Eq. \eqref{eq:3-1-2}), the parameter $\gamma$ represents the competition between the rate of energy dissipated in the form of BH luminosity (the BH light-to-mass ration in the unit of $m^2/s^3$) and the rate of energy dissipation in gas ($\varepsilon_a=\varepsilon_b$). Here, $\gamma\gg 1$ indicates the active quasar phase where energy is mostly dissipated in the form of BH luminosity, while $\gamma\ll 1$ represents the inactive phase (Fig. \ref{fig:3}). 

The physical meaning of these dimensionless parameters can also be found as the coefficients between luminosity $L_B$ and velocity dispersion on different scales. Using Eq. \eqref{eq:10} for velocity scale $\sigma_r^2$ and Eq. \eqref{eq:11} for the mass flow rates, we have
\begin{equation}
L_B \propto \beta \dot M_b \sigma_b^2 \propto {\gamma} \dot M_B \sigma_B^2 \propto \frac{1}{\eta}\dot M_p \sigma_p^2 \propto \eta \dot M_p c^2,
\label{eq:19}
\end{equation}
where $M_b$, $M_B\equiv M_{BH}$, and $M_p$ are the mass enclosed within the bulge size $r_b$, the BH sphere of influence $r_B$, and the radiation scale $r_p$. Here, $\dot M_b$, $\dot M_B$, and $\dot M_p$ are the mass flow rates on the scales $r_b$, $r_B$, and $r_p$. The same notation is also used for velocity dispersions $\sigma_b^2$, $\sigma_B^2$, and $\sigma_p^2$ on relevant scales $r_b$, $r_B$, and $r_p$. Using Eq. \eqref{eq:10} and the expression of $r_p$ in Eq. \eqref{eq:16}, 
\begin{equation}
\sigma_p^2\propto (\varepsilon_br_p)^{2/3} \quad \textrm{and} \quad \sigma_p\propto \eta c=\left({GL_B}/{c}\right)^{1/4}.
\label{eq:19-1}
\end{equation}
Therefore, we can also interpret the parameter $\eta$ as the ratio $\eta \propto \sigma_p/c$, where $\sigma_p$ is the typical velocity on the radiation scale $r_p$,


The ratio between different length scales can be conveniently expressed in terms of $\gamma$ and $\eta$:
\begin{equation}
\begin{split}
&\frac{r_B}{r_s}=\frac{1}{2\alpha_r^{3/5}}\left(\frac{\gamma}{\eta^4}\right)^{\frac{2}{5}}, \quad \frac{r_p}{r_s}=\frac{1}{2(3\alpha_r^2\gamma_r)^{3/4}}\left(\frac{\gamma}{\eta}\right),\\
&\frac{r_B}{r_p}=\frac{(3\alpha_r^2\gamma_r)^{3/4}}{\alpha_r^{3/5}}\left({\gamma}{\eta}\right)^{-\frac{3}{5}}, \quad \frac{r_x}{r_s}=\frac{z_r^{3/4}}{2^{1/4}}\left(\frac{\gamma}{\eta^4}\right)^{\frac{1}{4}},\\
&\frac{r_x}{r_p}=\left(\frac{6z_r\alpha_r^2\gamma_r}{\gamma}\right)^{\frac{3}{4}}, \quad \frac{r_x}{r_B}=(2z_r)^{3/4}\alpha_r^{3/5}\left(\frac{\eta^4}{\gamma}\right)^{\frac{3}{20}}.
\end{split}
\label{eq:20}
\end{equation}
In addition, ratios of mass and size between SMBH and bulge are
\begin{equation}
\begin{split}
\frac{M_{BH}}{M_b}=\frac{\beta}{\gamma}=\left(\frac{r_B}{r_b}\right)^{\frac{5}{3}}=\left(\frac{\sigma_B}{\sigma_b}\right)^5 \quad \textrm{and} \quad \frac{r_B}{r_b}=\left(\frac{\beta}{\gamma}\right)^{\frac{3}{5}}.
\end{split}
\label{eq:21}
\end{equation}
These relations will be used to study the SMBH distributions and evolution in the next section.

\section{SMBH demography in \texorpdfstring{$\gamma$-$\eta$}{} plane}
\label{sec:4-2}
The distribution and evolution of SMBHs can be described by parameters $\eta$ and $\gamma$ defined in Eq. \eqref{eq:18}. Figure \ref{fig:3} presents the distribution of local SMBHs, high redshift quasars, and the evolution path of a typical SMBH in the $\gamma$-$\eta$ plane. All data come from:
\begin{enumerate}
\item \noindent Table \ref{tab:A1} from a survey of local galaxies with known bulge mass $M_b$, size $r_b$, or velocity dispersion $\sigma_b$. The rate of energy flow $\varepsilon_b=\sigma_b^3/r_b$ can be explicitly calculated. With the luminosity $L_B$ and BH mass $M_{BH}$ for these galaxies, $\eta$ and $\gamma$ can be computed by Eq. \eqref{eq:18} and plotted in Fig. \ref{fig:3} as square symbols. Red squares represent active SMBHs, while blue squares are inactive SMBHs.

\item \noindent More than 100,000 quasars obtained from Sloan Digital Sky Survey Data Release 7 (SDSS DR7) \citep{Schneider:2010-THE-SLOAN-DIGITA-SKY-SURVEY-QUASAR-CATALOG, Shen:2011-A-Catalog-of-Quasar-Properties}. Quasars with redshift $z$ = 0, 0.6, 1.5, 3, and 4.6 (filled circles) are 
mapped onto the $\gamma$-$\eta$ plane in Fig. \ref{fig:3}, where the upper and lower limits can be identified. When computing the parameter $\gamma$ for each quasar, the parameter $\varepsilon_b=\varepsilon_{b0}(1+z)^{5/2}$ is calculated (Eq. \eqref{ZEqnNum9863112994}) based on the redshift $z$ of each quasar. The quasars are mostly distributed in the upper half-triangle, shifting to smaller $\eta$ (or lower luminosity) with time. \\

\item \noindent Quasars with a high redshift $z\approx 6$ (dark circles in Fig. \ref{fig:3}) obtained from the Canada-France High-z Quasar Survey (CFHQS) \citep{Willott:2010-EDDINGTON-LIMITED-ACCRETION-AND-THE-BLACK-HOLE}. Here, we have $\varepsilon_b \approx 0.013 m^2/s^3$ at this redshift (Eq. \eqref{ZEqnNum9863112994}). These high-z quasars have high $\eta$ or luminosity. \\

\item \noindent The evolution of $M_{BH}$ for a typical SMBH obtained from the evolution of the average comoving BH mass density. This particular path of evolution can be estimated from the quasar luminosity function (QLF) and used to track the mass accretion history of typical SMBHs. With the mass accretion history $\dot M_{BH}$ obtained from QLF, the next step is to compute the luminosity for typical SMBHs,
\begin{equation}
\frac{L_B}{M_0}=\frac{\dot M_{BH}}{M_0} \frac{\epsilon c^2}{1-\epsilon}=\frac{\partial (M_{BH}/M_0)}{\partial a} {H_0 a^{-1/2}} \frac{\epsilon c^2}{1-\epsilon},
\label{eq:22}
\end{equation}
where $M_0$ is the BH mass at $z=0$. Here, $\epsilon=0.1$ is the radiative efficiency, and $H_0 \approx$70km/s/Mpc is the Hubble constant. With $L_B$ solved by Eq. \eqref{eq:22} and $\varepsilon_b$ from Eq. \eqref{ZEqnNum9863112994}, the evolution of $\gamma$ and $\eta$ can be obtained from the definition in Eq. \eqref{eq:18}. This "mean" or "typical" path of evolution obtained from quasar luminosity functions provides insights into the evolution of observed SMBHs in Section \ref{sec:8}.  

Figure \ref{fig:4} plots the time variation of the BH mass $M_{BH}$ (normalized by $M_0=10^9M_{\odot}$) that is derived from the quasar luminosity function of the 2dF Redshift Survey \citep{Yu:2002-Observational-constraints}. The mass accretion rate $\dot{M}_{BH}$ is obtained from the time derivative of $M_{BH}$. The luminosity $L_B$ (normalized by $L_0=4.45\times 10^{43}\textrm{erg/s}$ at $z=0$) is obtained from Eq. \eqref{eq:22}. The time evolution of $L_B$ and $M_{BH}$ in Fig. \ref{fig:4} is then transformed into the evolution path (solid black line) in terms of the parameters $\gamma$ and $\eta$ in Fig. \ref{fig:3} ($\gamma\propto \eta^{-1}$ and $\gamma\propto \eta^2$). The luminosity $L_B$ first increases and then decreases after reaching the maximum luminosity at approximately $a^*$ = 0.28. The peak luminosity (point P1) divides the entire evolution into a rising stage E1 with increasing luminosity and a declining stage E2 with decreasing luminosity. This is important because we will derive analytical solutions for the BH mass function, the AGN duty cycle, and the evolution of observed high redshift SMBHs, all based on these two stages of evolution (Sections \ref{sec:7-1} to \ref{sec:9-2}).
\end{enumerate}

\section{SMBH distributions in \texorpdfstring{$\gamma$-$\eta$}{} plane}
\label{sec:6}
With length scales defined in Eqs. \eqref{eq:15}, \eqref{eq:16}, and \eqref{eq:17}, now we can identify the upper and lower limits of the SMBH distribution in the $\gamma-\eta$ plane and the boundary for active/inactive SMBHs (Fig. \ref{fig:3}). 

\begin{figure}
\includegraphics*[width=\columnwidth]{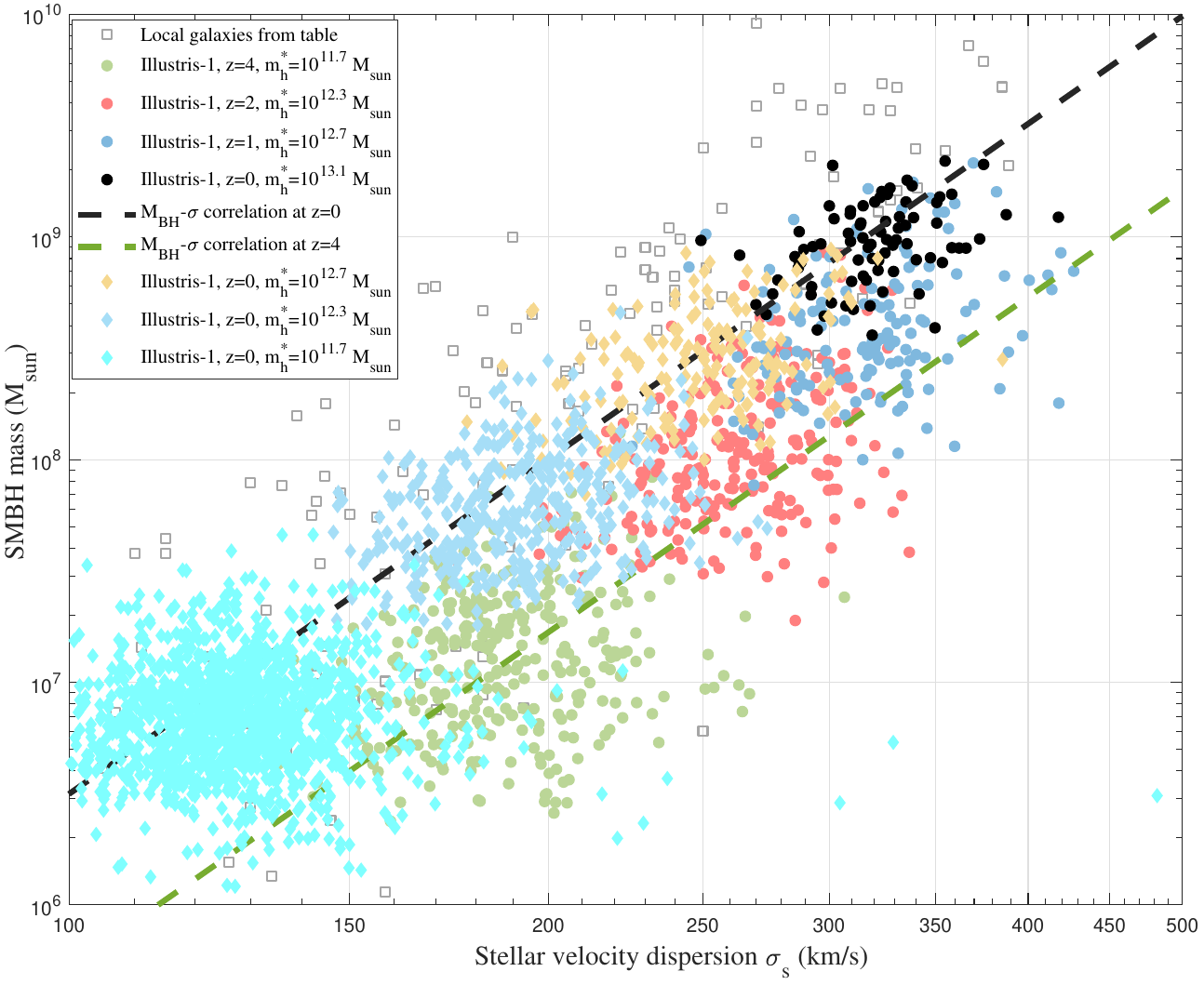}
\caption{The correlation between BH mass ($M_{BH}$) and stellar velocity dispersion ($\sigma_s$) for different halo mass $m_h$ and redshifts $z$. Gray squares plot the data for the local galaxies in Table \ref{tab:A1}. The colored circles represent the data for the haloes with a characteristic mass $m_h^*$ 
from Illustris simulations. The stellar velocity dispersion is calculated from the mean specific kinetic energy in Fig. \ref{fig:S4}. The colored diamonds represent the data for haloes of different masses in the current epoch $z=0$. For haloes of the same mass $m_h$ at different $z$, the BH mass ($M_{BH}$) is almost the same. However, the stellar velocity dispersion $\sigma_s$ in haloes of the same mass $m_h$ is much greater at high redshift $z$. A redshift-dependent correlation $M_{BH}-\sigma$ can be obtained from the simulations. The figure shows the best-fit correlations at $z=0$ (black dashed line from Eq. \eqref{eq:31}) and at $z=4$ (green dashed lines). At fixed velocity dispersion, $M_{BH}$ is slightly smaller at higher redshift.} 
\label{fig:S15}
\end{figure}

\begin{enumerate}
\item \noindent The upper limit is determined by setting the scales $r_p=r_B$. The maximum radiation scale $r_p$ cannot exceed $r_B$, the BH sphere of influence. Beyond that limit, the gravity of the SMBHs cannot hold the gas because of the radiation pressure. From Eq. \eqref{eq:20}, we have
\begin{equation}
{\gamma}{\eta}=\frac{(3\alpha_r^2\gamma_r)^{5/4}}{\alpha_r}\approx 10 \quad \textrm{(Solid red line in Fig. \ref{fig:3})},
\label{eq:23}
\end{equation}
where $\gamma_r\approx 1/3$ and $\alpha_r\approx 4$. Substituting Eq. \eqref{eq:18} into \eqref{eq:23}, the upper limit of BH luminosity $L_B$ is determined by $r_p=r_B$, 
\begin{equation}
L_B=3\alpha_r^{6/5}\gamma_r\varepsilon_b^{4/5}M_{BH}^{4/5}G^{-1/5}c,
\label{eq:24}
\end{equation}
where luminosity scales with mass as $L_B\propto (\varepsilon_b M_{BH})^{4/5}$ along the upper limit $\gamma\propto \eta^{-1}$. \\

\item \noindent The lower limit is determined by setting the scales $r_p=r_s$, that is, the minimum radiation scale $r_p$ cannot be less than the Schwarzschild radius $r_s$. From Eq. \eqref{eq:20}, we should have
\begin{equation}
\gamma={2(3\alpha_r^2\gamma_r)^{3/4}}\eta \quad \textrm{(Solid blue line in Fig. \ref{fig:3})}.
\label{eq:25}
\end{equation}
The lower limit of luminosity $L_B$ should be determined by $r_p=r_s$,
\begin{equation}
L_B=2^{4/3}(3\alpha_r^2\gamma_r)\varepsilon_b^{4/3}M_{BH}^{4/3}G^{1/3}c^{-5/3},
\label{eq:26}
\end{equation}
where luminosity scales with mass as $L_B\propto (\varepsilon_b M_{BH})^{4/3}$ along the lower limit $\gamma\propto \eta$.\\

\item \noindent Velocity dispersion on scale $r_p$ follows $\sigma_p^2 \propto (\varepsilon_b r_p)^{2/3}$ (from Eq. \eqref{eq:10}) and should read (using $r_p$ in Eq. \eqref{eq:16})
\begin{equation}
 \frac{\sigma_p}{c}=\left(\frac{\gamma_r}{3}\right)^{\frac{1}{4}}\eta \quad \textrm{or} \quad L_B = \frac{3c}{\gamma_r G} \sigma_p^4. 
\label{eq:27}
\end{equation}
The luminosity is proportional to the velocity dispersion on scale $r_p$ with a scaling of $L_B \propto \sigma_p^4$. Therefore, the constant $\eta$ in the $\eta-\gamma$ plane leads to a constant $\sigma_p$ (dashed blue lines in Fig. \ref{fig:3}).\\

\item \noindent Similarly, the velocity dispersion $\sigma_B$ on scale $r_B$ (BH sphere of influence) reads (using Eq. \eqref{eq:10} for $\sigma_B$ and Eq. \eqref{eq:15} for $r_B$)
\begin{equation}
\frac{\sigma_B}{c}=\frac{(\alpha_r\gamma_r)^{1/2}}{\alpha_r^{1/5}} \left(\frac{\eta^4}{\gamma}\right)^{\frac{1}{5}}, \quad M_{BH} = \alpha_r^{-3/2}\gamma_r^{-5/2}\frac{\sigma_B^5}{\varepsilon_b G},
\label{eq:30}
\end{equation}
such that we should have constant $\sigma_B$ along constant $\eta^4/\gamma$ lines (dashed black lines in Fig. \ref{fig:3}). The BH mass scales with velocity dispersion on scale $r_B$ as $M_{BH} \propto \sigma_B^5$ that can be further related to velocity dispersion $\sigma_b$ on the bulge scale $r_b$ as (using Eq. \eqref{eq:21} for the ratio $\sigma_B/\sigma_b$),
\begin{equation}
M_{BH} = \left[\alpha_r^{-3/2}\gamma_r^{-5/2}\right]\frac{B(z)\sigma_b^5}{\varepsilon_b(z) G},
\label{eq:31}
\end{equation}
where B(z) is the proportional coefficient of correlation $M_{BH}=B(z)M_b$ in Eq. \eqref{eq:2}. With the average mass ratio $B(z=0)$ = [0.002 0.003] \citep{Marconi:2003-The-relation-between-blac} and $\varepsilon_{b0}=10^{-4}m^2/s^3$ for local Universe (constants $\alpha_r\gamma_r=1$ and $\alpha_r=1.7$ from Fig. \ref{fig:S8}), the prediction of Eq. \eqref{eq:31} matches the widely accepted $M_{BH}$-$\sigma$ relation in Eq. \eqref{eq:1-2} \citep{Ferrarese:2005-Supermassive-Black-Holes-in-Galactic-Nuclei}, that is
\begin{equation}
\frac{M_{BH}}{10^8M_{\odot}} \approx \left(\frac{\sigma_b}{200km/s}\right)^5.
\label{eq:32}
\end{equation}
Recent study for JWST SMBHs supports a redshift-dependent coefficient $B(z)\propto (1+z)^{5/2}$ \citep{Pacucci:2024-The-Redshift-Evolution-of-the}. Since the key parameter $\varepsilon(b)\propto (1+z)^{5/2}$, in this case, Eq. \eqref{eq:31} leads to a relatively redshift-independent $M_{BH}-\sigma_b$ correlation.

\begin{figure}
\includegraphics*[width=\columnwidth]{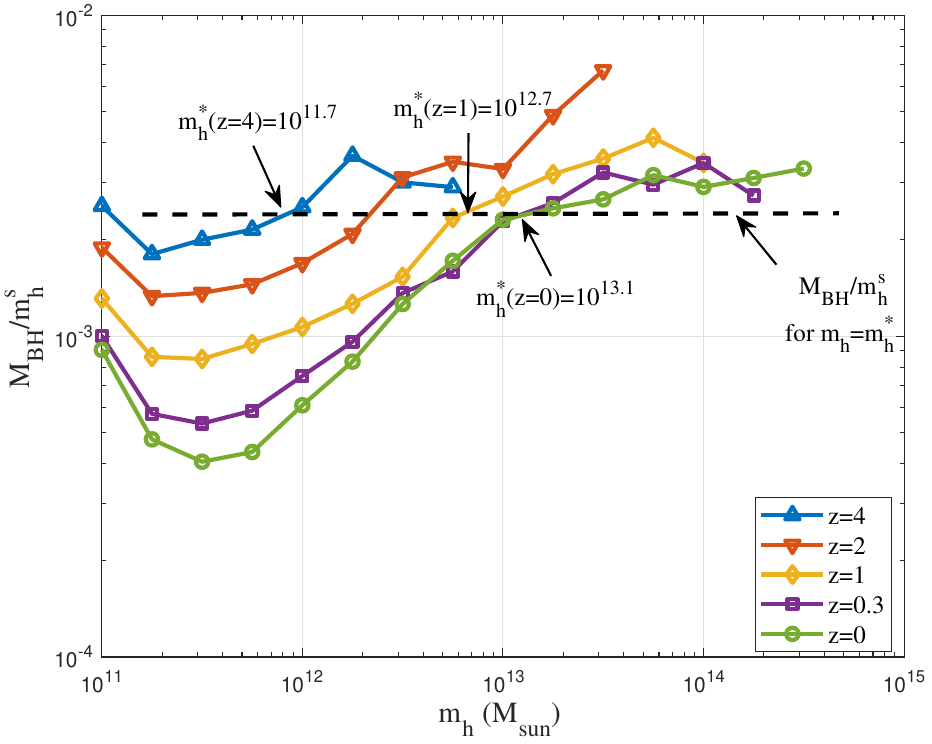}
\caption{The variation of the ratio between BH mass ($M_{BH}$) and stellar mass ($m_h^s$ in Fig. \ref{fig:S2}) from Illustris-1 simulations for haloes of different mass $m_h$. That ratio is averaged for all haloes of the same mass $m_h$. The mass ratio decreases with time for a fixed halo mass $m_h$ and increases with halo mass $m_h$ for a fixed redshift. For haloes of mass $m_h^*(z)$ 
at different redshifts, that ratio is almost constant with a value around 0.0025 that matches the BH-bulge mass ratio in local Universe \citep{Marconi:2003-The-relation-between-blac}.} 
\label{fig:S14}
\end{figure}

Figure \ref{fig:S14} presents results from Illustris-1 simulations for the evolution of the ratio between the BH mass and the stellar mass in haloes of different masses $m_h$. This ratio is around 0.0025 and is independent of $z$ for haloes of characteristic mass $m_h^*(z)$. Figures \ref{fig:S15} plot the $M_{BH}-\sigma$ correlation between the mass of BH and the stellar velocity dispersion. The simulation seems to point to a slightly redshift-dependent $M_{BH}-\sigma$ correlation. At fixed velocity dispersion $\sigma_b^2$, the BH mass $M_{BH}$ is slightly smaller at higher redshift. However, more observational data at high redshift are required as conclusive evidence. \\

\item \noindent The boundary of active and inactive SMBHs is established by setting two scales $r_p=r_x$. For $r_p\gg r_x$, the motion of gas on the scales $r>r_p$ does not feel the effect of the accretion disk (see $r_x$ in Eq. \eqref{eq:17}). The energy flow in the bulge can be at a rate of $\varepsilon_b$ down to the radiation scale $r_p$ for an active SMBH phase. However, for $r_p<r_x$, due to the dissipation effect of the accretion disk, the rate of energy flow that reaches the radiation scale $r_p$ can be much lower than $\varepsilon_b$, leading to less mass and energy supply and the inactive phase of SMBH. Therefore, from Eq. \eqref{eq:20}, a critical $\gamma$ reads
\begin{equation}
\gamma_c={6z_r\alpha_r^2\gamma_r}\approx 10,
\label{eq:33}
\end{equation}
such that $L_B=\gamma_c\varepsilon_b M_{BH}$ along this boundary. This critical value of $\gamma$ can be used to classify SMBHs as active and inactive (black dotted horizontal line in Fig. \ref{fig:3}). Most Seyfert galaxies (red squares) are above that boundary, while others (blue squares) are below that boundary. The distributions of quasars at different redshifts are also bounded between that boundary and the upper limit (solid red line in Fig. \ref{fig:3}). 
The lower limit of the Eddington ratio for active quasars ($\lambda_{min}$) can be determined from this critical value of $\gamma$ (Eq. \eqref{eq:7-1-4}).  

\end{enumerate}

\section{Deriving the SMBH evolution in \texorpdfstring{$\gamma$-$\eta$}{} plane}
\label{sec:7}
Power-law relations between BH luminosity $L_B$ and mass $M_{BH}$ were obtained when identifying the upper and lower limits for the distribution of SMBHs in $\gamma$-$\eta$ plane (Fig. \ref{fig:3}), i.e. $L_B \propto (\varepsilon_b M_{BH})^{4/5}$ in Eq. \eqref{eq:24} and $L_B \propto (\varepsilon_b M_{BH})^{4/3}$ in Eq. \eqref{eq:26}). The evolution of SMBH can also be formulated to follow these power-law scalings to be consistent with the upper and lower limits without breaking these limits. These power-law relations also reflect the effects of cosmic quenching on the evolution of SMBH through the key parameter $\varepsilon_b(z)$, which quantifies the rate of energy flow and the efficiency of gas cooling (Section \ref{sec:1-1-1}). The larger $\varepsilon_b$ at higher redshift $z$ means a more efficient gas cooling and a richer supply of cold gas for both star formation and SMBH mass accretion. For a general power law $L_B \propto (\varepsilon_b M_{BH})^b$, the BH mass $M_{BH}$ increases monotonically with time, while $\varepsilon_b\propto a^{-5/2}$ (Eq. \eqref{ZEqnNum9863112994}) decreases with time; a maximum rate of mass accretion (or luminosity $L_B$) naturally emerges around $z=2$ (see Fig. \ref{fig:4}). The smaller $\varepsilon_b$ at a lower redshift means less efficient gas cooling and a slower rate of mass flow $\dot{m}_r$ (Eq. \eqref{eq:11}), shuts down the cold gas supply and slows BH mass accretion. Similarly, controlled by the same parameter, the star formation rate peaks around the same redshift (Fig. \ref{fig:S37}).

This section proposes a three-phase evolution model based on these power-law relations. This evolution model will be applied later to derive the BH mass functions, the AGN duty cycle, and the AGN mass functions (Sections \ref{sec:7-1}, \ref{sec:7-3}, and \ref{sec:7-2-1}). Inspired by these power-law relations, we begin with the general solution of $M_{BH}$, assuming that the luminosity $L_B$ follows a general power-law as 
\begin{equation}
L_B = \alpha_0 \varepsilon_b^{p}M_{BH}^{1-\sigma},
\label{eq:34}
\end{equation}
where the pre-factor $\alpha_0$, exponents $p$ and $\sigma$ can be different at different stage (see Table \ref{tab:2}). The second equation relating the rate of mass accretion and BH luminosity reads (same as Eq. \eqref{eq:22}):
\begin{equation}
\frac{dM_{BH}}{dt} = L_B \frac{1-\epsilon}{\epsilon c^2},
\label{eq:35}
\end{equation}
where $\epsilon$ is the radiative efficiency and $c$ is the speed of light. Substituting $L_B$ from Eq. \eqref{eq:34} and the rate of energy flow $\varepsilon_b=\varepsilon_{b0} a^{-m}$ ($m=5/2$ in Eq. \eqref{ZEqnNum9863112994}), Eq. \eqref{eq:35} can be explicitly solved with respect to the scale factor $a$. 
The general solution for $M_{BH}$ reads
\begin{equation}
M_{BH} = M_{\infty} \left[1-\frac{\sigma}{|\sigma|} \left(\frac{a}{a_i}\right)^{-mp+\frac{3}{2}}\right]^{\frac{1}{\sigma}}.
\label{eq:38}
\end{equation}
The solution involves two key parameters: a mass scale $M_{\infty}=M_{BH}(t=\infty)$ and a characteristic scale factor $a_i$. Two parameters are related to other model parameters as
\begin{equation}
\begin{split}
M_{\infty}^{\sigma}a_i^{(mp-3/2)}=\frac{2|\sigma|}{2mp-3}\frac{\alpha_0(1-\epsilon)}{\epsilon c^2}\frac{\varepsilon_{b0}^p}{H_0},
\end{split}
\label{eq:38-1}
\end{equation}
where $H_0$ is the Hubble constant and $\varepsilon_{b0}\equiv \varepsilon_b(z=0)=10^{-4}m^2/s^3$.
This general solution of $M_{BH}$ (Eq. \eqref{eq:38}) depends only on three exponents ($\sigma$, $p$ and $m$), a mass scale $M_{\infty}$, and a scale factor $a_i$. This general solution can be applied to different stages of evolution. 
\begin{enumerate}
\item \noindent Co-evolution stage ("E1" of the dashed green line in Fig. \ref{fig:3} that is parallel to the upper limit, also shown in Fig. \ref{fig:4}). In this stage, the radiation scale co-evolves with the scale of the BH sphere of influence ($r_p \propto r_B$). SMBHs evolve along the line $\gamma\propto \eta^{-1}$. Assuming $r_p=\xi_r r_B$ with $\xi_r\le 1$, the evolution of SMBH follows
\begin{equation}
\gamma\eta = \frac{(3\alpha_r^2\gamma_r)^{5/4}}{\alpha_r}\xi_r^{5/3}
\label{eq:40}
\end{equation}
from Eq. \eqref{eq:23}, where $\xi_r$ indicates the distance between the evolution path and the upper limit (that is, $\xi_r=1$ means $r_p=r_B$). Since $r_p \propto r_B$, both SMBH and host galaxy evolve together with a rapid increase in both size and mass during this stage.
If $\beta \approx 1$ (BH luminosity $L_B$ is comparable to the energy flow into the bulge $\varepsilon_bM_b$), with $\gamma$ (Eq. \eqref{eq:18}) decreasing over time, the mass ratio $M_{BH}/M_b=\beta/\gamma$ increases over time (Eq. \eqref{eq:21}) and remains almost constant after this stage (Figs. \ref{fig:3} and \ref{fig:4}). The BH luminosity in this stage reads (from Eq. \eqref{eq:24})
\begin{equation}
L_B=3\alpha_r^{6/5}\gamma_r\xi_r^{4/3}\varepsilon_b^{4/5}M_{BH}^{4/5}G^{-1/5}c.
\label{eq:41}
\end{equation}
For $\beta$ in Eq. \eqref{eq:18} varying with time, the mass of the bulge reads
\begin{equation}
M_b = \frac{L_B}{\beta\varepsilon_b}=\frac{3}{\beta}\alpha_r^{6/5}\gamma_r\xi_r^{4/3}\varepsilon_b^{-1/5}M_{BH}^{4/5}G^{-1/5}c.
\label{eq:41-2}
\end{equation}
The evolution of the Eddington ratio should read
\begin{equation}
\lambda= \frac{L_B}{L_{Edd}} \propto \frac{\varepsilon_b^{4/5}c}{\varepsilon_{Edd} M_{BH}^{1/5}G^{1/5}} \propto a^{-2}M_{BH}^{-1/5},
\label{eq:42}
\end{equation}
where $\varepsilon_{Edd}=L_{Edd}/M_{BH}=6.3m^2/s^3$ is the rate of energy flow corresponding to the Eddington limit. The Eddington luminosity satisfies $L_{Edd}=1.26 \times 10^{38}(M_{BH}/M_{\odot})$ erg/s.

For $m=5/2$ from Eq. \eqref{ZEqnNum9863112994}, $p=4/5$ and $\sigma=1/5$ from Eq. \eqref{eq:41} (values are listed in Table \ref{tab:2}), the evolution of the BH mass at this stage can be obtained from the general solution in the Eq. \eqref{eq:38},
\begin{equation}
M_{BH} = M_{\infty1}\left[1-\left(\frac{a}{a_1}\right)^{-\frac{4}{5}m+\frac{3}{2}}\right]^5.
\label{eq:43}
\end{equation}
An initial scale factor $a_1$ exists when SMBH is formed and grows rapidly, i.e., $M_{BH}(a_1)=0$. 
The earlier SMBH is formed (the smaller $a_1$), the faster it grows. Individual SMBH may have different $M_{\infty 1}$ and $a_1$. However, for the evolution of a typical SMBH derived from the quasar luminosity function (solid black line in Fig. \ref{fig:3}), $\xi_r=r_p/r_B \approx 0.052$. The radiation scale $r_p$ is about 5\% of the BH sphere of influence. We also found $M_{\infty 1}=1.3\times 10^{11} M_{\odot}$ and $a_1 \approx 0.142$ to give the best fit (see Fig. \ref{fig:4}). The model (green dashed line) matches the evolution of the BH mass from the quasar luminosity function (solid blue line). At the end of this stage (point "P1"), the BH luminosity $L_B$ reaches its maximum (owing to the decreasing $\varepsilon_b$ and increasing $M_{BH}$ in Eq. \eqref{eq:41}). In this stage, we find the scaling $M_{BH} \propto \sigma_p^5/(\varepsilon_bG)$ (Eqs. \eqref{eq:27} and \eqref{eq:41}),
\begin{equation}
M_{BH} = \left[\alpha_r^{-3/2}\gamma_r^{-5/2}\xi_r^{-5/3}\right]\frac{\sigma_p^5}{\varepsilon_b G},
\label{eq:45}
\end{equation}
where $\sigma_p$ is the typical gas velocity on the radiation scale $r_p$.
\\

\item \noindent Transitional stage ("E2" of the dashed green line in Figs. \ref{fig:3} and \ref{fig:4}) following the scaling $\gamma \propto \eta^2$ in $\gamma$-$\eta$ plane. At this stage, the evolution of the BH sphere of influence $r_B$ is gradually decoupled from the radiation scale $r_p$. The ratio $r_B/r_p$ increases over time (see Eq. \eqref{eq:20}) due to the decreasing luminosity $L_B$ or the parameter $\eta$. The scale $r_B$ increases with increasing $M_{BH}$, while the radiation scale $r_p$ may decrease with time. The evolution of the two scales $r_B$ and $r_p$ is gradually decoupled at this stage. In this stage, the BH luminosity $L_B$ decreases with time, while there is only a slight increase in the BH mass $M_{BH}$ and the bulge mass $M_b$. The ratio $r_b/r_B$ or $M_b/M_{BH}$ remains almost constant (Eq. \eqref{eq:21}). 
 
This stage is well between the upper and lower limits in Fig. \ref{fig:3}, where the radiation scale $r_p$ satisfies $r_s\ll r_p\ll r_B$. In this range, the radiation scale is much smaller than the BH sphere of influence but still much larger than the Schwarzschild radius. The mass and energy flow are the dominant physics that impact the evolution of BH luminosity. First, the luminosity $L_B$ should depend on the rate of energy flow $\varepsilon_b$ and the BH mass $M_{BH}$. Second, the luminosity peaks at the transition time $a^*$ and declines after $a^*$ (Fig. \ref{fig:4}). The luminosity at a later time $a$ depends on the timespan between $a^*$ and $a$. Different SMBHs have different transition time $a^*$ and different values of $\varepsilon^*_b\equiv\varepsilon_b(a^*)$ at $a^*$. In principle, the earlier the transition or a smaller $a^*$, the larger $\varepsilon^*_b$ at $a^*$, the longer timespan between $a^*$ and $a$, and the smaller $L_B$ at a later time $a$. The luminosity $L_B$ should also depend on the value of $\varepsilon^*_b$ at the transition. Without loss of generality, we can express $L_B\propto (\varepsilon^*_b)^x \varepsilon_b^y M_{BH}^z$ for $a\gg a^*$. Based on a simple dimensional analysis, we find exponents $z=1$ and $x+y=1$. Therefore, the BH luminosity $L_B$ follows the power-law
\begin{equation}
L_B = \gamma^* (\varepsilon_b^*)^{1-p} \varepsilon_b^p M_{BH} \quad \textrm{and} \quad L_B^* = \gamma^* \varepsilon_b^* M_{BH}^*,
\label{eq:46}
\end{equation}
where $p$ is an exponent to be determined. Here $\gamma^*$ and $\varepsilon_b^*$ represent the values of $\gamma$ and $\varepsilon_b$ at the peak luminosity of point "P1" (the superscript '*' represents the values at "P1" in Figs. \ref{fig:3} and \ref{fig:4}). 

To determine the exponent $p$, substitution of $L_B$ in Eq. \eqref{eq:46} into Eqs. \eqref{eq:9} and \eqref{eq:16} leads to the evolution of radiation scale $r_p$ and relevant physical quantities on scale $r_p$ (density $\rho_p$, characteristic time $t_p=\sigma_p/r_p$, pressure $P_p$, and velocity dispersion $\sigma_p^2$),
\begin{equation}
\begin{split}
&r_p \propto \varepsilon_b^{\frac{3}{4}p-1}M_{BH}^{\frac{3}{4}}, \quad \rho_p \propto \varepsilon_b^{2-p}M_{BH}^{-1}, \quad t_p\propto  \varepsilon_b^{\frac{p}{2}-1}M_{BH}^{\frac{1}{2}},\\
&P_p\propto \varepsilon_b^{2-\frac{p}{2}}M_{BH}^{-\frac{1}{2}} \quad \textrm{and}\quad \sigma_p^2\propto \varepsilon_b^{\frac{p}{2}}M_{BH}^{\frac{1}{2}},
\end{split}
\label{eq:46-12}
\end{equation}
where parameter $\varepsilon_b\propto a^{-5/2}$. We expect these quantities on scale $r_p$ not to diverge with time going to infinity. This requires $p\le 2$ for density $\rho_p$ not to diverge and $p\ge 2$ for the characteristic time $t_p$ (or the gas cooling time on scale $r_p$ in Section \ref{sec:1-1-1}) not to diverge. Therefore, $p=2$ is required such that the gas density $\rho_p \equiv \rho_r(r=r_p)$ and the characteristic time $t_p$ remains constant during this stage of evolution. Other quantities on the scale $r_p$ reads
\begin{equation}
\sigma_p \propto \varepsilon_b^{1/2}M_{BH}^{1/4}, \quad r_p \propto \varepsilon_b^{1/2}M_{BH}^{3/4} \quad \textrm{and} \quad P_p\propto \varepsilon_bM_{BH}^{-1/2}. 
\label{eq:46-11}
\end{equation}
All of these quantities decrease over time.
For $p=2$ and an almost constant BH mass in this stage, Eq. \eqref{eq:46} is equivalent to the evolution path of $\gamma\propto \eta^2$ in the $\gamma$-$\eta$ plane. This is confirmed by the evolution of a typical SMBH in Fig. \ref{fig:3} (solid black line) from the observed quasar luminosity function in Fig. \ref{fig:4}. In addition, the luminosity of BH decreases with time as $L_B\propto \varepsilon_b^2\propto a^{-5}$ in this stage. 

With BH luminosity $L_B$ from Eq. \eqref{eq:46}, the solution of the BH mass $M_{BH}$ can also be directly obtained from the general solution Eq. \eqref{eq:38} (with $p=2$ and $\sigma=0$)
\begin{equation}
M_{BH} = M_{\infty2} \exp \left[-\left(\frac{a}{a_2}\right)^{-mp+\frac{3}{2}}\right].
\label{eq:48}
\end{equation}
Here, we invoke the identity for exponent function $\textbf{lim}(1+x/n)^n=e^x$ with $n\rightarrow \infty$. Similarly to the mass evolution in the first stage E1 (Eq. \eqref{eq:38}), the evolution in stage E2 involves a mass scale $M_{\infty 2}$ and a characteristic scale factor $a_2$. Similarly to Eq. \eqref{eq:38-1}, parameter $a_2$ is related to other parameters such as:
\begin{equation}
a_2^{(mp-3/2)}=\frac{2}{2mp-3}\frac{\alpha_2(1-\epsilon)}{\epsilon c^2}\frac{\varepsilon_{b0}^p}{H_0},
\label{eq:48-1}
\end{equation}
where $\alpha_2 =  \gamma^* (\varepsilon_b^*)^{1-p}$ is the pre-factor of the general power-law $L_B=\alpha_2\varepsilon_b^p M_{BH}$ in Eq. \eqref{eq:34}. 

The individual SMBH may have different parameters $\gamma^*$ and $\varepsilon_b^*$. For the evolution of a typical SMBH in Figs. \ref{fig:3} and \ref{fig:4}, the time to reach maximum luminosity is $a^* \approx 0.28$ with $\gamma^* \approx 250$ and $\varepsilon_b^* \approx 0.0025m^2/s^3$, such that $a_2 \approx 0.31$. With $M_{\infty2}=10^9 M_{\odot}$, the mass evolution in this stage (Eq. \eqref{eq:48}) is also presented in Fig. \ref{fig:4}, which is in agreement with the BH mass accretion from the quasar luminosity function. At this stage, SMBH is initially active and becomes inactive when $\gamma$ is less than the critical value $\gamma_c$ in Eq. \eqref{eq:33} (or when $r_p<r_x$). Finally, at this stage, the BH mass $M_{BH}$ evolves as $M_{BH} \propto \sigma_p^4$ (from Eqs. \eqref{eq:27} and \eqref{eq:46}),
\begin{equation}
M_{BH} = \left(\frac{3}{\gamma_r\gamma^*}\right)\left(\frac{\varepsilon_b}{\varepsilon_b^*}\right)^{1-p}\frac{\sigma_p^4c}{\varepsilon_b G},
\label{eq:50}
\end{equation}
where $\sigma_p$ is the typical velocity on the radiation scale $r_p$.

The power-law relation for BH luminosity (Eq. \eqref{eq:34}) and the evolution of the SMBH mass (Eqs. \eqref{eq:43} and \eqref{eq:48}) will be applied to formulate the evolution of BH mass functions, the AGN duty cycle, the AGN mass functions, and the Eddington ratio distribution in Sections \ref{sec:7-1} to \ref{sec:7-2}.\\

\item \noindent Completely dormant stage with $\gamma \propto \eta$ ("E3" of the dashed green line in Fig. \ref{fig:3}). This is the limiting stage. During this stage, BH mass accretion is extremely slow with vanishing luminosity $L_B$. The radiation scale $r_p$ is fully decoupled from the BH sphere of influence $r_B$ and equals the Schwarzschild radius, that is, $r_p=r_s$ (the lower limit in Fig. \ref{fig:3}). In this stage, the evolution of the BH mass $M_{BH}$ can be obtained from the general solution (Eq. \eqref{eq:38}) with $m=5/2$, $p=4/3$, and $\sigma=-1/3$ (see $L_B$ in Eq. \eqref{eq:26}), 
\begin{equation}
M_{BH} = M_{\infty3}\left[1+\left(\frac{a}{a_3}\right)^{-\frac{4}{3}m+\frac{3}{2}}\right]^{-3}.
\label{eq:51}
\end{equation}
Furthermore, the BH mass $M_{BH}$ evolves as $M_{BH} \propto \sigma_p^3$ in this stage (from Eqs. \eqref{eq:26} and \eqref{eq:27}),
\begin{equation}
M_{BH} = \left(\frac{\alpha_r^{-3/2}\gamma_r^{-3/2}}{2}\right)\frac{\sigma_p^3c^2}{\varepsilon_b G}.
\label{eq:53}
\end{equation}
Again, $\sigma_p$ is the typical velocity on the radiation scale $r_p$.

\end{enumerate}

\begin{table}
    \caption{Estimated values of $m$, $p$, and $\sigma$ for each evolution stage (values of $M_{\infty}$ and $a_i$ are for the evolution of a typical SMBH in Figs. \ref{fig:3} and \ref{fig:4}).}
    \label{tab:2}
\centering
    \begin{tabular}{p{0.5in}p{0.7in}p{0.5in}p{0.5in}}
    \hline
     Quantity             &  Stage "E1"                       & Stage "E2"                    & Stage "E3"     \\
    \hline
    $L_B$                 &  Eq. \eqref{eq:41}                & Eq. \eqref{eq:46}             & Eq. \eqref{eq:26}         \\
    \hline   
    $M_B$                 &  Eq. \eqref{eq:43}                & Eq. \eqref{eq:48}             & Eq. \eqref{eq:51}         \\
    \hline   
    $m$                   & 5/2                               & 5/2                           & 5/2                       \\
    \hline
    $p$                   & 4/5                               & 2                             & 4/3                       \\
    \hline
    $\sigma$              & 1/5                               & 0                             & -1/3                      \\
    \hline
    $M_{\infty}$          & $1.3\times 10^{11} M_{\odot}$    & $10^9 M_{\odot}$              & $10^9 M_{\odot}$                \\
    \hline
    $a_i$                 & 0.142                             & 0.31                          & $1.8\times 10^{-4}$                      \\
    \hline
    \end{tabular}
\end{table}

\section{Deriving the BH mass function}
\label{sec:7-1}
In this section, we apply the three-phase evolution model to derive the evolution of the SMBH population over time, which is often described by a continuity equation in mass space \citep{Shankar:2013-Accretion-driven-evolution-of-black-holes, Tucci:2017-Constraining-supermassive-black-hole-evolution}, 
\begin{equation}
\frac{\partial \Phi_{BH}}{\partial t}(M,z) + M \frac{\partial}{\partial M} \left[\frac{\langle\dot M\rangle (M,z)} {M} \Phi_{BH}(M,z) \right]=0,
\label{eq:7-1-1}
\end{equation}
where $\Phi_{BH}(M,z)=\Phi^*_{BH}(M,z)M\ln(10)$ is the BH mass function defined in the logarithmic units of BH mass $M$, while $\Phi^*_{BH}(M,z)$ is the usual mass function
defined as the SMBH number density per co-moving volume with a mass in the interval $M$ and $M+dM$. Here $\langle\dot M\rangle (M,z)$ is the average accretion rate for all SMBHs of the same mass $M$ at any redshift $z$, a key quantity for driving the evolution of the BH mass function. 

To compute $\langle\dot M\rangle$, we first define the Eddington ratio
\begin{equation}
\lambda \equiv \frac{L}{L_{Edd}}=\frac{L}{\varepsilon_{Edd}M}=\gamma\frac{\varepsilon_b}{\varepsilon_{Edd}}\quad \textrm{and} \quad \gamma=\frac{L}{\varepsilon_bM},
\label{eq:7-1-2}
\end{equation}
where $L_{Edd}=\varepsilon_{Edd}M$ is the Eddington luminosity, the constant $\varepsilon_{Edd}=L_{Edd}/M=6.3m^2/s^3$ is the rate of $\varepsilon$ corresponding to the Eddington limit, and parameter $\gamma$ is introduced in Eq. \eqref{eq:18}. For SMBHs with a mass $M$ and luminosity $L$, the parameter $\gamma$ has a physical meaning similar to the Eddington ratio $\lambda$. However, unlike the Eddington ratio,  $\gamma$ also reflects the effect of cosmic quenching through the redshift variation of $\varepsilon_b$. 

Next, the probability distribution of the Eddington ratio $P(\lambda\vert M,z)$ defines the fraction of SMBH of mass $M$ that accretes at the Eddington ratio $\lambda$ per unit $\log\lambda$ at redshift $z$. The average accretion rate $\dot M$ can be calculated by integrating the accretion of all active SMBHs at a fixed mass $M$,
\begin{equation}
\begin{split}
\langle\dot M\rangle &= \varepsilon_{Edd}\frac{1-\epsilon}{\epsilon c^2} \int_{\lambda_{min}}^{\lambda_{max}}P(\lambda\vert M,z)\lambda MU(M,z)d\log\lambda,\\
&=\varepsilon_{Edd}\frac{1-\epsilon}{\epsilon c^2}MU(M,z)\langle \lambda \rangle(M,z),
\end{split}
\label{eq:7-1-3}
\end{equation}
where $\epsilon$ is the radiative efficiency, $U(M,z)$ is the AGN duty cycle, i.e., the fraction of SMBHs of mass $M$ that are active with an Eddington ratio $\lambda > \lambda_{min}$ at redshift $z$. The integral is extended to all active SMBHs with $\lambda$ greater than a minimum value $\lambda_{min}$ and less than a maximum value $\lambda_{max}$. The average Eddington ratio reads
\begin{equation}
\langle \lambda \rangle(M,z) = \int_{\lambda_{min}}^{\lambda_{max}}P(\lambda\vert M,z)\lambda d\log\lambda. 
\label{eq:7-1-3-1}
\end{equation}

The average accretion rate $\langle\dot M\rangle$ depends on the Eddington ratio distribution and the duty cycle. The critical value of $\gamma_c \approx 10$ (see Fig. \ref{fig:3} and Eq. \eqref{eq:33}) can be related to the minimum Eddington ratio for active SMBHs as (using Eq. \eqref{eq:7-1-2})
\begin{equation}
\lambda_{min}=\gamma_c\frac{\varepsilon_b}{\varepsilon_{Edd}}=\gamma_c\frac{\varepsilon_{b0}}{\varepsilon_{Edd}}a^{-5/2}\approx 10^{-4}a^{-5/2}.
\label{eq:7-1-4}
\end{equation}

By introducing the duty cycle $U(M,z)$, the mass functions of all active SMBHs (referred to as the AGN mass function) can be related to the total BH mass function as 
\begin{equation}
\Phi_{AGN}(M,z)\equiv \Phi_{BH}(M,z) U(M,z). 
\label{eq:7-1-4-1}
\end{equation}
Finally, the AGN mass function can be related to the quasar luminosity function by the convolution equation
\begin{equation}
\Phi_{L}(L,z) = \int_{\lambda_{min}}^{\lambda_{max}} P(\lambda\vert M,z)\Phi_{AGN}(M,z) d\log\lambda.
\label{eq:7-1-5}
\end{equation}

Integrating the continuity Eq. \eqref{eq:7-1-1} with respect to the BH mass $M$ leads to the evolution of BH mass density $\rho_{BH}(z)$ (using Eq. \eqref{eq:7-1-3}),
\begin{equation}
\begin{split}
&\rho_{BH}(z) = \int \Phi_{BH}(M,z)Md\log M,\\
&\frac{\partial \rho_{BH}}{\partial t} = \int {\langle\dot{M}\rangle}\Phi_{BH}(M,z)d\log M,\\
&\frac{\partial \rho_{BH}}{\partial t} = \varepsilon_{Edd} \frac{1-\epsilon}{\epsilon c^2}\int \Phi_{AGN}(M,z)M\langle \lambda \rangle d\log M.\\
\end{split}
\label{eq:7-1-5-1}
\end{equation}
The mean Eddington ratio can also be computed from the AGN mass function and quasar luminosity function as
\begin{equation}
\langle \lambda \rangle = \frac{\int \Phi_{L}(L,z)L d\log L }{\varepsilon_{Edd}\int \Phi_{AGN}(M,z)M d\log M}.
\label{eq:7-1-5-2}
\end{equation}

By inserting Eq. \eqref{eq:7-1-5-2} into  Eq. \eqref{eq:7-1-5-1}, we can directly relate the evolution of BH mass density with the quasar luminosity function
\begin{equation}
\frac{\partial \rho_{BH}}{\partial t} = \frac{1-\epsilon}{\epsilon c^2} \int \Phi_{L}(L,z)L d\log L.
\label{eq:7-1-5-3}
\end{equation}
This equation is often used to estimate the BH density evolution based on the observed quasar luminosity function. 

With a given luminosity function $\Phi_L(L,z)$ and the Eddington ratio distribution $P(\lambda\vert M,z)$ from observations, Eqs. \eqref{eq:7-1-1} to \eqref{eq:7-1-5} provide a self-closed set of equations for the evolution of the SMBH and AGN mass functions and the AGN duty cycle. Although very complex, this set of equations can be numerically solved if a double power law is assumed for either the AGN mass function $\Phi_{AGN}(M,z)$ \citep{Cao:2010-Cosmological-Evolution-of-Massive-Black-Holes} or the AGN duty cycle $U(M,z)$ \citep{Tucci:2017-Constraining-supermassive-black-hole-evolution}. 

It is noted that the average mass accretion rate $\langle\dot M\rangle$ is a central quantity in the current formulation. With a known model for $\langle\dot M\rangle$ and the local BH mass function as a boundary condition at $z=0$ for continuity Eq. \eqref{eq:7-1-1}, the evolution of the BH mass function $\Phi_{BH}(M,z)$ can be solved from the continuity Eq. \eqref{eq:7-1-1}. On the other hand, with the same known model for $\langle\dot M\rangle$ and a model for the average Eddington ratio $\langle \lambda \rangle$, the AGN duty cycle $U(M,z)$ is completely known from Eq. \eqref{eq:7-1-3}. Therefore, the evolution of the AGN mass function $\Phi_{AGN}(M,z)$ can be obtained with the known model for $\langle\dot M\rangle$. The resulting AGN mass function should be consistent with the observed luminosity function $\Phi_L(L,z)$ through Eq. \eqref{eq:7-1-5}.

In this work, without numerically solving these complicated equations, we attempt to apply the power law relation for luminosity $L$ (or $\langle\dot M\rangle$) we established (Eq. \eqref{eq:34}) to derive analytical solutions for the evolution of the SMBH and AGN mass functions and the AGN duty cycle. First, these analytical solutions will provide physical insight into the complex evolution of the SMBH population. Second, it is hoped that these analytical solutions can be compared with numerical solutions and observations to test the validity of the power-law evolution model of Eq. \eqref{eq:34}. With this in mind, we first model the average accretion rate $\langle\dot M\rangle$ that can be related to the average luminosity $\langle L \rangle$ for all SMBHs of the same mass $M$ (using Eq. \eqref{eq:35}), 
\begin{equation}
\langle \dot M\rangle = \left\langle \frac{dM}{dt} \right\rangle = \langle L \rangle \frac{1-\epsilon}{\epsilon c^2}.
\label{eq:7-1-6}
\end{equation}

The evolution of active SMBHs involves two separate stages: the rising stage "E1" with a rising luminosity before reaching the peak luminosity $L_p$ and the declining stage "E2" with a declining luminosity after reaching the peak $L_p$ (Figs. \ref{fig:3} and \ref{fig:4}). To solve the analytical BH mass function, since most SMBHs are in either stage E1 or stage E2, we focus on the luminosity in stages "E1" and "E2". A two-stage luminosity model from Eqs. \eqref{eq:34} and \eqref{eq:46} reads 
\begin{equation}
\begin{split}
&\textrm{Model 1:}\quad \langle L \rangle = \alpha_1 \varepsilon_b^p M^{1-\sigma} \quad \textrm{Stage E1 (Figs. \ref{fig:3} and \ref{fig:4}}), \\
&\textrm{Model 2:}\quad \langle L \rangle = \alpha_2 \varepsilon_b^p M \quad \quad \quad \textrm{Stage E2 (Figs. \ref{fig:3} and \ref{fig:4}}).
\end{split}
\label{eq:7-1-7}
\end{equation}
Without loss of generality, we use $\varepsilon_b=\varepsilon_{b0} a^{-m}$. The predicted values of $m$, $p$, and $\sigma$ are listed in Table \ref{tab:2}. Stage E1 corresponds to the evolution $\gamma\propto \eta^{-1}$ in the $\gamma$-$\eta$ space, while stage E2 corresponds to the evolution $\gamma\propto \eta^2$ in $\gamma$-$\eta$ space (see Fig. \ref{fig:3}).  

The average mass accretion rate $\langle\dot M\rangle$ is related to the average luminosity $\langle L\rangle$ (Eq. \eqref{eq:7-1-6}). Insert the evolution model (Eq. \eqref{eq:7-1-7}) into Eq. \eqref{eq:7-1-6} and the continuity Eq. \eqref{eq:7-1-1}, analytical solutions of the BH mass function can be obtained for Model 1 and Model 2, respectively. The local BH mass function at $z=0$ is used as a boundary condition, that is, $\Phi_{BH0}(M)\equiv \Phi_{BH}(M,z=0)$. Since most small active SMBHs at low redshift are still in stage E1 with a rising luminosity and fast mass accretion, while large active SMBHs are already in stage E2 with a declining luminosity and slow mass accretion, the final solution of the BH mass function can be constructed as an interpolation of two solutions. 

\begin{figure}
\includegraphics*[width=\columnwidth]{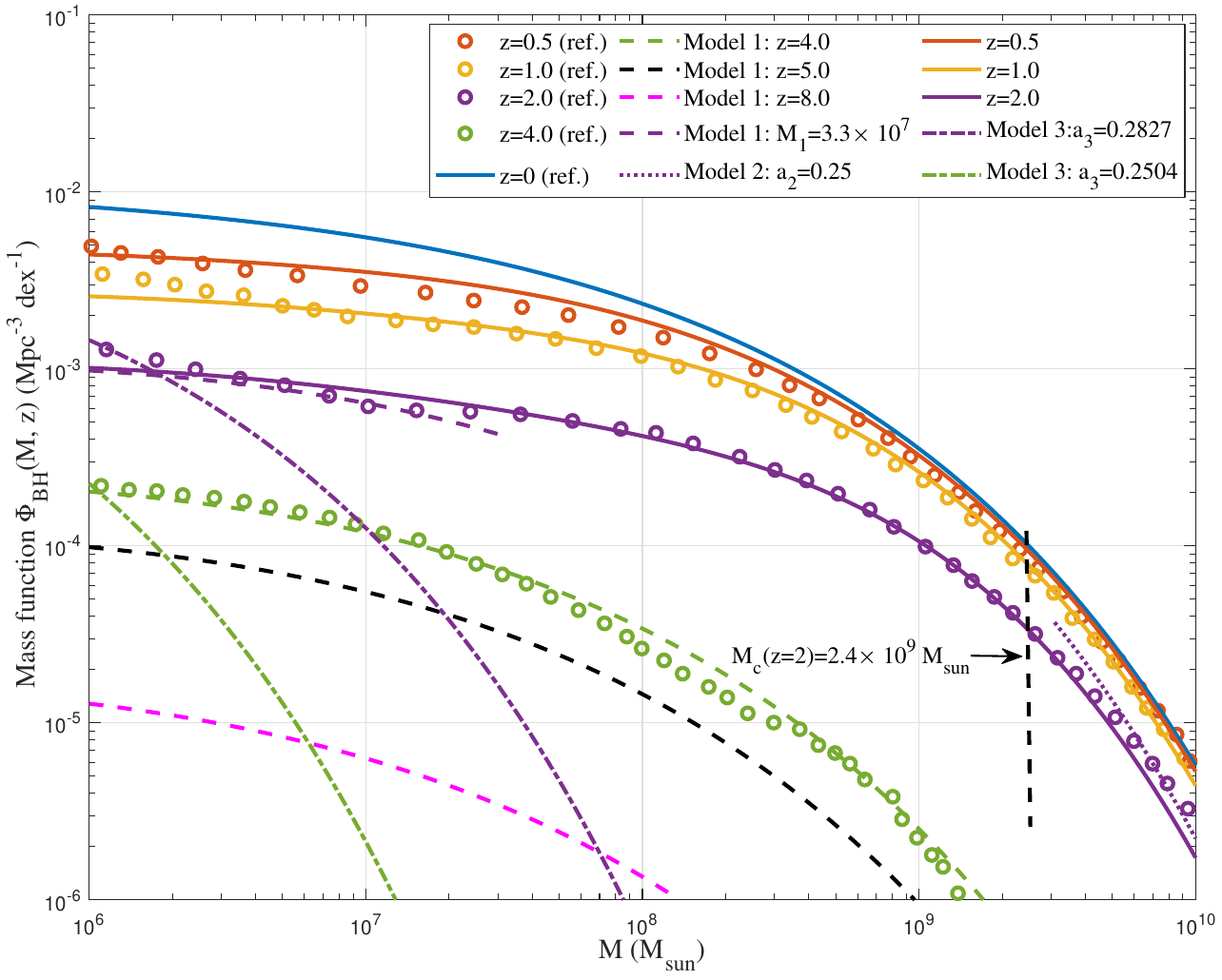}
\caption{The variation of BH mass function $\Phi_{BH}(M,z)$. The solid blue line plots the local mass function at $z=0$ in Eq. \eqref{eq:7-1-9-1}. Symbols plot the numerical results of the BH mass function at different redshifts \citep{Tucci:2017-Constraining-supermassive-black-hole-evolution}. Other solid lines present the analytical solutions built by interpolation between two solutions (Eq. \eqref{eq:7-1-9-2}). The dashed lines present the analytical solution $\Phi_{BH}^1(M,z)$ according to Model 1 (Eq. \eqref{eq:7-1-8}) with $M_1=3.3\times 10^7 M_{\odot}$. It predicts the evolution at high redshift or the low mass end at low redshift. The dotted lines represent the analytical solution $\Phi_{BH}^2(M,z)$ according to Model 2 (Eq. \eqref{eq:7-1-9}). It describes the evolution at the high-mass end and low redshift. Complete solutions (solid lines) were built by interpolating two solutions. A good agreement with the reference validates the analytical solutions. For comparison, the evolution according to Model 3 at a constant Eddington ratio $\lambda_f=0.01$ is also presented as dashed-dotted lines. However, no good agreement can be obtained for Model 3. SMBHs do not appear to evolve at a fixed Eddington ratio at high redshift.}
\label{fig:S22}
\end{figure}

\begin{figure}
\includegraphics*[width=\columnwidth]{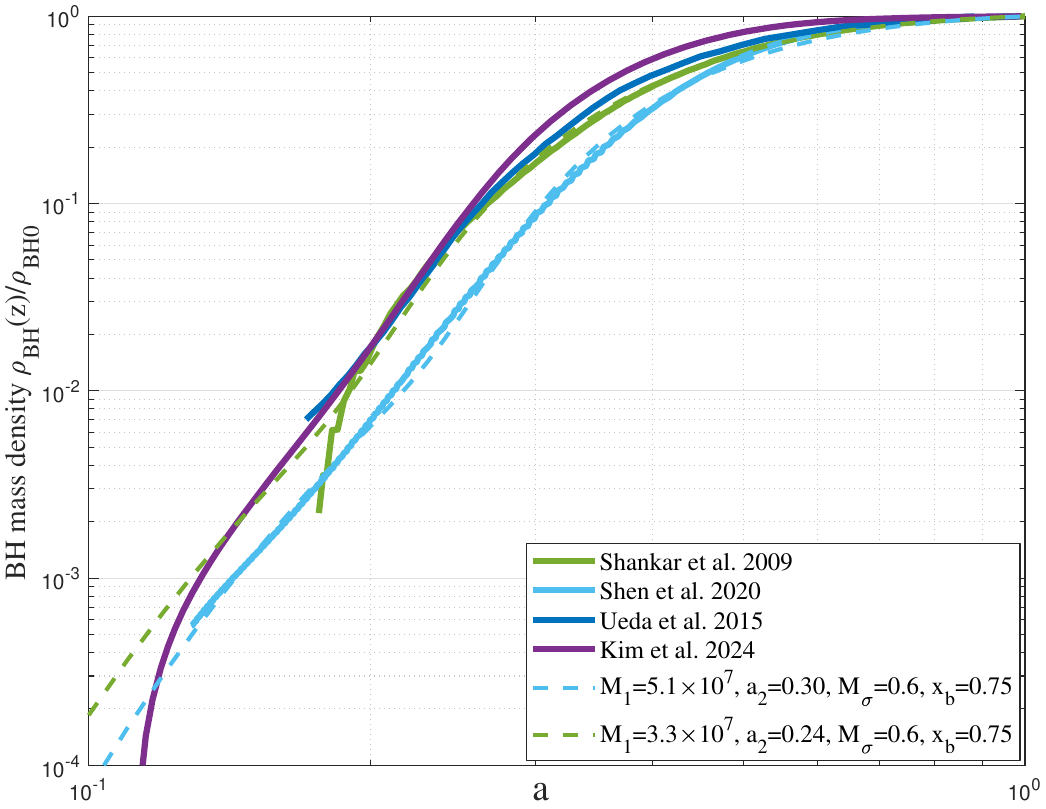}
\caption{The evolution of normalized BH mass density $\rho_{BH}(z)/\rho_{BH0}$ with scale factor $a$ estimated from the luminosity function by Shankar et al. \citep{Shankar:2009-Self-Consistent-Models-of-the-AGN-and-Black-Hole}, Shen et al. \citep{shen:2020-The-bolometric-quasar-luminosity-function}, Ueda et al. \citep{UEDA:2015-Cosmological-evolution-of-supermassive-black-holes}, and Kim et al. \citep{Kim:2024-Cosmic-star-formation-history-and-black-hole-accretion-history}. The dashed lines present the evolution of the BH mass density according to the complete solution of the BH mass function in Eq. \eqref{eq:7-1-9-2}. To fit the data from different authors, the analytical model requires a mass scale $M_1$ between $2.5\times 10^7$ and $5\times10^7M_{\odot}$ for $\Phi_{BH}^1$ in Eq. \eqref{eq:7-1-8} and a characteristic scale factor $a_2$ between 0.24 and 0.3 for $\Phi_{BH}^2$ in Eq. \eqref{eq:7-1-9}.} 
\label{fig:S21}
\end{figure}

The first analytical solution of the small-mass-end BH mass function based on the evolution Model 1 of Eq. \eqref{eq:7-1-7} reads
\begin{equation}
\begin{split}
&\Phi^1_{BH}(M,z) = \left(M/Y_1\right)^{\sigma} \Phi_{BH0}(Y_1), \\
& Y_1 = \left[M^{\sigma}+(a^{3/2-mp}-1)M_1^{\sigma}\right]^{1/\sigma},\\
& M_1^{\sigma}=\frac{2\sigma}{2mp-3}\frac{\alpha_1(1-\epsilon)}{\epsilon c^2}\frac{\varepsilon_{b0}^p}{H_0}=M_{\infty1}^{\sigma}a_1^{(mp-3/2)}.
\end{split}
\label{eq:7-1-8}
\end{equation}
The mass function $\Phi_{BH}^1$ at any redshift $z$ can be calculated easily via a nonlinear time-dependent mapping between BH mass $M$ and variable $Y_1$. At $z=0$ or $a=1$, we have $Y_1=M$, which recovers the local mass function $\Phi_{BH0}$. Since most active SMBHs at high redshift or small SMBHs at low redshift are in stage E1 (in their early stage of life), Eq. \eqref{eq:7-1-8} can be used to estimate the BH mass function at a high redshift or the mass function at the low mass end. The only parameter involved in this solution is a lumped mass parameter $M_1$ that can be related to $\alpha_1$ and $\epsilon$, or the parameters $M_{\infty1}$ and $a_1$ for the evolution of the BH mass in stage "E1" (Eq. \eqref{eq:43}). The value of the parameter $M_1$ can be identified by matching the analytical solution with the numerical solutions or observations (Figs.\ref{fig:S22} and \ref{fig:S21}). 

Next, the second analytical solution of the large-mass-end BH mass function based on the evolution Model 2 in Eq. \eqref{eq:7-1-7} reads
\begin{equation}
\begin{split}
&\Phi^2_{BH}(M,z) = \Phi_{BH0}(e^{Y_2} M), \\
& Y_2 = {a_2}^{mp-3/2}\left(a^{3/2-mp}-1\right),\\
& a_2^{(mp-3/2)}=\frac{2}{2mp-3}\frac{\alpha_2(1-\epsilon)}{\epsilon c^2}\frac{\varepsilon_{b0}^p}{H_0},
\end{split}
\label{eq:7-1-9}
\end{equation}
where $a_2$ is the redshift parameter for the evolution of BH mass in stage "E2" (Eq. \eqref{eq:48}) that can also be determined by matching to observations (Figs.\ref{fig:S22} and \ref{fig:S21}). Since most large BHs at the low redshift are in stage E2, Eq. \eqref{eq:7-1-9} can be used to model the BH mass function at low redshift or the mass function at the large mass end. 

Instead of the evolution Model 1 for stage E1 in Eq. \eqref{eq:7-1-7}, we also present the analytical solution for SMBHs evolving at a constant Eddington ratio $\lambda_f$ in stage E1:
\begin{equation}
\begin{split}
&\textrm{Model 3:}\quad \langle L \rangle = \lambda_f \varepsilon_{Edd} M \quad \quad \textrm{for stage "E1"},
\end{split}
\label{eq:7-1-9-3}
\end{equation}
where $\varepsilon_{Edd}=6.3m^2/s^3$. The solution for Model 3 can be easily obtained from Eq. \eqref{eq:7-1-9} with $m=0$, $p=0$, and $\alpha_2=\lambda_f \varepsilon_{Edd}$. For this scenario, the BH mass function evolves as
\begin{equation}
\begin{split}
&\Phi^3_{BH}(M,z) = \Phi_{BH0}(e^{Y_3} M), \\
& Y_3 = a_3^{-3/2}\left(1-a^{3/2}\right),\\
& a_3^{-3/2} = \frac{2}{3} \lambda_f \varepsilon_{Edd} \frac{(1-\epsilon)}{\epsilon c^2 H_0},
\end{split}
\label{eq:7-1-9-4}
\end{equation}
where $a_3$ is the only parameter involved that is related to the fixed Eddington ratio $\lambda_f$. The comparison between $\Phi_{BH}^1(M,z)$ and $\Phi_{BH}^3(M,z)$ for Model 1 and Model 3 provides important information on the SMBH evolution. Model 3 with a fixed Eddington ratio does not appear to be a feasible evolution at high redshift (Fig. \ref{fig:S22}). 

Finally, since $\Phi_{BH}^1$ is valid for small mass $M$, while $\Phi_{BH}^2$ is good for large mass $M$, a complete model for BH mass functions can be obtained by interpolating two mass functions, 
\begin{equation}
\begin{split}
&\Phi_{BH}(M,z) = \Phi^1_{BH}(M,z)(1-S(M))+\Phi^2_{BH}(M,z)S(M),\\
&S(M) = \frac{1}{1+x_b\exp\left(-\frac{\log_{10}(M)-\log_{10}(M_c)}{M_{\sigma}}\right)},
\end{split}
\label{eq:7-1-9-2}
\end{equation}
where the interpolation function satisfies $S(M\gg M_c)=1$ and $S(M\ll M_c)=0$ to smoothly interpolate two mass functions. 

To obtain the evolution of the BH mass function, we still need a local BH mass function $\Phi_{BH0}(M)$ as a boundary condition. Similarly to an existing local mass function \citep{Tucci:2017-Constraining-supermassive-black-hole-evolution}, we adopt a more general Schechter fitting function 
\begin{equation}
\begin{split}
\Phi_{BH0}^*(M)=\phi^*_B\left(\frac{M}{\hat{M}}\right)^{-\lambda_1}\exp\left[-\beta_B\left(\frac{M}{\hat{M}}\right)^{2-2\lambda_2}\right],
\end{split}
\label{eq:7-1-9-1}
\end{equation}
with parameters $\phi^*_B=10^{-9.715}$Mpc$^{-3}$$M_{\odot}^{-1}$, $\hat{M}=10^{7.411}M_{\odot}$, $\beta_B=1$, $\lambda_1=1$, and $\lambda_2=0.83$. This BH mass function shares the same form as the double-$\lambda$ mass function we developed for dark matter haloes \citep{Xu:2023-Dark-matter-halo-mass-functions-and}. For the halo mass function, the values of $\lambda$ are related to the halo mass accretion. Similarly, the $\lambda_1$ and $\lambda_2$ in the BH mass function may also come from the BH mass accretion ($\dot M  \propto M^\lambda$).

The BH mass function in logarithmic units of mass should be $\Phi_{BH0}(M)=\Phi_{BH0}^*(M)M\ln(10)$. The x-th order moment reads
\begin{equation}
\begin{split}
\langle M^x \rangle &= \int \Phi_{BH0}(M)M^xd\log M =\int \Phi^*_{BH0}(M) M^x d M\\
&= \frac{\phi^*_B\hat{M}^{1+x}}{2-2\lambda_2}\beta_B^{-\frac{x-\lambda_1+1}{2-2\lambda_2}}\Gamma\left(\frac{x-\lambda_1+1}{2-2\lambda_2}\right).
\end{split}
\label{eq:7-1-9-3-2}
\end{equation}

Figure \ref{fig:S22} presents the evolution of the BH mass function $\Phi_{BH}(M,z)$. The symbols show the numerical results at different redshifts \citep{Tucci:2017-Constraining-supermassive-black-hole-evolution}. The solid line presents the analytical solutions by interpolating two solutions involving a break mass $M_c(z)$ (see Eq. \eqref{eq:7-1-9-2}). The interpolation parameters used are $M_{\sigma}=0.6$, $x_b=0.75$, and $M_c=10^7a^{-5} M_{\odot}$. The dashed lines present the analytical solution $\Phi_{BH}^1(M,z)$ according to Model 1 (Eq. \eqref{eq:7-1-8}) with parameter $M_1=3.3\times 10^7 M_{\odot}$. It predicts the mass function at high redshift or the low-mass end at low redshift. The dotted line represents the analytical solution $\Phi_{BH}^2(M,z)$ according to Model 2 (Eq. \eqref{eq:7-1-9}) with $a_2=0.25$. Good agreement with the numerical solutions supports the analytical solutions. With $\lambda_1=1$, the BH mass function $\Phi_{BH}(M)\propto M^0$ at the small-mass end. 

For comparison, the evolution according to Model 3 at a fixed Eddington ratio $\lambda_f$ is also presented as dashed-dotted lines for $\Phi_{BH}^3(M,z)$. However, no good agreement can be obtained with numerical solutions. The parameter $a_3$ is selected to match the numerical value of the BH mass function at $M=10^6M_{\odot}$, which corresponds to an Eddington ratio $\lambda_f=0.01$ for $\epsilon=0.1$ (Eq. \eqref{eq:7-1-9-4}). Model 3 appears to be inappropriate at high redshift, i.e., high-redshift SMBHs should not follow the evolution at a fixed Eddington ratio.  

With BH mass functions explicitly obtained, the evolution of the BH mass density can be derived analytically. For Model 2, the total BH mass density evolves as (using Eq. \eqref{eq:7-1-9}),
\begin{equation}
\begin{split}
\rho_{BH}(z) &= \int \Phi_{BH}(M,z)Md\log M  \\
&= e^{-Y_2}\int \Phi_{BH0}(M)Md\log M = e^{-Y_2}\rho_{BH0},
\end{split}
\label{eq:7-1-10}
\end{equation}
where $\rho_{BH0}\equiv \rho_{BH}(z=0)$ is the local BH mass density. Therefore, the evolution of the BH mass density at low redshift is determined by the variation of $Y_2$ in Eq. \eqref{eq:7-1-9}. Similarly, the BH mass density at a high redshift can be derived using Eq. \eqref{eq:7-1-8} according to Model 1,  
\begin{equation}
\begin{split}
\rho_{BH}(z) &= \int \Phi_{BH0}(Y_1)M^{1+\sigma}Y_1^{-\sigma}d\log M\\
            &= \int \Phi_{BH0}(Y_1)Md\log Y_1. \\
\end{split}
\label{eq:7-1-12}
\end{equation}

The complete evolution of the BH mass density can also be obtained by integrating the complete BH mass function in Eq. \eqref{eq:7-1-9-2} (solid lines in Fig. \ref{fig:S22}). Figure \ref{fig:S21} plots the time variation of the normalized BH mass density computed from the quasar luminosity function by different authors \citep{Shankar:2009-Self-Consistent-Models-of-the-AGN-and-Black-Hole,shen:2020-The-bolometric-quasar-luminosity-function,UEDA:2015-Cosmological-evolution-of-supermassive-black-holes,Kim:2024-Cosmic-star-formation-history-and-black-hole-accretion-history}. Analytical results of the BH mass density are also presented for comparison. To fit data from different authors, the analytical model requires a mass scale $M_1$ between $2.5\times 10^7$ and $5\times10^7M_{\odot}$ and a characteristic scale factor $a_2$ between 0.24 and 0.3. Again, good agreement validates analytical models.


\section{AGN luminosity and mass functions}
\label{sec:7-3}
In this section, we derive the analytical AGN mass function for a given quasar luminosity function, which will provide an independent check of the AGN mass function obtained from the BH mass function and AGN duty cycle (Eq. \eqref{eq:7-1-4-1}). The quasar and AGN luminosity functions are related by the Eddington ratio distribution $P(\lambda\vert M,z)$,
\begin{equation}
\Phi_{L}(L,z) = \int_{\lambda_{min}}^{\lambda_{max}}d\log\lambda P(\lambda\vert M,z)\Phi_{AGN}(M,z).
\label{eq:7-3-1}
\end{equation}
When luminosity function $\Phi_L$ and Eddington ratio distribution $P$ are given, the analytical AGN mass function $\Phi_{AGN}$ can be derived. We will use the Eddington ratio distribution explicitly derived in Section \ref{sec:7-2} (Eq. \eqref{eq:7-2-5}). Based on a large set of observations in different wavebands, the luminosity function adopts a double power-law \citep{shen:2020-The-bolometric-quasar-luminosity-function} with parameters determined from observations, 
\begin{equation}
\Phi_{L}(L,z) = \frac{\phi^*}{(L/L^*)^{\gamma_1}+(L/L^*)^{\gamma_2}}.
\label{eq:7-3-2}
\end{equation}
Here $\phi^*$ is the comoving number density for normalization, $L^*$ is the break luminosity, $\gamma_1$ and $\gamma_2$ are the faint-end and bright-end slopes. The redshift variation of these parameters is discussed in \citep{shen:2020-The-bolometric-quasar-luminosity-function} and presented as a function of the scale factor $a$,
\begin{equation}
\begin{split}
&\gamma_1 = 0.3653(3a)^{0.6006}, \\
&\gamma_2 = \frac{4.9418}{(3a)^{0.9963}+(3a)^{-1.0716}}, \\
&\log \phi^*=-3.6276-0.3444a^{-1} \quad \textrm{dex}^\textrm{-1} \textrm{Mpc}^\textrm{-3},\\
&\log L^* = \frac{25.9312}{(3a)^{0.5758}+(3a)^{-0.4698}} \quad \textrm{L}_{\odot}.
\end{split}
\label{eq:7-3-2-2}
\end{equation}

It was noted that the faint-end slope $\gamma_1$ could be well approximated by a simple power law and approaching 1/5 at high redshift $z=7$ \citep{shen:2020-The-bolometric-quasar-luminosity-function}. This leads to an AGN mass function $\propto M^{-1/5}$ at the low mass end and high redshift (Fig. \ref{fig:S32}), which can be predicted by the analytical quasar duty cycle in Eq . \eqref{eq:8-1}. The faint-end slope of the luminosity function or the small-mass-end slope of the AGN mass function can be related to the mass accretion model in stage E1 (Eq. \eqref{eq:7-1-7}), i.e., $\gamma_1\approx \sigma$ in Table \ref{tab:2}.

Break luminosity $L^*(z)$ corresponds to the maximum (or mode) in the intrinsic distribution of the peak luminosity $L_p$ of all quasars at any given redshift $z$ \citep{Hopkins:2006-The-Evolution-in-the-Faint-End-Slope-of-the-Quasar}. The redshift variation of $L^*$ represents the evolution of the peak luminosity $L_p$ of all quasars. Therefore, the variation in the break luminosity $L^*$ can also be modeled using a two-stage model. Substituting the mass evolution (Eqs. \eqref{eq:43} and \eqref{eq:48}) and the parameters in Table \ref{tab:2} into the two-stage model (Eq. \eqref{eq:7-1-7}), we can model the evolution of $L^*$ as,
\begin{equation}
\begin{split}
&\textrm{Model 1:}\quad L^*_1 = L_1\left[\frac{1}{(aa_1)^{1/2}}-\frac{1}{a}\right]^4 \quad \textrm{for stage "E1"}, \\
&\textrm{Model 2:}\quad L^*_2 = L_2 a^{-5}\exp\left[-\left(\frac{a}{a_2}\right)^{-7/2}\right] \quad \textrm{for stage "E2"},\\
&\textrm{where}\quad L_1 = \alpha_1 (\varepsilon_{b0} M_1)^{4/5} \quad \textrm{and} \quad L_2 = \alpha_2 \varepsilon_{b0}^2 M_{\infty2}.
\end{split}
\label{eq:7-3-2-3}
\end{equation}
Here $M_1$ is the mass scale for the analytical evolution of the BH mass function (Eq. \eqref{eq:7-1-8}). The evolution of the break luminosity can be described for two separate stages, i.e., $L^*_1$ modeled by the parameters $L_1$ and $a_1$ in the rising stage "E1", and $L^*_2$ modeled by the parameters $L^*_2$ and $a_2$ in the declining stage "E2". The complete evolution can be similarly constructed via interpolating $L^*_1$ and $L^*_2$,
\begin{equation}
\begin{split}
&L^*(a) = L^*_1(a)(1-S(a))+L^*_2(a)S(a),\\
&S(a) = \frac{1}{1+x_a\exp\left(-\frac{\log_{10}(a)-\log_{10}(a_c)}{a_{\sigma}}\right)}.
\end{split}
\label{eq:7-3-2-5}
\end{equation}

\begin{figure}
\includegraphics*[width=\columnwidth]{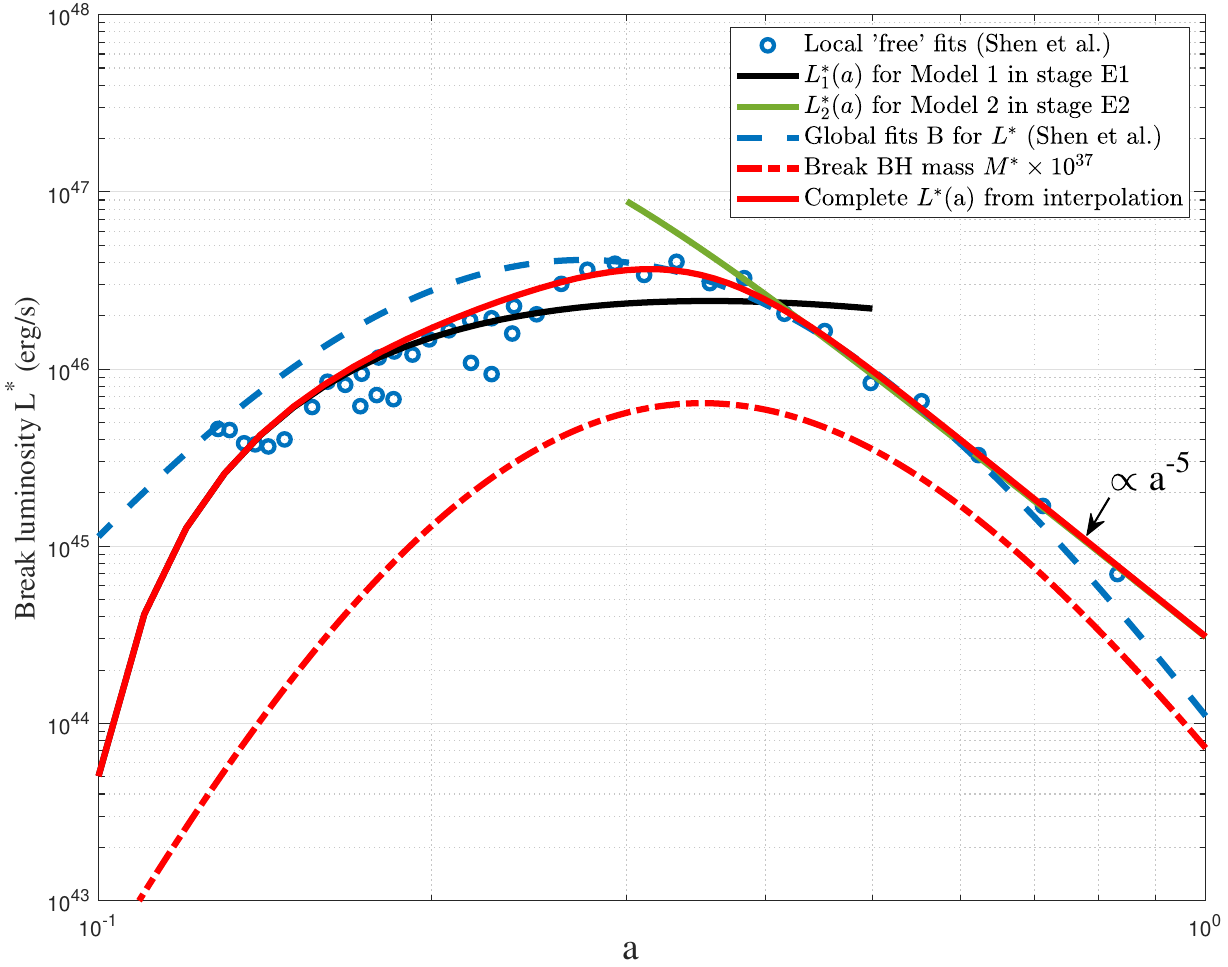}
\caption{The variation of break luminosity $L^*$ with the scale factor $a$. Symbols represent the data of $L^*$ obtained from the local fits for each redshift \citep{shen:2020-The-bolometric-quasar-luminosity-function}. The dashed blue line plots the best global fit of $L^*$ for all redshifts from the same reference. The two-stage model in Eq. \eqref{eq:7-3-2-3} is plotted with the best parameters of $L_1=10^{44.59}$erg/s and $a_1=0.089$ (or $z=10$) for stage E1 (solid black line for $L^*_1$) and the best parameters of $L_2=10^{44.49}$erg/s and $a_2=0.2244$ for stage E2 (solid green line for $L^*_2$ ). A power-law evolution of $L^*\propto a^{-5}$ can be observed at low redshift, as predicted by Eq. \eqref{eq:7-3-2-3}. This is relevant for the quasar lifetime and the Eddington ratio distribution (Section \ref{sec:7-2}). The complete solution of $L^*$ by interpolating $L^*_1$ and $L^*_2$ is plotted as a solid red line (Eq. \eqref{eq:7-3-2-5}). The evolution of the break mass $M^*$ is obtained from the fit $L^*$ of Shen et al. using Eq. \eqref{eq:7-3-2-4} (red dashed line rescaled by $\times 10^{37}$).} 
\label{fig:S28}
\end{figure}

The break mass $M^*$ corresponding to the break luminosity $L^*$ is the maximum in the intrinsic distribution of the BH mass when the quasars are at their maximum luminosity. From Eq. \eqref{eq:24}, we have
\begin{equation}
\begin{split}
M^*=\left(\frac{L^*G^{1/5}}{c}\right)^{5/4}\frac{1}{\varepsilon_b} \quad \textrm{and} \quad \lambda_0 = \frac{L^*}{\varepsilon_{Edd}M^*},
\end{split}
\label{eq:7-3-2-4}
\end{equation}
such that the evolution of $L^*$ and $M^*$ follows the $\gamma \propto \eta^{-1}$ in $\gamma-\eta$ plane. Similarly, $\lambda_0$ is the corresponding break Eddington ratio for the Eddington ratio distribution (Eq. \eqref{eq:7-2-5}). Break mass $M^*$ sets a mass scale for the AGN mass function in Eq. \eqref{eq:7-3-3}.

Figure \ref{fig:S28} plots the variation of the break luminosity. Symbols represent the data of $L^*$ obtained from the local fits for each redshift, while the blue dashed line plots the global fit of $L^*$ for all redshifts \citep{shen:2020-The-bolometric-quasar-luminosity-function}. The two-stage model in Eq. \eqref{eq:7-3-2-3} is also presented. The best-fit parameters of $L^*_1=10^{44.59}$erg/s and $a_1=0.089$ can be obtained for stage "E1" (solid black line). The best-fit parameters of $L^*_2=10^{44.49}$erg/s and $a_2=0.2244$ can be obtained for stage "E2" (solid green line). The power law $L^*\propto a^{-5}$ is clearly shown at low redshift, which agrees with Eq. \eqref{eq:7-3-2-3}. This will be used to derive the quasar lifetime and the Eddington ratio distribution (Section \ref{sec:7-2}).  Complete solution $L^*$ by interpolation in Eq. \eqref{eq:7-3-2-5} is plotted as a solid red line with interpolation parameters $x_a=0.4525$, $a_{\sigma}=0.0539$, and $a_c=0.3778$. The evolution of the break mass $M^*$($M_{\odot}$) is obtained from $L^*$ by Eq. \eqref{eq:7-3-2-4} (red dashed line rescaled by $\times 10^{37}$). 

\begin{figure}
\includegraphics*[width=\columnwidth]{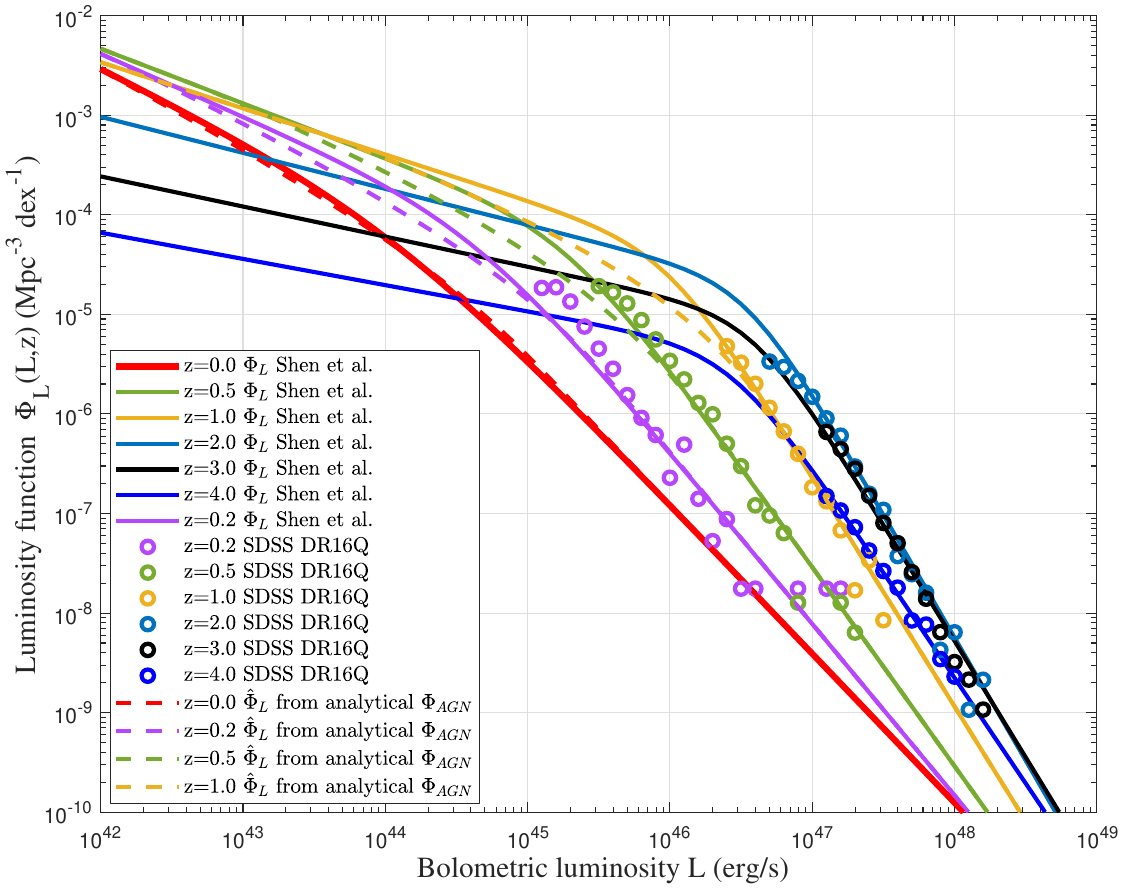}
\caption{The quasar luminosity function $\Phi_L(L,z)$ at different redshift $z$ from Eq. \eqref{eq:7-3-2} (solid lines). For comparison, symbols represent the data from the recent Sloan Digital Sky Survey (SDSS DR16Q) \citep{Wu:2022-A-Catalog-of-Quasar-Properties-from-Sloan}. General agreement between the fitted luminosity function and the SDSS data can be found. Dashed lines present the luminosity function $\hat\Phi_L$ obtained using Eq. \eqref{eq:7-3-1} with the analytical AGN mass function $\hat\Phi_{AGN}$ in Eq. \eqref{eq:7-3-3} and the Eddington ratio distribution $P(\lambda)$ in Eq. \eqref{eq:7-2-5}. The agreement between $\hat\Phi_L$ and the fitted luminosity function $\Phi_L$ at both the faint and bright ends of luminosity validates the analytical solutions for AGN mass function.} 
\label{fig:S29}
\end{figure}

With all parameters determined, the double power-law luminosity function $\Phi_L(L,z)$ in Eq. \eqref{eq:7-3-2} is plotted in Fig. \ref{fig:S29}. For comparison, symbols represent data from the recent Sloan Digital Sky Survey Data Release 16 quasar catalog (SDSS DR16Q) that contains more than 750k quasars \citep{Wu:2022-A-Catalog-of-Quasar-Properties-from-Sloan}. General agreement between the fitted luminosity function and the SDSS data can be found.  

For the double power-law luminosity function in Eq. \eqref{eq:7-3-2} and the Eddington ratio distribution in Eq. \eqref{eq:7-2-5} (discussed in Section \ref{sec:7-2}), it is possible to derive the analytical AGN mass function $\hat\Phi_{AGN}$ from the convolution Eq. \eqref{eq:7-3-1}. Mathematically, $\hat\Phi_{AGN}$ should have the same double power-law form as the luminosity function $\Phi_L$ with the same slope but involving a break mass $M^*$. We first assume the AGN mass function with a piece-wise power-law,
\begin{equation}
\begin{split}
&\hat\Phi_{AGN}(M,z) = B_1\left({M}/{M^*}\right)^{-\gamma_1} \quad \textrm{for} \quad {M}/{M^*}<\hat{x},\\
&\hat\Phi_{AGN}(M,z) = B_2\left({M}/{M^*}\right)^{-\gamma_2} \quad \textrm{for} \quad {M}/{M^*}>\hat{x},\\
&\hat{x} = \left(B_2/B_1\right)^{\frac{1}{\gamma_2-\gamma_1}},
\label{eq:7-3-3}
\end{split}
\end{equation}
where $\hat{x}$ is the location where two power-law functions meet. Here, $B_1$ and $B_2$ are two pre-factors to be determined. The value of $\hat{x}$ can be found from the continuity condition at $\hat{x}$. 

Substituting the Eddington ratio distribution (Eq. \eqref{eq:7-2-5}) and the AGN mass function (Eq. \eqref{eq:7-3-3}) into the convolution Eq. \eqref{eq:7-3-1}, we can derive the luminosity function $\hat{\Phi}_L$ analytically. By matching $\hat{\Phi}_L$ with the fitted luminosity function $\Phi_L$ in Eq. \eqref{eq:7-3-2} at both the faint and the bright end, the constants $B_1$ and $B_2$ can be found
\begin{equation}
\begin{split}
&B_1 = \phi^*\beta^{\frac{\gamma_1}{\tau}}\frac{\Gamma\left[-\frac{\alpha}{\tau},\beta\left(\frac{\lambda_{min}}{\lambda_0}\right)^{\tau}\right]-\Gamma\left[-\frac{\alpha}{\tau},\beta\left(\frac{\lambda_{max}}{\lambda_0}\right)^{\tau}\right]}{\Gamma\left[\frac{\gamma_1-\alpha}{\tau},\beta\left(\frac{\lambda_{min}}{\lambda_0}\right)^{\tau}\right]-\Gamma\left[\frac{\gamma_1-\alpha}{\tau},\beta\left(\frac{\lambda_{max}}{\lambda_0}\right)^{\tau}\right]},\\
&B_2 = \phi^*\beta^{\frac{\gamma_2}{\tau}}\frac{\Gamma\left[-\frac{\alpha}{\tau},\beta\left(\frac{\lambda_{min}}{\lambda_0}\right)^{\tau}\right]-\Gamma\left[-\frac{\alpha}{\tau},\beta\left(\frac{\lambda_{max}}{\lambda_0}\right)^{\tau}\right]}{\Gamma\left[\frac{\gamma_2-\alpha}{\tau},\beta\left(\frac{\lambda_{min}}{\lambda_0}\right)^{\tau}\right]-\Gamma\left[\frac{\gamma_2-\alpha}{\tau},\beta\left(\frac{\lambda_{max}}{\lambda_0}\right)^{\tau}\right]},
\end{split}
\label{eq:7-3-4}
\end{equation}
where $\alpha$, $\beta$, $\tau$ are parameters of Eddington ratio distribution (Eq. \eqref{eq:7-2-5}). Therefore, the AGN mass function can be completely determined by the parameters ($\phi^*$, $\gamma_1$, and $\gamma_2$) from the quasar luminosity function, the parameters ($\alpha$, $\beta$, $\tau$, $\lambda_0$) from the Eddington ratio distribution, and the range of the Eddington ratio for AGN ($\lambda_{min}$, $\lambda_{max}$). Since the upper incomplete gamma function $\Gamma(x,y)=0$ for $y\rightarrow \infty$, terms involving $\lambda_{max}$ can be neglected for a sufficiently large $\lambda_{max}$. 

\begin{figure}
\includegraphics*[width=\columnwidth]{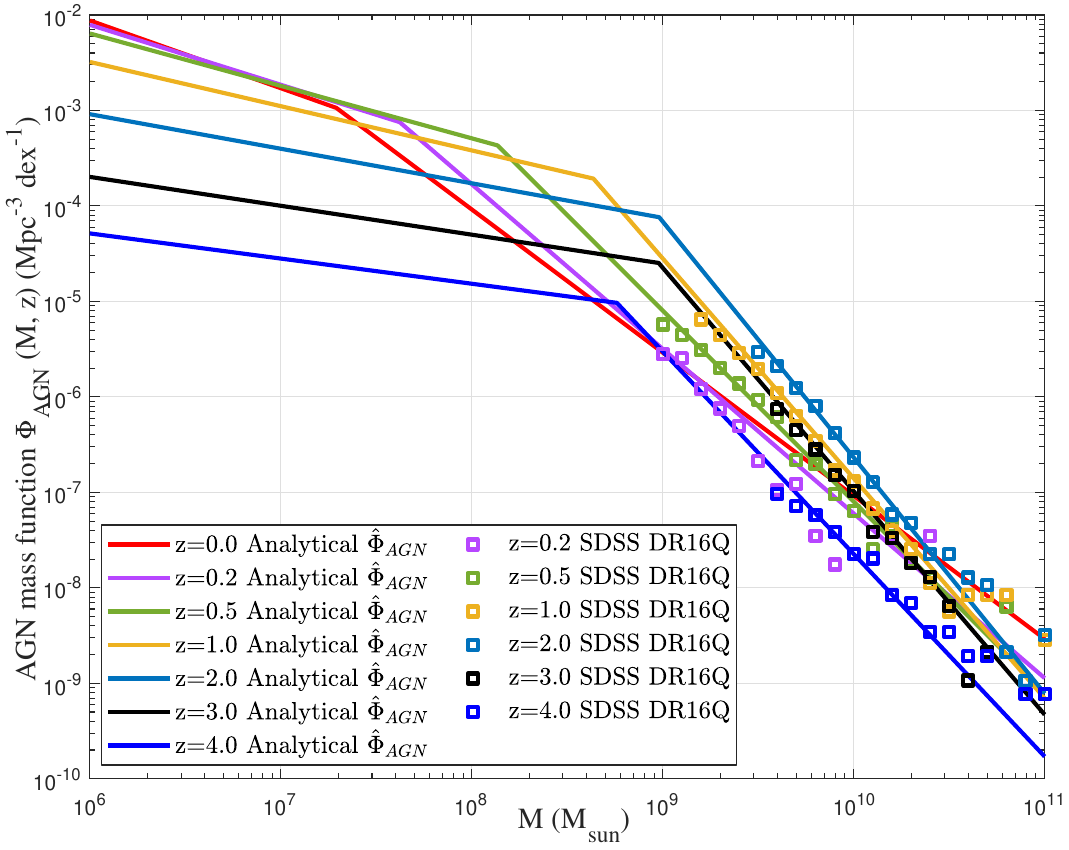}
\caption{The double power-law analytical AGN mass function $\hat{\Phi}_{AGN}(M,z)$ at different redshift $z$ (Eq. \eqref{eq:7-3-3}) derived with a given double power-law luminosity function $\Phi_L(L,z)$. For comparison, symbols represent the data from the recent Sloan Digital Sky Survey Data Release (SDSS DR16Q) \citep{Wu:2022-A-Catalog-of-Quasar-Properties-from-Sloan}.} 
\label{fig:S30}
\end{figure}

Figure \ref{fig:S29} presents the luminosity function $\hat\Phi_L(L,z)$ calculated with $\hat{\Phi}_{AGN}(M,z)$ in Eq. \eqref{eq:7-3-3}, the Eddington ratio distribution in Eq. \eqref{eq:7-2-5}, and the convolution Eq. \eqref{eq:7-3-1}, where $\hat\Phi_L$ captures $\Phi_L$ at both the faint and the bright ends of the luminosity. The derived Eddington ratio distribution in Eq. \eqref{eq:7-2-5} only includes the Type II AGNs that are more dominant than Type I AGNs. The discrepancy between $\hat\Phi_L$ and $\Phi_L$ around the break luminosity $L^*$ may be attributed to the missing Type I AGNs in the Eddington ratio distribution. Figure \ref{fig:S30} plots the analytical AGN mass function $\hat\Phi_{AGN}(M,z)$, compared to the data from the SDSS DR16Q release. Good agreement validates the analytical AGN mass function $\hat{\Phi}_{AGN}(M,z)$ at different redshifts.

\section{Deriving the quasar duty cycle}
\label{sec:7-2-1}

The quasar duty cycle, defined as the fraction of active AGNs in all SMBHs of the same mass, can be explicitly derived from the two-stage model. Since the average rate of mass accretion $\langle \dot M\rangle$ is dependent on the duty cycle (Eq. \eqref{eq:7-1-3}), we can express the duty cycle as a function of the average luminosity $\langle L \rangle$ (Eq. \eqref{eq:7-1-7}). With parameters listed in Table \ref{tab:2}, the two-stage duty cycle model can be formulated analytically,
\begin{equation}
\begin{split}
&U(M,z) = {\langle L \rangle}/{\left(\varepsilon_{Edd} M \langle \lambda \rangle\right)},\\
&\textrm{Model 1:}\quad U^1(M,z) = \frac{\alpha_1 \varepsilon_b^{4/5} }{\varepsilon_{Edd} \langle \lambda \rangle} M^{-1/5} \quad \textrm{for stage "E1"}, \\
&\textrm{Model 2:}\quad U^2(M,z) = \frac{\alpha_2 \varepsilon_b^2}{\varepsilon_{Edd} \langle \lambda \rangle} \quad \textrm{for stage "E2"}.
\end{split}
\label{eq:8-1}
\end{equation}
Here constant $\varepsilon_{Edd}=6.3m^2/s^3$. In stage E1 with a rising luminosity, that is, at high redshift $z$ or low mass end with small $M$, the duty cycle $U^1(M,z) \propto a  M^{-1/5}$ increases with time due to the rate of energy flow $\varepsilon_b\propto a^{-5/2}$ and the mean Eddington ratio $\langle \lambda \rangle \propto a^{-3}$ (see Eq. \eqref{eq:8-4}). This leads to a small-mass end AGN mass function $\Phi_{AGN}\propto M^{-1/5}$ (Eq. \eqref{eq:7-1-4-1}) and a faint-end quasar luminosity function $\Phi_{L}\propto L^{-1/5}$ (Eq. \eqref{eq:7-3-3}) based on the analytical solutions. In stage E2 with a decreasing luminosity, the duty cycle $U^2(M,z) \propto a^{-2}$ is independent of $M$ and decreases with time. 

Similarly to the BH mass function $\Phi_{BH}$ in Eq. \eqref{eq:7-1-9-2}, the complete AGN duty cycle can be modeled via the interpolation between two duty cycles in two separate stages,
\begin{equation}
\begin{split}
&U(M,z) = U^1(M,z)(1-S(M))+U^2(M,z) S(M),\\
&S(M) = \frac{1}{1+x_b\exp\left(-\frac{\log_{10}(M)-\log_{10}(M_c)}{M_{\sigma}}\right)},
\end{split}
\label{eq:8-2}
\end{equation}
where $S(M)$ is an interpolation function. Finally, the AGN mass function can be obtained from the BH mass function $\Phi_{BH}(M,z)$ in Section \ref{sec:7-1} and the duty cycle model in Eq. \eqref{eq:8-2},
\begin{equation}
\begin{split}
\Tilde{\Phi}_{AGN}(M,z) = \Phi_{BH}(M,z) U(M,z),\\
\end{split}
\label{eq:8-3}
\end{equation}
which can be directly compared with the AGN mass function $\hat\Phi_{AGN}$ obtained from quasar luminosity function in Fig. \ref{fig:S30} to check the validity of the two-stage evolution model. 

\begin{figure}
\includegraphics*[width=\columnwidth]{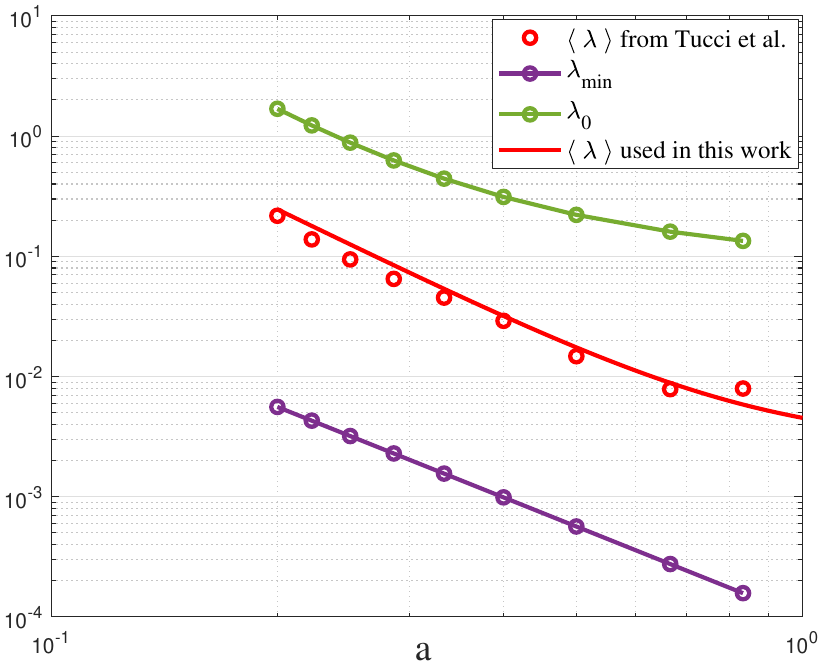}
\caption{The redshift variation of the minimum Eddington ratio $\lambda_{min}(z)$ (Eq. \eqref{eq:7-1-4}), the mean Eddington ratio $\langle \lambda \rangle(z)$ (Eq. \eqref{eq:8-4}), and the break Eddington ratio $\lambda_0(z)$ (Eq. \eqref{eq:7-3-2-4}) for the Eddington ration distribution.} 
\label{fig:S26}
\end{figure}

The complete duty cycle model only involves two parameters, $\alpha_1$ and $\alpha_2$, and the interpolation parameters $x_b$, $M_c$ and $M_{\sigma}$. Furthermore, the mean Eddington ratio $\langle \lambda \rangle(z)$ is required to model the duty cycle. We adopt the mean Eddington ratio in \cite{Tucci:2017-Constraining-supermassive-black-hole-evolution} that was plotted in Fig. \ref{fig:S26}. The empirical redshift dependence of $\langle \lambda \rangle(z)$ can be written as:
\begin{equation}
\begin{split}
\langle \lambda \rangle(z) = 1.762\times 10^{-3} \left(1.573+a^{-3.062}\right). 
\end{split}
\label{eq:8-4}
\end{equation}
This empirical mean Eddington ratio includes contributions from both Type-1 and Type-2 AGNs.

\begin{figure}
\includegraphics*[width=\columnwidth]{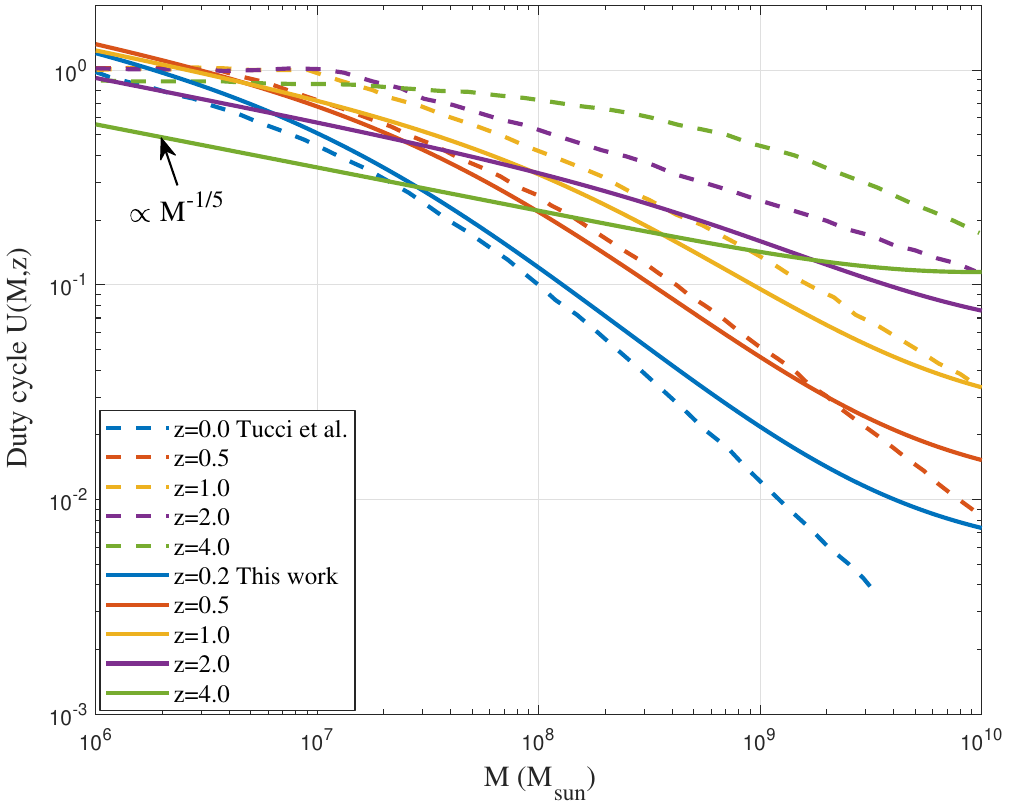}
\caption{The analytical duty cycle $U(M,z)$ as predicted by the two-stage model in Eqs. \eqref{eq:8-1} and \eqref{eq:8-2}. Model parameters of $\alpha_1=10^9$ and $\alpha_2=7.9\times 10^3$ were used to generate the plot, along with the interpolation parameters $x_b=0.75$, $M_{\sigma}=0.6$, and $M_c=10^7a^{-5}M_{\odot}$. Dashed lines present the numerical solutions of $U(M,z)$ in \citep{Tucci:2017-Constraining-supermassive-black-hole-evolution} for comparison. Our prediction generally agrees with numerical solutions, while the analytical model predicts a lower duty cycle at high redshift. At high redshift, the duty cycle follows a simple scaling $\propto M^{-1/5}$.} 
\label{fig:S31}
\end{figure}

Figure \ref{fig:S31} illustrates the variation of the analytical duty cycle $U(M,z)$. The model parameters of $\alpha_1=10^9$ and $\alpha_2=7.9\times 10^3$ were used, along with the interpolation parameters $x_b=0.75$, $M_{\sigma}=0.6$ and $M_c=10^7a^{-5}M_{\odot}$ (similar scaling as $L^*\propto a^{-5}$). At low redshift, the duty cycle approaches one, and almost all small SMBHs of $M<10^7M_{\odot}$ are active. The duty cycle decreases with $M$ to less than 0.01 at $M=10^{10}M_{\odot}$, and most large SMBHs are inactive at low redshift. At high redshift, the duty cycle is relatively independent of the BH mass $M$, and most larger BHs are active. This is expected with the downsizing behavior and anti-hierarchical growth of SMBHs. Low-mass SMBHs actively accrete mass at low redshift, while high-mass SMBHs actively grow at high redshift. Cosmic quenching plays an important role in the evolution of SMBH through the rate of energy flow $\varepsilon_b$, which regulates the cooling and supply rate of cold gas. The rapidly decreasing $\varepsilon_b$ means efficient gas cooling and fast SMBH growth at high redshift such that most AGNs are active. Numerical solutions of $U(M,z)$ \citep{Tucci:2017-Constraining-supermassive-black-hole-evolution} are also presented for comparison. The general agreement can be obtained at low redshift. The analytical model predicts a lower duty cycle at high redshift.

\begin{figure}
\includegraphics*[width=\columnwidth]{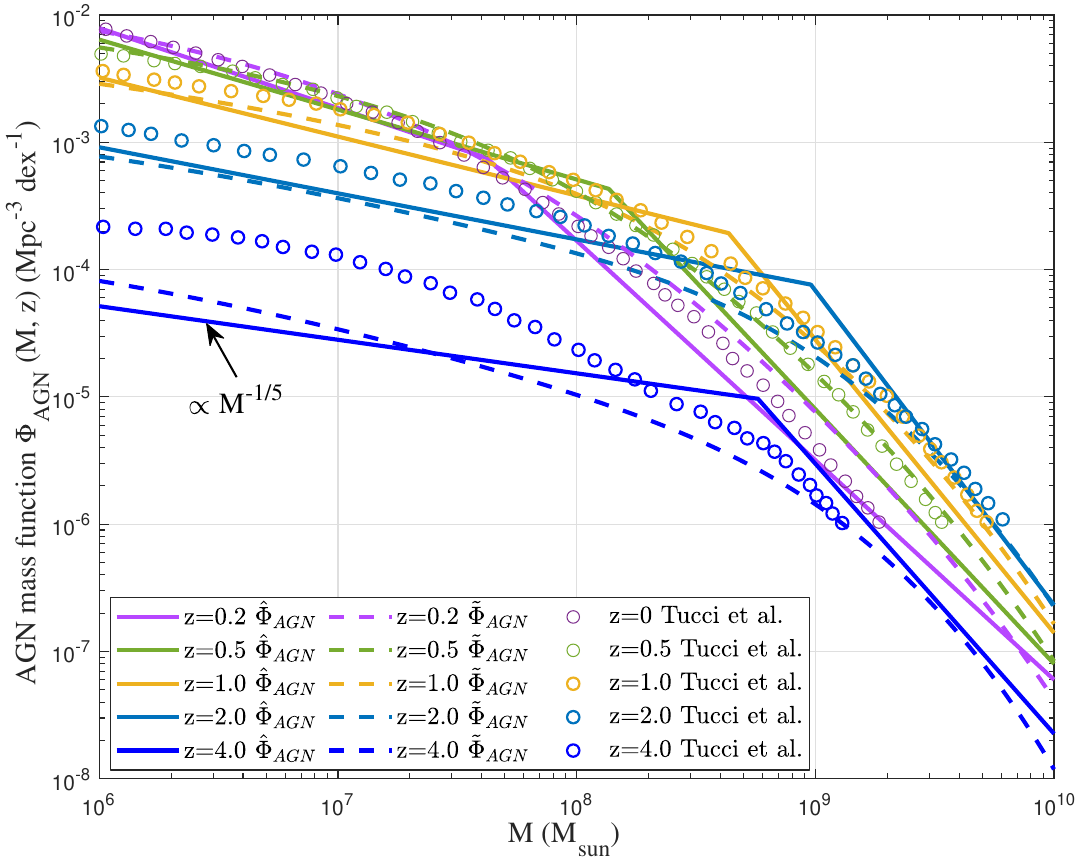}
\caption{The analytical AGN mass function $\Tilde{\Phi}_{AGN}(M,z)$ (dashed lines) as predicted by the two-stage models for both BH mass function (Eq. \eqref{eq:7-1-9-2}) and duty cycle (Eq. \eqref{eq:8-2}). The model parameters $M_1=5.1\times 10^7M_{\odot}$ and $a_2=0.2$ were used for the BH mass function model. The model parameters of $\alpha_1=10^9$ and $\alpha_2=7.9\times 10^3$ were used for the duty cycle model. Solid lines represent the analytical AGN mass function $\hat{\Phi}_{AGN}(M,z)$ obtained from the quasar luminosity function $\Phi_L(L,z)$. The agreement between $\Tilde{\Phi}_{AGN}(M,z)$ and $\hat{\Phi}_{AGN}(M,z)$ validates the two-stage evolution model. The numerical solutions (symbols) \citep{Tucci:2017-Constraining-supermassive-black-hole-evolution} are also presented. Good agreement is found at low redshift, while the analytical model predicts a lower number density for small-mass SMBHs at high redshift.} 
\label{fig:S32}
\end{figure}

Figure \ref{fig:S32} illustrates the analytical AGN mass function $\Tilde{\Phi}_{AGN}$ derived from the analytical BH mass function $\Phi_{BH}$ and the analytical duty cycle $U(M,z)$ (Eq. \eqref{eq:8-3}). The model parameters used for $\Phi_{BH}(M,z)$ (Eqs. \eqref{eq:7-1-8}, \eqref{eq:7-1-9}, and \eqref{eq:7-1-9-2}) are: $M_1=5.1\times 10^7M_{\odot}$ and $a_2=0.2$, together with the interpolation parameters $x_b=0.75$, $M_{\sigma}=0.6$, and $M_c=10^7a^{-5}M_{\odot}$. The model parameters for the duty cycle (Eqs. \eqref{eq:8-1} and \eqref{eq:8-2}) are: $\alpha_1=10^9$ and $\alpha_2=7.9\times 10^3$, along with the same interpolation parameters. The number density of active small-mass SMBHs increases steadily with time. In contrast, the number density of active large SMBHs peaks at the redshift $z=2$ and then decreases sharply at low redshift (cosmic downsizing). The analytical AGN mass function $\hat{\Phi}_{AGN}(M,z)$ obtained directly from the quasar luminosity function $\Phi_L(L,z)$ (see Fig. \ref{fig:S30}) is also presented for comparison. In principle, the AGN mass function $\hat{\Phi}_{AGN}(M,z)$ depends only on the Eddington ratio distribution $P(\lambda)$ and the luminosity function $\Phi_L(L,z)$. The AGN mass function $\Tilde{\Phi}_{AGN}(M,z)$ is obtained independently from the BH mass function and duty cycle based on the two-stage evolution in Eq. \eqref{eq:7-1-7}. The agreement between two AGN mass functions $\Tilde{\Phi}_{AGN}(M,z)$ and $\hat{\Phi}_{AGN}(M,z)$ supports the two-stage evolution for average luminosity (Eq. \eqref{eq:7-1-7}), the BH mass function (Eqs. \eqref{eq:7-1-8} and \eqref{eq:7-1-9}), and the quasar duty cycle (Eq. \eqref{eq:8-1}). With these validations, the two-stage evolution (Eq. \eqref{eq:7-1-7}) is further applied to predict the redshift evolution of some observed SMBHs in Section \ref{sec:8}.

\section{Deriving the Eddington ratio distribution}
\label{sec:7-2}
The Eddington ratio distribution is a critical component in relating the AGN mass function $\Phi_{AGN}$ to the observed quasar luminosity function $\Phi_L$ (Eq. \eqref{eq:7-1-5}). As suggested by Hopkins et al. \citep{Hopkins:2009-Quasars-Are-Not-Light-Bulbs-Testing-Models}, the Eddington ratio distribution $P(\lambda\vert M,z)$ can be directly related to a quasar lifetime or light curve model. This section will derive the quasar light curves and the Eddington ratio distribution based on the BH luminosity and mass evolution model (Model 2 in Eq. \eqref{eq:7-1-7}). 

Inserting the BH mass evolution (Eq. \eqref{eq:48}) into Model 2 of Eq. \eqref{eq:7-1-7}, we obtain the redshift variation of luminosity in stage E2 with a decreasing luminosity from its peak ($p=2$ and $m=5/2$)
\begin{equation}
L(M,z) = (\alpha_2\varepsilon_{b0}^p M_{\infty2}) a^{-mp} \exp\left[-\left(\frac{a}{a_2}\right)^{-mp+3/2}\right].
\label{eq:7-2-1}
\end{equation}
where $L$ represents a typical luminosity evolution for SMBHs of mass $M$. Assuming quasars reaching a peak luminosity $L_p$ at a scale factor $a_p$ or time $t_p$, i.e. $L(a=a_p)=L_p$, we can write the luminosity ratio $L/L_p$ as a function of scale factor $a$,
\begin{equation}
\frac{L}{L_p} =\left(\frac{a}{a_p}\right)^{-mp}\exp\left[\left(\frac{a_p}{a_2}\right)^{-mp+3/2}-\left(\frac{a}{a_2}\right)^{-mp+3/2}\right].
\label{eq:7-2-2}
\end{equation}
For $a\gg a_p$, a power-law decay of the quasar light curve can be obtained with $L\propto a^{-mp}\propto t^{-2mp/3}$. For $p=2$ and $m=5/2$ in stage E2, $L\propto a^{-5}\propto t^{-10/3}$. This power-law light curve is consistent with the self-regulated growth of BH, where the BH feedback expels gas and shuts down accretion \citep{Hopkins:2009-Quasars-Are-Not-Light-Bulbs-Testing-Models}.

We want to model the time the quasar takes in the declining phase evolving from the peak luminosity $L_p$ to the current luminosity $L$, i.e., $\Delta a=a-a_p$. For $a\gg a_p$ or Hubble time $t_H\gg t_p$, by inverting Eq. \eqref{eq:7-2-2}, we express the quasar lifetime $t_q$ as a function of $L$,
\begin{equation}
\begin{split}
&\frac{t_q}{t_p} = \left(1+\frac{\Delta a}{a_p}\right)^{\frac{3}{2}}-1\\
&\approx {\alpha_q} \left(\frac{L}{L_p}\right)^{-\frac{3}{2mp}}\exp\left[-\ln(\alpha_q){\alpha_q}^{1-\frac{2mp}{3}}\left(\frac{L}{L_p}\right)^{1-\frac{3}{2mp}}\right],\\
&\textrm{where} \quad \alpha_q = \exp\left[\frac{3}{2mp}\left(\frac{a_p}{a_2}\right)^{-mp+3/2}\right].
\end{split}
\label{eq:7-2-3}
\end{equation}

The distribution of quasar lifetimes $t_q$ can be translated to the Eddington ratio distribution \citep{Hopkins:2009-Quasars-Are-Not-Light-Bulbs-Testing-Models}. Assuming a variable $\tau_q=1$ exists when a quasar is in the active stage and $\tau_q=0$ when it is in the inactive stage. By the ergodic hypothesis, the average of $\tau_q$ for a given quasar over a sufficiently long time (that is, the ratio of total time in active stage to Hubble time $\langle \tau_q \rangle =t_q/t_H$) should equal the average of $\tau_q$ over the ensemble of all quasars at time $t_H$ (that is, the fraction of active quasars $\delta(z)$). This leads to the approximation of $\delta(z) \approx t_q/t_H$. With this approximation, we can write the Eddington ration distribution as
\begin{equation}
\begin{split}
&P(\lambda\vert M,z) \propto \Phi_{\lambda}(\lambda \vert M,z) \propto \frac{d\delta}{d\log \lambda} \approx \frac{t_q}{t_H(z)}\frac{d\log t_q}{d\log L},
\end{split}
\label{eq:7-2-4}
\end{equation}
where $\Phi_{\lambda}(\lambda \vert M,z)$ is the number density function of BHs with an Eddington ratio $\lambda$ at a fixed mass $M$ and redshift $z$. 

\begin{figure}
\includegraphics*[width=\columnwidth]{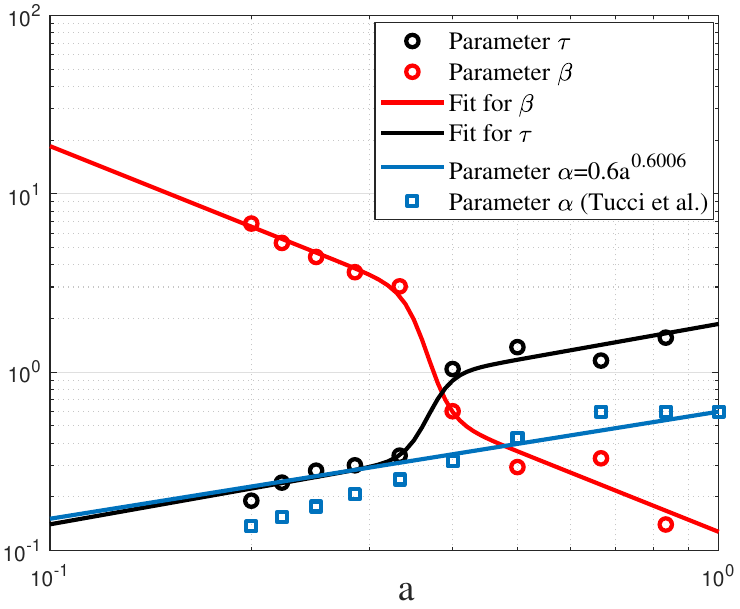}
\caption{The variation of parameters $\alpha$, $\beta$, and $\tau$ in Eddington ratio distribution $P(\lambda,z)$ (Eq. \eqref{eq:7-2-5}). The blue squares represent the values of $\alpha$ (power law slope) \citep{Tucci:2017-Constraining-supermassive-black-hole-evolution}. The solid blue line represents the $\alpha$ used in this work with the same redshift dependence as the faint-end slope $\gamma_1$ in the quasar luminosity function. The red and black circles represent the best local fit of $\beta$ and $\tau$ to give the best match with the number density function $\Phi_{\lambda}(\lambda,z)$ at each redshift. The red and black solid lines plot the best global fit of $\beta$ and $\tau$ at all redshifts. A sharp transition between $a=0.3$ and $a=0.4$ (or $z\approx2$) can be observed for $\beta$ and $\tau$.} 
\label{fig:S24}
\end{figure}

For a power-law quasar light curve, $P(\lambda\vert M,z) \propto t_q$ from Eq. \eqref{eq:7-2-4}, that is, quasars with a longer lifetime $t_q$ are more likely to be observed with greater probability $P$. Substituting $L=\lambda \varepsilon_{Edd}M$ and $L_p=\lambda_0 \varepsilon_{Edd}M$ into the equation for $t_q$ (Eq. \eqref{eq:7-2-3}), we can express the Eddington ratio distribution in a general form along with the normalization condition,
\begin{equation}
\begin{split}
&P(\lambda,z) = A \left(\frac{\lambda}{\lambda_0}\right)^{-\alpha}\exp\left[-\beta\left(\frac{\lambda}{\lambda_0}\right)^{\tau}\right],\\ 
&\int_{\lambda_{min}}^{\lambda_{max}}P(\lambda,z)d\log \lambda=1,
\end{split}
\label{eq:7-2-5}
\end{equation}
where $A$ is a normalization factor. In the quasar literature, $\tau=1$ and $\beta=1$ are usually taken while $\alpha$ is a free parameter. In this work, we take all three as free parameters since this form (Eq. \eqref{eq:7-2-5}) is directly derived from the light curve in Eq. \eqref{eq:7-2-3}. The $\gamma$th moment of distribution $P(\lambda\vert M,z)$ is also provided in an analytical form,
\begin{equation}
\begin{split}
&\int_{y_1}^{y_2} Ax^{-\alpha} \exp(-\beta x^{\tau})x^{\gamma} d\log x\\
&=\frac{A}{\ln(10)\tau}\beta^{\frac{\alpha-\gamma}{\tau}}\left[\Gamma \left(\frac{\gamma-\alpha}{\tau},\beta y_1^{\tau}\right)-\Gamma\left(\frac{\gamma-\alpha}{\tau},\beta y_2^{\tau}\right)\right].
\end{split}
\label{eq:7-2-6}
\end{equation}
From this equation, the normalization factor $A$ reads (with $\gamma=0$)
\begin{equation}
\begin{split}
&A = \frac{\ln(10)\tau \beta^{-(\alpha/\tau)}}{\Gamma\left[-\frac{\alpha}{\tau},\beta\left(\frac{\lambda_{min}}{\lambda_0}\right)^{\tau}\right]-\Gamma\left[-\frac{\alpha}{\tau},\beta\left(\frac{\lambda_{max}}{\lambda_0}\right)^{\tau}\right]}.\\
\end{split}
\label{eq:7-2-7}
\end{equation}

Here, the minimum Eddington ratio $\lambda_{min}$ is provided in Eq. \eqref{eq:7-1-4}, and $\lambda_{max}$ is the maximum Eddington ratio. $\Gamma(x,y)$ is the upper incomplete gamma function. With $\beta=1$ and $\tau=1$, Eq. \eqref{eq:7-2-5} reduces to the Schechter function form of the Eddington ratio distribution that is often adopted in the literature \citep{Hopkins:2009-Quasars-Are-Not-Light-Bulbs-Testing-Models,Cao:2010-Cosmological-Evolution-of-Massive-Black-Holes,Tucci:2017-Constraining-supermassive-black-hole-evolution}, where $\lambda_{min}=10^{-4}$ and $\lambda_0=1.5$ are often assumed. However, at high redshifts, the quasar's lifetime is comparable to the Universe's age. Most quasars are still very luminous; $\lambda_{min}$ and $\lambda_0$ should be larger at higher redshift \citep{Cao:2010-Cosmological-Evolution-of-Massive-Black-Holes}, as we show in Fig. \ref{fig:S26}. This work adopts a more general Eq. \eqref{eq:7-2-5} for Eddington ratio distribution and a redshift-dependent $\lambda_{min}$ and $\lambda_0$. Figure \ref{fig:S26} presents the variation of $\lambda_{min}$ and $\lambda_0$ (Eqs. \eqref{eq:7-1-4} and \eqref{eq:7-3-2-4}).

As suggested by Hopkins et al., the quasar lifetimes provide a physical interpretation of the break luminosity and the faint and bright end slopes of the luminosity function \citep{Hopkins:2006-The-Evolution-in-the-Faint-End-Slope-of-the-Quasar}. The bright end consists of quasars radiating near their peak luminosities. In contrast, the faint end consists of quasars passing over their peak luminosity and in a less luminous phase of evolution (stage E2). Therefore, the faint-end slope $\gamma_1$ is determined by the behavior of the quasar lifetimes and, hence, the Eddington ratio distribution. This suggests that the parameter $\alpha$ for $P(\lambda,z)$ should have a redshift dependence similar to that of $\gamma_1$ in the luminosity function. Furthermore, $\alpha=0.6$ is fixed at a low redshift by the observations \citep{Hopkins:2009-Quasars-Are-Not-Light-Bulbs-Testing-Models,Tucci:2017-Constraining-supermassive-black-hole-evolution,Kauffmann:2009-Feast-and-Famine-regulation-of-black-hole}. This leads to a power law variation for the slope $\alpha$ with the scale factor $a$, as shown in Eq. \eqref{eq:7-2-7-1} and Fig. \ref{fig:S24}. The parameter $\alpha$ adopted by \citep{Tucci:2017-Constraining-supermassive-black-hole-evolution} is plotted as blue squares. 
\begin{equation}
\begin{split}
&\alpha = 0.6a^{0.6006},\\
&\log \beta = -\frac{3}{2} \log a -\left(\frac{0.663}{1+6\exp\left[-\frac{\log(a)+0.467}{0.018}\right]}+0.233\right), \\
&\log \tau = \frac{2}{3} \log a -\left(\frac{-0.457}{1+6\exp\left[-\frac{\log(a)+0.467}{0.018}\right]}+0.187\right).
\end{split}
\label{eq:7-2-7-1}
\end{equation}

\begin{figure}
\includegraphics*[width=\columnwidth]{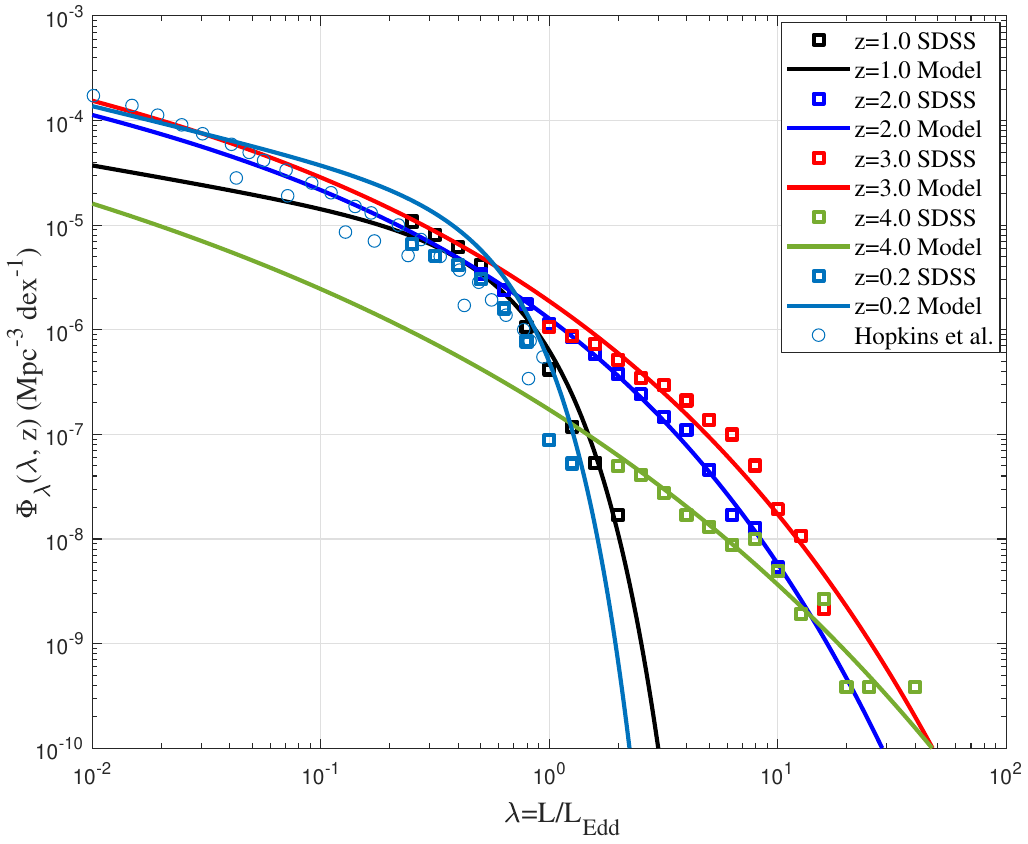}
\caption{The variation of number density function $\Phi_{\lambda}(\lambda,z)$ that is proportional to the Eddington ratio distribution $P(\lambda,z)$ (Eq. \eqref{eq:7-2-4}). Solid lines plot the $\Phi_{\lambda}(\lambda,z)$ from the general model for the Eddington ratio distribution (Eq. \eqref{eq:7-2-5}). The parameters $\alpha$, $\beta$, and $\tau$ are taken from the best local fit in Fig. \ref{fig:S24}. Square symbols represent data from the Sloan Digital Sky Survey (SDSS DR16Q) \citep{Wu:2022-A-Catalog-of-Quasar-Properties-from-Sloan} with $\lambda$ greater than the break Eddington ratio $\lambda_0$. The circles represent the data from Hopkins et al. \citep{Hopkins:2009-Quasars-Are-Not-Light-Bulbs-Testing-Models} at $z=0.2$ that extends to small $\lambda$.} 
\label{fig:S25}
\end{figure}

For parameters $\beta$ and $\tau$, we determine the best fit to SDSS data (Fig. \ref{fig:S25}) at each redshift (local fit) and present them as circles in Fig. \ref{fig:S24}. The global fit of $\beta$ and $\tau$ at all redshifts (Eq. \eqref{eq:7-2-7-1}) are presented as solid lines in Fig. \ref{fig:S24}. The parameter $\beta$ decreases consistently, while $\tau$ increases with time. Both have a sharp transition around $z=2$, where $\beta\approx 1$ and $\tau\approx 1$. The standard Schechter function form of the Eddington ratio distribution with $\beta=1$ and $\tau=1$ might be over-constrained. Equations \eqref{eq:7-2-5} and \eqref{eq:7-2-7-1} give a general model for the Eddington ratio distribution that is applied to derive the AGN mass function through convolution Eq. \eqref{eq:7-3-1}. It should noted that the Eddington ratio distribution in this section does not include the contributions from the Type I AGN, which is usually assumed to be log-normal \citep{Tucci:2017-Constraining-supermassive-black-hole-evolution}.

\section{New limit for Super-Eddington growth}
\label{sec:9-2}
In this work, we present a two-stage model for the evolution of SMBH luminosity (Eq. \eqref{eq:7-1-7}) with a redshift-dependent parameter $\varepsilon_b(z)$. It has been applied to derive the evolution of the BH mass functions and the AGN duty cycle and was compared with numerical solutions and observations for model validations (Sections \ref{sec:7-1} to \ref{sec:7-2}). This evolution model is independent of the mechanism by which the initial BH seeds are formed. Instead, it predicts the path of evolution once the BH seeds have formed (Section \ref{sec:7}). Unlike the Eddington accretion with a fixed Eddington ratio, SMBHs along this path of evolution have an initial rapid growth and a varying Eddington ratio that involves a super-Eddington growth in a relatively short time when BHs are still relatively small (stage E1), followed by stage E2 with a decreasing luminosity $L_B\propto a^{-5}$. 

To better understand this, we compare the Eddington accretion and our evolution model and present a new luminosity limit for BH super-Eddington growth. For Eddington accretion at Eddington luminosity $L_{Edd}$, the pressure (or force $L_{Edd}/c$) due to BH luminosity is exactly balanced by the static weight of gases surrounding the SMBH, 
\begin{equation}
\frac{L_{Edd}}{4\pi c r^2} = \frac{G M_{BH} m_p}{r^2 \sigma_T} \quad \textrm{or} \quad a_{Edd}=\frac{L_{Edd}}{M_{BH}c} =4\pi \frac{Gm_p}{\sigma_T},
\label{eq:54}
\end{equation}
where $\sigma_T\approx 6.65\times 10^{-29}m^2$ is the Thomson scattering cross-section for electron, $m_p \approx 1.67 \times 10^{-27}kg$ is the mass of a proton, $a_{Edd}= 2.1\times 10^{-8}{m}/{s^2}$ is a characteristic (Eddington) acceleration resulting from the gravity of static gas. Therefore, the Eddington limit is the maximum luminosity with static gas surrounding an SMBH. 

In our evolution model, gas is never a static medium. Instead, the gas forms a turbulent medium with random dynamic motions. Along the radial direction, an equivalent dynamic pressure can be related to the random motion $P_r \propto \rho_r \sigma_r^2 $, where $\sigma_r^2$ is the velocity dispersion (Eq. \eqref{eq:10}). This dynamic pressure mimics the pressure term in the Jeans equation due to the random velocity \citep{Mo:2010-Galaxy-formation-and-evolution}. 
The force associated with the gradient of this dynamic pressure is $F_r\propto \sigma_r^4/G$ (Eq. \eqref{eq:10}). The difference between our model and the Eddington limit is that the radiation force $L_B/c$ from the SMBH luminosity must balance the force $F_r$ due to this turbulent and dynamic motion, instead of the static weight in Eq. \eqref{eq:54}. We now consider the force balance on the radiation scale $r_p$. The mean flow of gases rotating around BH provides the centrifugal force that balances the BH gravity. While the random motion of gases provides the force $F_r$ that balances the force $L_B/c$ from the BH luminosity. Therefore, the forces exerted on the spherical surface of $4\pi r_p^2$ read 
\begin{equation}
\begin{split}
&\frac{L_{B}}{c}=\frac{\sigma_p^4}{G} \quad \textrm{where} \quad \sigma_p^2=(\varepsilon_br_p)^{2/3}.
\end{split}
\label{eq:54a-2}
\end{equation}
Here $\sigma_p^2$ is the velocity dispersion on the radiation scale $r_p$ (Eq. \eqref{eq:16}) (see the scaling laws involving $\varepsilon_b$ in Eq. \eqref{eq:10}). On the scale of BH influence $r_B$ (Eq. \eqref{eq:15}), the BH mass reads
\begin{equation}
\begin{split}
&M_{BH} \propto \varepsilon_b^{2/3} G^{-1} r_B^{5/3}.
\end{split}
\label{eq:54a-3}
\end{equation}
The radiation scale $r_p$ cannot exceed the BH scale of influence $r_B$. Beyond that limit, the gravity of the SMBHs cannot hold the gas repelled by the radiation pressure. By setting $r_p=r_B$, we can obtain the maximum BH luminosity, a new limit different from the Eddington limit. First, with $r_p=r_B$ in Eqs. \eqref{eq:54a-2} and \eqref{eq:54a-3}, the velocity dispersion is related to the BH mass as
\begin{equation}
\begin{split}
&\sigma_p^2 \propto \left(M_{BH}\varepsilon_b G\right)^{2/5},
\end{split}
\label{eq:54a-4}
\end{equation}
where $\sigma_p^2$ can be much larger at high redshift with larger $\varepsilon_b$. This means a larger BH luminosity or mass accretion rate is required to balance that dynamic pressure or force due to $\sigma_p^2$. Substituting this into Eq. \eqref{eq:54a-2}, the maximum luminosity or acceleration reads
\begin{equation}
\begin{split}
&L_{X}=x_r\varepsilon_b^\frac{4}{5}G^{-\frac{1}{5}}cM_{BH}^{\frac{4}{5}}, \quad a_X = \frac{L_X}{M_{BH}c}=\frac{x_r\varepsilon_b^{\frac{4}{5}}}{(M_{BH}G)^{\frac{1}{5}}},
\end{split}
\label{eq:54a}
\end{equation}
where $x_r$ is a numerical factor on the order of ten. With $\varepsilon_b=\varepsilon_{b0}a^{-5/2}$, the new luminosity limit finally reads
\begin{equation}
\begin{split}
&L_{X}=x_r(1+z)^2\varepsilon_{b0}^{{4}/{5}}G^{-{1}/{5}}cM_{BH}^{{4}/{5}}.
\end{split}
\label{eq:54a-5}
\end{equation}

\begin{figure}
\includegraphics*[width=\columnwidth]{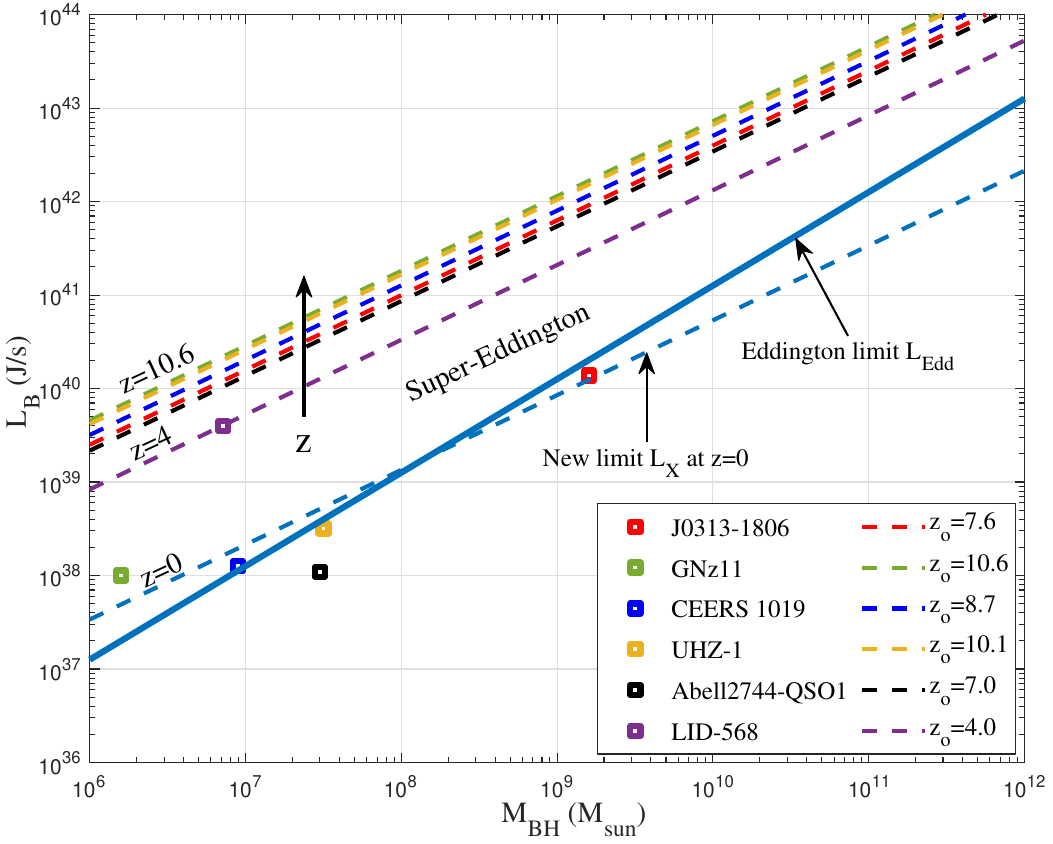}
\caption{The comparison between the new luminosity limit $L_X$ and the Eddington limit $L_{Edd}$. Solid blue line plots the Eddington limit $L_{Edd}=\varepsilon_{Edd}M_{BH}$ with $\varepsilon_{EDD}=6.3$m$^2$/s$^3$. This is obtained from the force balance between the radiation pressure and the static weight of surrounding gases (Eq. \eqref{eq:54}). Square symbols represent the observed SMBHs with known mass $M_{ob}$ and luminosity $L_{ob}$ at given redshift $z_{ob}$ (Table \ref{tab:2-2}). Dashed lines plot the new limit $L_X$ for different SMBHs at different redshifts with $x_r=15$. The new limit is obtained from the force balance between the radiation pressure and the dynamic pressure from the random motion of gases (Eq. \eqref{eq:54a}). The new limit $L_X$ is higher at higher redshift. Super-Eddington growth is allowed for SMBHs at high redshift and for low-mass SMBHs at low redshift. The recently discovered LID-568 at redshift $z_o=4$ was found to accrete at 40 times the Eddington ratio. It is right on the boundary of the new limit $L_X$.} 
\label{fig:S38}
\end{figure}

The new limit depends on the key parameter $\varepsilon_b$ and is redshift dependent. Figure \ref{fig:S38} presents the comparison between the new time-dependent limit $L_X$ and the Eddington limit $L_{Edd}$ (Eq. \eqref{eq:54}). Square symbols plot the observed SMBHs with known mass $M_{ob}$ and luminosity $L_{ob}$ at given redshift $z_{ob}$ (Table \ref{tab:2-2}). Dashed lines plot the new limit $L_X$ for SMBHs observed at different redshifts $z_{ob}$. Since the rate of energy flow $\varepsilon_b$ decreases with time, this means more efficient gas cooling and a richer supply of cold gas in the early Universe to allow rapid initial growth. Similarly, compared to $a_{Edd}$, the limit acceleration $a_X$ is also time-dependent and is much higher in the early Universe. Since the BH mass $M_{BH}$ also increases with time, $a_X$ is much higher in the early Universe and monotonically decreases with time. There exists a period with $a_X>a_{Edd}$ or $L_B>L_{Edd}$ in stage E1, compared to the Eddington limit in Eq. \eqref{eq:54}. This is clearly shown in Fig. \ref{fig:S38}. All SMBHs are within the new limit, with GNz11 and CEERS 1019 beyond the Eddington limit. The recently discovered LID-568 accreting at 40 times Eddington ratio seems right on the new limit \citep{suh:2024-feeding-hidden-monsters-super-eddington}. 

During the super-Eddington growth, the radiation pressure from the BH luminosity must support the radial pressure resulting from the random motion in the surrounding gas, which can be much higher than the pressure from the static weight of the gas. This requires fast mass accretion and high luminosity. Therefore, SMBH luminosity may exceed the Eddington limit in its early stage. SMBHs may evolve with super-Eddington accretion for a short period (depending on the radiative efficiency $\epsilon$) when they are still relatively small. The predicted super-Eddington evolution of some observed high-redshift SMBHs is presented in the Appendix (Section \ref{sec:8} and Fig. \ref{fig:S36}).

\section{Conclusion}
\label{sec:9}
\rev{
Observations strongly suggest a coevolution of supermassive black holes (SMBHs) and host galaxies. In this paper, we consider the mass and energy flow in a near-equilibrium bulge suffused by gases of different temperatures ranging from cold at small scales to warm and hot at large scales. By assuming that i) the rate of energy flow is independent of the scale $r$ and ii) the permeated gases are in local virial equilibrium, a key parameter $\varepsilon_b$ (unit: $m^2$/$s^3$) was identified to quantify the mass and energy flow in gases. The permeated gas is self-regulated in a way that the net energy accumulated on any scale $r$ due to the energy flow always balances the energy dissipated on the same scale. Therefore, $\varepsilon_b$ is also relevant to the gas cooling and the supply of cold gas and thus regulates the synchronized evolution of SMBHs and hosts. Since parameter $\varepsilon_b\propto (1+z)^{5/2}$, a larger $\varepsilon_b$ in the early Universe means faster mass and energy flow and more efficient gas cooling that allows a rapid evolution of SMBHs and hosts and fast star formation. At lower redshifts, a smaller $\varepsilon_b$ means slower mass and energy flow, less efficient gas cooling, less cold gas supply, and slower star formation and SMBH growth. Therefore, the rapid decrease in $\varepsilon_b$ represents a global "quenching" process on the cosmic scale that slows down the evolution of SMBHs and hosts. Since parameter $\varepsilon_b$ is the central quantity of the cosmic quenching and the SMBH-host coevolution, a "key-$\varepsilon$ theory" can be termed to represent the cosmic quenching and the associated scaling laws involving $\varepsilon_b$ that quantifies the coevolution.

This relatively simple theory, characterized by a single parameter $\varepsilon_b$, neglects all the transient phenomena. When properly calibrated by the mass-size relations from simulations and observations, this simple theory gives rise to the dominant mean cosmic evolution of SMBHs and host galaxies. By contrast, the transient phenomena are rapid, short-lived, and high-energy events that occur over a short time scale compared to the overall galaxy formation (i.e., the massive merging, the merging-induced disruptions, and AGN jets and winds, etc). These transient phenomena may trigger start formation bursts, disrupt the existing gases, and impact the structure evolution. The effects of these transient phenomena may be helpful in explaining the dispersion around the mean cosmic evolutions. However, the mean cosmic evolutions of SMBHs and hosts are mostly regulated by the mass and energy flow and the energy dissipation in gases, all characterized by the single key parameter $\varepsilon_b$.} 

Based on relevant assumptions, scaling laws involving parameter $\varepsilon_b$ were identified that govern the evolution of both SMBHs and host galaxies. For host galaxies, we identify the bulge mass-size relation $M_b\propto \varepsilon_b^{2/3}r_b^{5/3}G^{-1}$ and the dispersion-size relation $\sigma_b^2\propto(\varepsilon_b r_b)^{2/3}\propto a^{-1}$. These scaling laws agree well with observations, especially true for early-type galaxies. Similar scaling laws were also proposed for dark matter haloes, which are relevant to halo structures and dark matter particle mass and properties \citep{Xu:2023-Universal-scaling-laws-and-density-slope,Xu:2022-Postulating-dark-matter-partic}.

For SMBHs, an initial rapid growth stage is identified with a sharp increase in luminosity $L_B\propto (\varepsilon_b M_{BH})^{4/5}G^{-1/5}c$, followed by a transition stage with a decrease in luminosity $L_B\propto \varepsilon_b^2 M_{BH} \propto a^{-5}$, and a dormant stage with $L_B\propto (\varepsilon_b M_{BH})^{4/3}G^{1/3}c^{-5/3}$. For SMBH-galaxy coevolution, the observed $M_b$-$\sigma_b$ correlation is analytically derived as $M_{BH}\propto \sigma_b^5/(\varepsilon_b G)$ (Eq. \eqref{eq:31}). The luminosity naturally peaks at $z\approx 2$ due to the increase $M_{BH}$ and the decrease $\varepsilon_b$ from cosmic quenching. By introducing two dimensionless parameters $\gamma={L_B}/(M_{BH}\varepsilon_b)$ and $\eta=({GL_B}/{c^5})^{{1}/{4}}$, the distribution and evolution of SMBHs is concisely mapped in the $\gamma$-$\eta$ plane. The upper and lower limits of the SMBH distribution are $\gamma \eta =10$ and $\gamma \eta^{-1}=10$, respectively, together with $\gamma_c=10$ as the boundary of active and inactive SMBH. The three-phase evolution of SMBH follows $\gamma \propto \eta^{-1}$, $\gamma \propto \eta^2$, and $\gamma \propto \eta$, respectively.

Based on these scaling laws, analytical solutions are formulated for the evolution of the BH mass function, the AGN mass function $\Phi_{AGN}(M,z)$, the duty cycle $U(M,z)$, and the Eddington ratio distribution. The model predicts $\Phi_L\propto L^{-1/5}$ for the faint-end quasar luminosity function, $\Phi_{AGN}\propto M^{-1/5}$ for a small mass $M$, and $U\propto M^{-1/5}$ at high redshift. Finally, for high-redshift SMBHs with observed luminosity and BH mass, complete redshift evolution is predicted by these scaling laws (Section \ref{sec:8}). The results reveal an initial super-Eddington growth in a short period when the SMBHs are still small, followed by a slow growth due to cosmic quenching when the SMBHs become large. A new luminosity limit is obtained from the balance between the radiation force from the SMBH luminosity and the effective force due to the random motion in dynamic gases. This new redshift-dependent limit $L_X\propto\varepsilon_b^{4/5}M_{BH}^{4/5}G^{-1/5}c$ allows a super-Eddington growth during the early stage of SMBH evolution. 

\section*{Data Availability}
Datasets for this article are available on Zenodo \citep{Xu:2022-The-cosmic-quenching-and-scaling-laws-for}.

\section*{Acknowledgments}
This research was supported by Laboratory Directed Research and Development at Pacific Northwest National Laboratory (PNNL). PNNL is a multiprogram national laboratory operated for the U.S. Department of Energy (DOE) by Battelle Memorial Institute under contract no. DE-AC05-76RL01830. The author is grateful to Prof. Curtis Struck for his constructive comments. 

\bibliographystyle{mnras}
\bibliography{Papers}

\appendix
\addtocontents{toc}{\protect\setcounter{tocdepth}{1}}
\renewcommand\nomgroup[1]{%
  \item[\bfseries
  \ifstrequal{#1}{A}{Physics Constants}{%
  \ifstrequal{#1}{B}{Cosmology Variables}{%
  \ifstrequal{#1}{J}{N-body Simulations}{%
  \ifstrequal{#1}{C}{SMBHs and Host Galaxies and Dark Matter Haloes}{%
  \ifstrequal{#1}{D}{BH mass functions}{%
  \ifstrequal{#1}{E}{Quasar Luminosity, AGN Mass Functions and Duty Cycle}{%
  \ifstrequal{#1}{G}{Eddington Ratio Distribution}{%
  \ifstrequal{#1}{H}{Scaling Laws in Bulge}{%
  \ifstrequal{#1}{K}{Other Symbols}{}}}}}}}}}%
]}
\newcommand{\nomunit}[1]{%
\renewcommand{\nomentryend}{\hspace*{\fill}#1}}

\nomenclature[A, 01]{$c$}{Speed of light \nomunit{$3\times 10^8$m/s}}
\nomenclature[A, 02]{$G$}{Gravitational constant \nomunit{$6.67\times 10^{-11}$m$^3$kg$^{-1}$s$^{-2}$}}
\nomenclature[A, 04]{$\sigma_T$}{Thomson scattering cross-section \nomunit{$6.65\times 10^{-29}$m$^2$}}
\nomenclature[A, 05]{$m_p$}{Mass of proton \nomunit{$1.67 \times 10^{-27}$kg}}
\nomenclature[A, 06]{$\quad$}{{\parbox[t][0.05cm]{10cm}{}}}

\nomenclature[B, 01]{$z$, $a$, $t$}{Redshift, scale factor, and comic time}
\nomenclature[B, 02]{$\Omega_{DE}$}{Dark energy fraction \nomunit{$0.7274$}}
\nomenclature[B, 03]{$\Omega_m$}{Total matter fraction \nomunit{$0.2726$}}
\nomenclature[B, 04]{$\Omega_b$}{Baryonic matter fraction \nomunit{$0.0456$}}
\nomenclature[B, 05]{$H(z)$}{Hubble parameter}
\nomenclature[B, 06]{$H_0$}{Hubble constant \nomunit{$70$km/s/Mpc}}
\nomenclature[B, 07]{$h$}{Dimensionless Hubble parameter \nomunit{0.7}}
\nomenclature[B, 08]{$\epsilon$}{Black hole radiative efficiency}
\nomenclature[B, 09]{$t_{sal}$}{The Salpeter time \nomunit{$4.5\times 10^8 \epsilon/(1-\epsilon)$years}}
\nomenclature[B, 10]{$L_{Edd}$}{Eddington luminosity} 
\nomenclature[B, 11]{$a_{Edd}$}{Eddington acceleration \nomunit{$2.1\times 10^{-8}$m/s$^2$}} 
\nomenclature[B, 12]{$\varepsilon_b(z)$}{Central quantity of cosmic quenching that quantifies the mass and energy flow in bulge \nomunit{$\varepsilon_b=\varepsilon_{b0}a^{-5/2}$}}
\nomenclature[B, 13]{$\varepsilon_{Edd}$}{Eddington limit of $\varepsilon_b$ \nomunit{$\varepsilon_{Edd}=L_{Edd}/M_{BH}=6.3$m$^2$/s$^3$}} 
\nomenclature[B, 14]{$\varepsilon_{b0}$}{The rate of energy flow at $z=0$ \nomunit{$10^{-4}$m$^2$/s$^3$}}
\nomenclature[B, 15]{$\varepsilon(m_h,z)$}{Rate of energy flow in haloes of mass $m_h$ in Eq. \eqref{ZEqnNum9863112993}}
\nomenclature[B, 17]{$\varepsilon_a(z)$}{The rate of baryonic energy dissipation in bulge in Eq. \eqref{eq:3-1-2}}
\nomenclature[B, 18]{$\quad$}{{\parbox[t][0.05cm]{10cm}{}}}

\nomenclature[C, 01]{$L_b$}{Bulge luminosity}
\nomenclature[C, 02]{$L_B$}{SMBH luminosity}
\nomenclature[C, 03]{$M_b$}{Bulge mass}
\nomenclature[C, 04]{$M_{BH}$}{SMBH mass}
\nomenclature[C, 05]{$m_h$}{Dark matter halo mass}
\nomenclature[C, 06]{$m_h^*(z)$}{Characteristic mass of dark matter haloes}
\nomenclature[C, 07]{$P_{rad}$}{SMBH radiation pressure \nomunit{$P_{rad}=L_B/(4\pi r^2c)$}}
\nomenclature[C, 08]{$r_b$}{The scale of bulge (bulge size) in Eq. \eqref{eq:15}}
\nomenclature[C, 09]{$r_B$}{The scale of BH sphere influence in Eq. \eqref{eq:15}}
\nomenclature[C, 10]{$r_s$}{The scale of Schwarzschild radius in Eq. \eqref{eq:15}}
\nomenclature[C, 11]{$r_p$}{The radiation scale in Eq. \eqref{eq:16} where the radiation pressure $P_{rad}$ balances the dynamic gas pressure $P_r$ }
\nomenclature[C, 10]{$r_x$}{The dissipation scale in Eq. \eqref{eq:17} below which the effects of accretion disk become important}
\nomenclature[C, 11]{$\sigma_b^2$}{Bulge velocity dispersion in Eq. \eqref{ZEqnNum98631129992}}
\nomenclature[C, 13]{$\gamma$, $\beta$, $\eta$}{Key dimensionless parameters defined in Eq. \eqref{eq:18} for the SMBH distribution and evolution}
\nomenclature[C, 14]{$\quad$}{{\parbox[t][0.05cm]{10cm}{}}}

\nomenclature[D, 01]{$\Phi_{BH}(M,z)$}{BH mass function in logarithmic units of BH mass $M$ in Eq. \eqref{eq:7-1-1}. The analytical model is presented in Eq. \eqref{eq:7-1-9-2}}
\nomenclature[D, 02]{$\Phi^*_{BH}(M,z)$}{The usual BH mass function for the number density of SMBHs per co-moving volume in Eq. \eqref{eq:7-1-1}}
\nomenclature[D, 03]{$\Phi_{BH0}(M)$}{BH mass function at $z=0$}
\nomenclature[D, 04]{$\Phi^*_{BH0}(M)$}{Schechter type mass function at $z=0$ in Eq. \eqref{eq:7-1-9-1}}
\nomenclature[D, 05]{$\Phi^1_{BH}(M,z)$}{Mass function at small mass end in Eq. \eqref{eq:7-1-8}}
\nomenclature[D, 06]{$\Phi^2_{BH}(M,z)$}{Mass function at large mass end in Eq. \eqref{eq:7-1-9}}
\nomenclature[D, 07]{$\Phi^3_{BH}(M,z)$}{Mass function for fixed Eddington ratio in Eq. \eqref{eq:7-1-9-4}}
\nomenclature[D, 08]{$\rho_{BH}(z)$}{BH mass density increasing with $z$ in Eqs. \eqref{eq:7-1-10} and \eqref{eq:7-1-12}}
\nomenclature[D, 09]{$\rho_{BH0}$}{The local BH mass density at $z=0$}
\nomenclature[D, 10]{$\quad$}{{\parbox[t][0.05cm]{10cm}{}}}

\nomenclature[E, 01]{$\Phi_{L}(L,z)$}{Quasar luminosity function in Eq. \eqref{eq:7-3-1}}
\nomenclature[E, 02]{$\Phi_{AGN}(M,z)$}{AGN mass function}
\nomenclature[E, 03]{$\hat\Phi_{AGN}(M,z)$}{Analytical AGN mass function (Eq. \eqref{eq:7-3-3}) obtained from luminosity function and Eddington ratio distribution}
\nomenclature[E, 04]{$\Tilde\Phi_{AGN}(M,z)$}{AGN mass function (Eq. \eqref{eq:8-3}) obtained from BH mass function and quasar duty cycle}
\nomenclature[E, 05]{$L^*(z)$}{Break luminosity for quasar luminosity function in Eq. \eqref{eq:7-3-2}}
\nomenclature[E, 06]{$L_1^*(z)$}{Analytical break luminosity in rising stage in Eq. \eqref{eq:7-3-2-3}}
\nomenclature[E, 07]{$L_2^*(z)$}{Analytical Break luminosity in declining stage in Eq. \eqref{eq:7-3-2-3}}
\nomenclature[E, 08]{$\phi^*(z)$}{Number density for quasar luminosity function in Eq. \eqref{eq:7-3-2}}
\nomenclature[E, 09]{$M^*(z)$}{Break mass in AGN mass function in Eqs. \eqref{eq:7-3-2-4} and \eqref{eq:7-3-3}}
\nomenclature[E, 10]{$U(M,z)$}{Duty cycle. The analytical model is presented in Eq. \eqref{eq:8-2}}
\nomenclature[E, 11]{$U^1(M,z)$}{AGN duty cycle at small-mass end in Eq. \eqref{eq:8-1}}
\nomenclature[E, 12]{$U^2(M,z)$}{AGN duty cycle at large-mass end in Eq. \eqref{eq:8-1}}
\nomenclature[E, 13]{$\quad$}{{\parbox[t][0.05cm]{10cm}{}}}

\nomenclature[G, 01]{$P(\lambda,z)$}{Eddington ratio distribution}
\nomenclature[G, 02]{$\alpha$, $\beta$, $\tau$}{Parameters in Eddington ratio distribution in Eq. \eqref{eq:7-2-5}}
\nomenclature[G, 03]{$\lambda$}{Eddington ratio}
\nomenclature[G, 04]{$\lambda_0$}{Break Eddington ratio in Eqs. \eqref{eq:7-3-2-4} and \eqref{eq:7-2-5}}
\nomenclature[G, 05]{$\langle \lambda \rangle$}{Mean Eddington ratio in Eq. \eqref{eq:7-1-3-1}}
\nomenclature[G, 06]{$\lambda_{min}$}{Minimum quasar Eddington ratio in Eq. \eqref{eq:7-1-4} and \eqref{eq:7-2-5}}
\nomenclature[G, 07]{$\lambda_{max}$}{Maximum quasar Eddington ratio in Eq. \eqref{eq:7-2-5}}
\nomenclature[G, 08]{$\Phi_{\lambda}(\lambda,z)$}{Number density function that is proportional to $P(\lambda,z)$}
\nomenclature[G, 09]{$\quad$}{{\parbox[t][0.05cm]{10cm}{}}}

\nomenclature[J, 01]{$M_h$ $M_g$ $M_s$}{Total mass of dark matter, gas, and star in all haloes}
\nomenclature[J, 02]{$\Lambda^i_m(m_h,r,z)$}{Cumulative mass function in Eq. \eqref{ZEqnNum98631129}}
\nomenclature[J, 03]{$\Lambda^i_{pv}(m_h,r,z)$}{Cumulative kinetic energy in Eq. \eqref{ZEqnNum986311299}}
\nomenclature[J, 04]{$\rho_h^i(m_h,r,z)$}{Mean mass density of specie $i$ in Eq. \eqref{ZEqnNum98631129}}
\nomenclature[J, 05]{$m_h^i$}{Total mass of specie $i$ in haloes of mass $m_h$}
\nomenclature[J, 06]{$\sigma_g^2(r)$}{Velocity dispersion of gas in bulge in Eq. \eqref{ZEqnNum98631129992}}
\nomenclature[J, 07]{$\quad$}{{\parbox[t][0.05cm]{10cm}{}}}

\nomenclature[H, 01]{$r$}{Scale for the distance to the center of bulge}
\nomenclature[H, 02]{$m_r$}{Total mass enclosed within scale $r$ in Eq. \eqref{eq:9}}
\nomenclature[H, 03]{$\rho_r$}{Average mass density enclosed within scale $r$ in Eq. \eqref{eq:9}}
\nomenclature[H, 04]{$t_r$}{Characteristic time on scale $r$ in Eq. \eqref{eq:10}}
\nomenclature[H, 05]{$\sigma_r$}{Characteristic velocity on scale $r$ in Eq. \eqref{eq:10}}
\nomenclature[H, 06]{$P_r$}{Dynamic pressure at $r$ due to gas random motion (Eq. \eqref{eq:10})}
\nomenclature[H, 07]{$F_r$}{Force as the gradient of dynamic pressure at $r$ in Eq. \eqref{eq:10}}
\nomenclature[H, 08]{$\dot m_r$}{Characteristic rate of mass flow on scale $r$ in Eq. \eqref{eq:11}}
\nomenclature[H, 09]{$\dot M_b$}{Rate of mass flow at bulge scale $r=r_b$ in Eq. \eqref{eq:19}}
\nomenclature[H, 10]{$\dot M_B$}{Rate of mass flow at BH influence scale $r=r_B$ (Eq. \eqref{eq:19})}
\nomenclature[H, 11]{$\dot M_p$}{Rate of mass flow at the radiation scale $r=r_p$ in Eq. \eqref{eq:19}}
\nomenclature[H, 12]{$\quad$}{{\parbox[t][0.05cm]{10cm}{}}}

\nomenclature[K, 01]{$c_s$}{Sound speed in accretion disk in Eq. \eqref{eq:17-1}}
\nomenclature[K, 02]{$H_s$}{Thickness of the accretion disk at $r_s$ in Eq. \eqref{eq:17-1}}
\nomenclature[K, 03]{$\nu_s$}{Kinematic viscosity of the accretion disk at $r_s$ in Eq. \eqref{eq:17-1}}
\nomenclature[K, 04]{$L_{ob}$,$M_{ob}$}{Luminosity and mass of observed SMBHs in Eq. \eqref{eq:54-1}}
\nomenclature[K, 05]{$a_{ob}$,$z_{ob}$}{Scale factor and redshift of observed SMBHs in Eq. \eqref{eq:54-1}}
\nomenclature[K, 06]{$\lambda_{ob}$}{Eddington ratio of observed SMBHs}
\nomenclature[K, 07]{$\theta$}{Parameter for SMBH evolution in Eq. \eqref{eq:54-6} \nomunit{$\theta=\eta^4/\gamma$} }
\nomenclature[K, 08]{$\Gamma(T,z)$}{Gas cooling function in Eq. \eqref{eq:3-1-5}}
\nomenclature[K, 09]{$n_H$}{Number density of gas particles in Eq. \eqref{eq:3-1-5}}
\nomenclature[K, 10]{$\Upsilon$}{Mass-to-light ratio}
\printnomenclature

\section{The redshift evolution of SMBH}
\label{sec:8}

\begin{table}
    \caption{Five observed high redshift SMBHs in $\gamma$-$\eta$ plane.}
    \label{tab:2-2}
    \tabletypesize{\scriptsize}
\centering
    \begin{tabular}{p{0.6in}p{0.2in}p{0.45in}p{0.5in}p{0.15in}p{0.4in}}
    \hline
     Name             & $z_{ob}$              & $M_{ob}$($M_{\odot}$)            & $L_{ob}$(J/s)     & $\gamma_{ob}$ & $\eta_{ob}$ \\
    \hline
    J0313-1806        &  7.642                & $1.6\times 10^9$              & $1.38\times 10^{40}$ & 196.4 & 0.78E-3         \\
    \hline   
    Ceers 1019        &  8.679                & $0.89\times 10^7$             & $1.26\times 10^{38}$   & 242.5 & 0.24E-3     \\
    \hline   
    UHZ-1             & 10.073                & $3.16\times 10^7$             & $3.18\times 10^{38}$     &123.2 & 0.31E-3                 \\
    \hline
    Abell2744         & 7.0451                & $3\times 10^7$                & $1.1\times 10^{38}$        & 99.9 & 0.23E-3               \\
    \hline
    GNz11             & 10.6                  & $1.6\times 10^6$              & $1\times 10^{38}$            & 688.4 &  0.23E-3       \\
    \hline
    \end{tabular}
\end{table}

Sections \ref{sec:7-1} to \ref{sec:7-2} provide an evolution model of the BH mass and luminosity in two separate stages (Eqs. \eqref{eq:38} and \eqref{eq:7-1-7}), along with applications to the SMBH and AGN mass functions and duty cycle for model validation. In this section, we apply the theory to model the complete evolution of individual SMBHs observed at a given redshift with a given mass and luminosity. This is achieved by modeling their evolution in the $\gamma-\eta$ plane first and converting it to the corresponding evolution in redshift space. Although the predicted evolution cannot be directly compared to observations, high-resolution simulations may provide potential validations.  

\begin{figure}
\includegraphics*[width=\columnwidth]{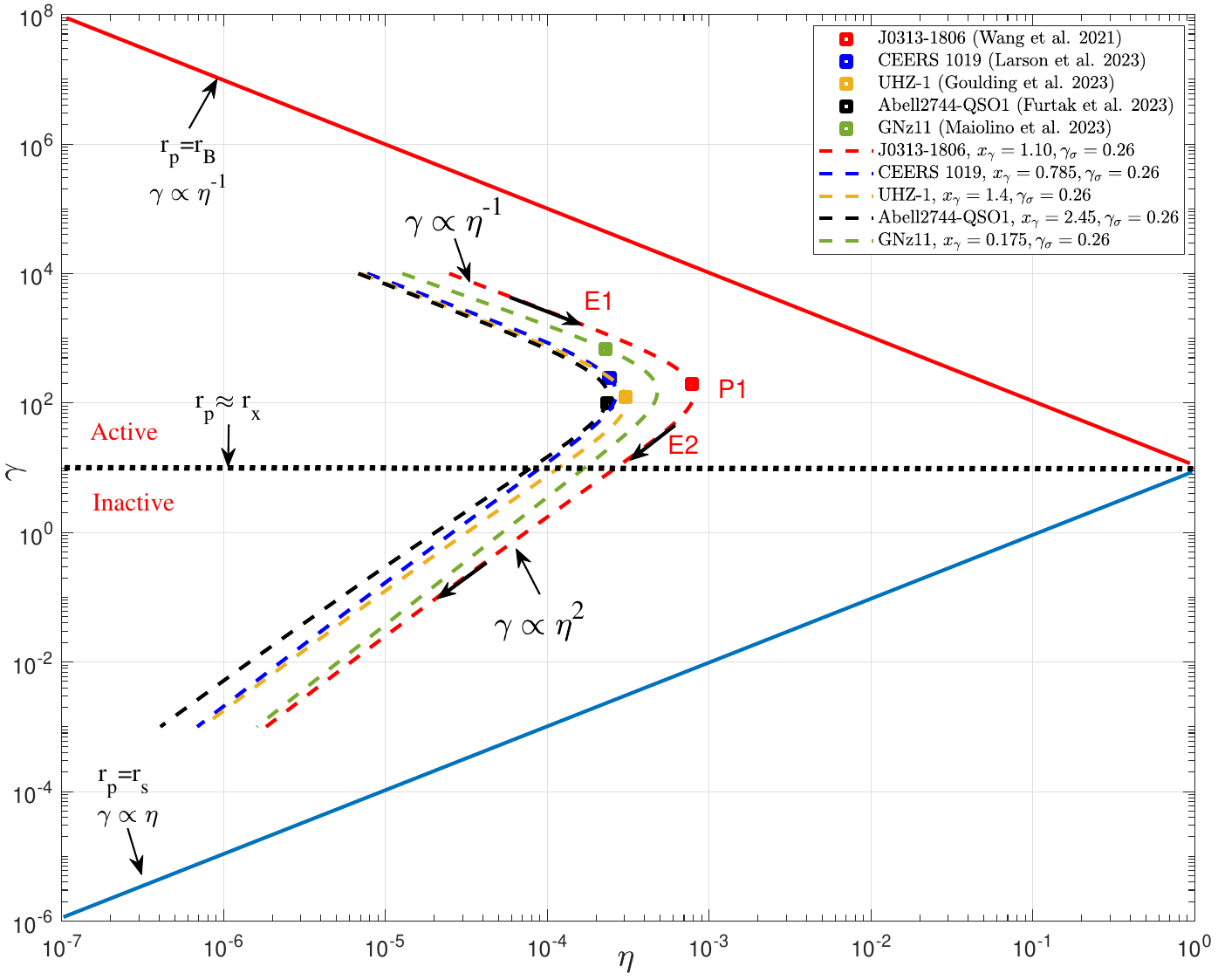}
\caption{The evolution of five high-redshift SMBHs in $\gamma$-$\eta$ plane. Square symbols represent the observed SMBHs in the $\gamma$-$\eta$ plane (Eq. \eqref{eq:54-1}). These active SMBHs are in the upper half-triangle. The solid red line represents the upper limit of the distribution of SMBHs (See Fig. \ref{fig:3}). The solid blue line represents the lower limit. The black dotted line represents the boundary between active and inactive SMBHs (Fig. \ref{fig:3}). Dashed lines represent the evolution of observed SMBHs with two stages ($\gamma\propto \eta^{-1}$ for E1 and $\gamma\propto \eta^{2}$ for E2) from Eq. \eqref{eq:54-2}. Parameters $x_{\gamma}$ and $\gamma_{\sigma}$ are given for each SMBH, as shown in Figure. The turning point P1 denotes the location of peak luminosity.} 
\label{fig:S34}
\end{figure}

For any SMBH observed at redshift $z_{ob}$ with a given mass $M_{ob}$ and luminosity $L_{ob}$, the first step is to map that mass and luminosity onto the $\gamma$-$\eta$ plane based on the definition of $\gamma$ and $\eta$ in Eq. \eqref{eq:18},
\begin{equation}
\gamma_{ob} = \frac{L_{ob}}{\varepsilon_b M_{ob}} =\frac{L_{ob} (a_{ob})^{5/2}}{\varepsilon_{b0} M_{ob}} \quad \textrm{and} \quad \eta_{ob} = \left(\frac{G L_{ob}}{c^5}\right)^{1/4},
\label{eq:54-1}
\end{equation}
where $a_{ob}=1/(1+z_{ob})$ is the corresponding scale factor. Figure \ref{fig:S34} presents five high-redshift SMBHs from recent studies (square symbols): J0313-1806 \citep{Wang:2021-A-Luminous-Quasar-at-Redshift}, CEERS 1019 \citep{Larson:2023-A-CEERS-Discovery-of-an-Accreting-Supermassive-Black-Hole}, UHZ-1 \citep{Goulding:2023-UNCOVER-The-Growth-of-the-First-Massive-Black-Holes}, Abell2744-QSO1 \citep{Furtak:2023-UNCOVERing-the-extended-strong-lensing-structures}, and GNz11 \citep{Maiolino:2023-A-small-and-vigorous-black-hole}. Table \ref{tab:2-2} lists these five observed SMBHs and the corresponding $\gamma_{ob}$ and $\eta_{ob}$. These SMBHs are expected to be near their peak luminosities, either in stage E1 with a rising luminosity or in stage E2 with a declining luminosity. According to the evolution model in $\gamma$-$\eta$ plane (Figs. \ref{fig:3} and \ref{fig:S34}), $\gamma\propto \eta^{-1}$ in stage E1 and $\gamma\propto \eta^{2}$ in stage E2 (Section \ref{sec:7}). 

\begin{figure}
\includegraphics*[width=\columnwidth]{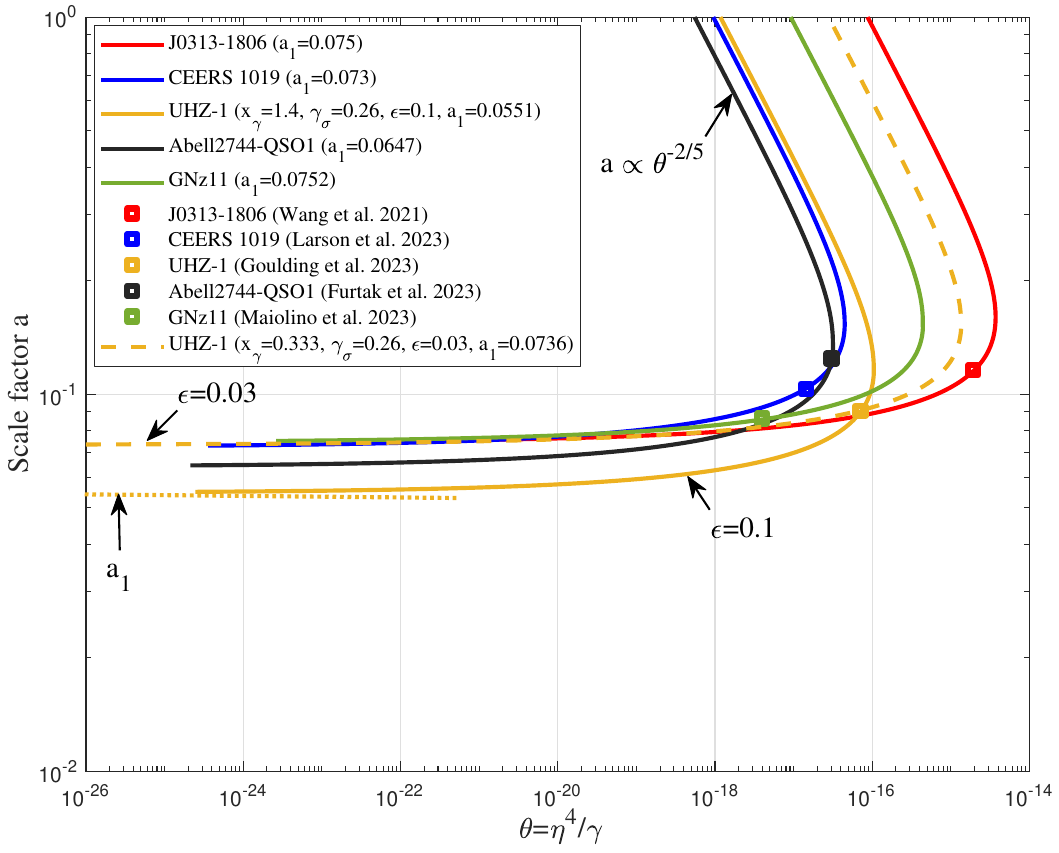}
\caption{The variation of scale factor $a$ as a function of parameter $\theta=\eta^4/\gamma$ for the evolution of SMBHs in Fig. \ref{fig:S34}. This is obtained by solving Eq. \eqref{eq:54-6} numerically with constants $\epsilon=0.1$, $\varepsilon_{b0}=10^{-4}$m$^2$/s$^3$, and $H_0=70$km/s/Mpc. The limiting scaling $\theta\propto a^{-5/2}$ is found at large $a$. Every SMBH has an initial scale factor $a_1$, as required for evolution in stage E1 (Eq. \eqref{eq:43}). Solutions with the right choice of $a_1$ should pass exactly those square symbols from observations. For UHZ-1, two different radiative efficiencies were used with $\epsilon=0.03$ and $\epsilon=0.1$. The smaller $\epsilon$ leads to larger $a_1$, i.e., a later formation of BH seed.} 
\label{fig:S35}
\end{figure}

Next, we need to construct a complete evolution via interpolating two stages (similar to the interpolation in mass space in Eq. \eqref{eq:8-2}),
\begin{equation}
\begin{split}
&\eta(\gamma) = c_1 \gamma^{-1}S(\gamma)+c_2\gamma^{1/2}(1-S(\gamma)),\\
&S(\gamma) = \frac{1}{1+x_{\gamma}\exp\left(-\frac{\log_{10}(\gamma)-\log_{10}(\gamma_c)}{\gamma_{\sigma}}\right)},
\end{split}
\label{eq:54-2}
\end{equation}
where $S(\gamma)$ is an interpolation function. The parameter $\gamma_c$ denotes the approximate location of the transition. The parameter $\gamma_{\sigma}$ controls the smoothness of the transition between two stages. The parameter $x_{\gamma}$ controls the degree of asymmetry of the transition. The smaller $\gamma_{\sigma}$, the sharper the transition. The two-stage evolution are $\eta = c_1\gamma^{-1}$ when $\gamma\gg \gamma_c$ and $\eta = c_2\gamma^{1/2}$ when $\gamma\ll \gamma_c$, where $c_1$ and $c_2$ are two coefficients. To determine the complete evolution of SMBH from a single observation, we take $\gamma_c \approx  \gamma_{ob}$. We would expect a small $x_{\gamma}$ for SMBHs in stage E1 and a large $x_{\gamma}$ for SMBHs in stage E2. Using the condition $\eta(\gamma_{ob})=\eta_{ob}$, two coefficients $c_1$ and $c_2$ is obtained
\begin{equation}
\begin{split}
&\eta_{ob} =c_1(\gamma_{ob})^{-1}\frac{1}{1+x_{\gamma}}+c_2(\gamma_{ob})^{1/2}\frac{x_{\gamma}}{1+x_{\gamma}},\\
& c_1 = x_{\gamma}\eta_{ob}\gamma_{ob} \quad \textrm{and} \quad c_2=x_{\gamma}^{-1}\eta_{ob}(\gamma_{ob})^{-1/2}.
\end{split}
\label{eq:54-3}
\end{equation}
Now the only two free parameters are $x_{\gamma}$ and $\gamma_{\sigma}$. Two parameters $x_{\gamma}\approx 1.08$ and $\gamma_{\sigma}\approx 0.26$ is obtained by fitting the evolution Eq. \eqref{eq:54-2} to the typical evolution in Fig. \ref{fig:3} (solid black curve). Assuming that the smoothness of the transition is universal, we should expect $\gamma_{\sigma}=0.26$ for different individual SMBHs. However, $x_{\gamma}$ cannot be uniquely determined and differs for different SMBHs. For $\gamma_{\sigma}=0.26$, taking the derivation $d\eta/d\gamma=0$ in Eq. \eqref{eq:54-2}, we may find the relation $x_{\gamma}\approx 2(\gamma_m/\gamma_{ob})^{3/2}$, where $\gamma_m$ is $\gamma$ at the turning point P1 with maximum $\eta$ or luminosity. Since $\gamma_{ob}$ is proportional to the Eddington ratio $\lambda_{ob}$, a rough estimate is $x_{\gamma} \approx 1/\lambda_{ob}$. 

The third step involves converting the evolution in the $\gamma$-$\eta$ plane to the evolution in redshift $z$ or scale factor $a$. From the definition of $\gamma$ and $\eta$ in Eq. \eqref{eq:18}, the luminosity and mass are
\begin{equation}
L_B = \frac{\eta^4c^5}{G} \quad \textrm{and} \quad M_{BH}= \frac{\eta^4}{\gamma} a^{5/2} \frac{c^5}{\varepsilon_{b0}G}.
\label{eq:54-4}
\end{equation}
Substituting this into the equation between $L_B$ and $\dot M_{BH}$ (Eq. \eqref{eq:22})
\begin{equation}
L_B=\dot M_{BH} \frac{\epsilon c^2}{1-\epsilon}=\frac{\partial M_{BH}}{\partial a} {H_0 a^{-1/2}} \frac{\epsilon c^2}{1-\epsilon},
\label{eq:54-5}
\end{equation}
we obtain a differential equation for $a$ as a function of $\eta$ and $\gamma$,
\begin{equation}
\frac{\partial \log a}{\partial \log \theta} = \frac{1}{\frac{\gamma}{a}\frac{1-\epsilon}{\epsilon} \frac{\varepsilon_{b0}}{H_0c^2}-\frac{5}{2}},
\label{eq:54-6}
\end{equation}
where the dimensionless variable $\theta=\eta^4/\gamma$ that is related to the velocity dispersion $\sigma_B$ on scale $r_B$ (Eq. \eqref{eq:30}). With a given evolution in the $\gamma$-$\eta$ plane, Eq. \eqref{eq:54-6} needs to be solved numerically to determine the scale factor $a$ at given $\gamma$ and $\eta$. The solution also depends on the radiative efficiency $\epsilon$ and the Hubble constant $H_0$. For $\gamma$ decreasing with $a$, the limiting scaling $\theta\propto a^{-5/2}$ or $L_B\propto a^{-5}$. 

Figure \ref{fig:S35} plots the numerical solutions of $a(\theta)$ for five SMBHs in Fig. \ref{fig:S34}. Square symbols represent the data from the observations. For each SMBH, there exists an initial scale factor $a_1$ representing the formation time of the initial seed BH, as required for the evolution in stage E1 (Eq. \eqref{eq:43}). The initial scale factor $a_1$ is calculated by try-and-error. Solutions with the right choice of $a_1$ should pass exactly the square symbols of the observations. The different choice of $\epsilon$ affects the initial scale factor $a_1$. Smaller $\epsilon$ (or radiatively inefficient accretion) is generally associated with super-Eddington accretion \citep{Massonneau:2023-How-the-super-Eddington-regime-regulates-black-hole-growth}. From our model, a smaller $\epsilon$ leads to the later formation of the BH seed or to a larger initial scale factor $a_1$ in Fig. \ref{fig:S35}, which means a shorter rapid growth period of stage E1.

\begin{figure}
\includegraphics*[width=\columnwidth]{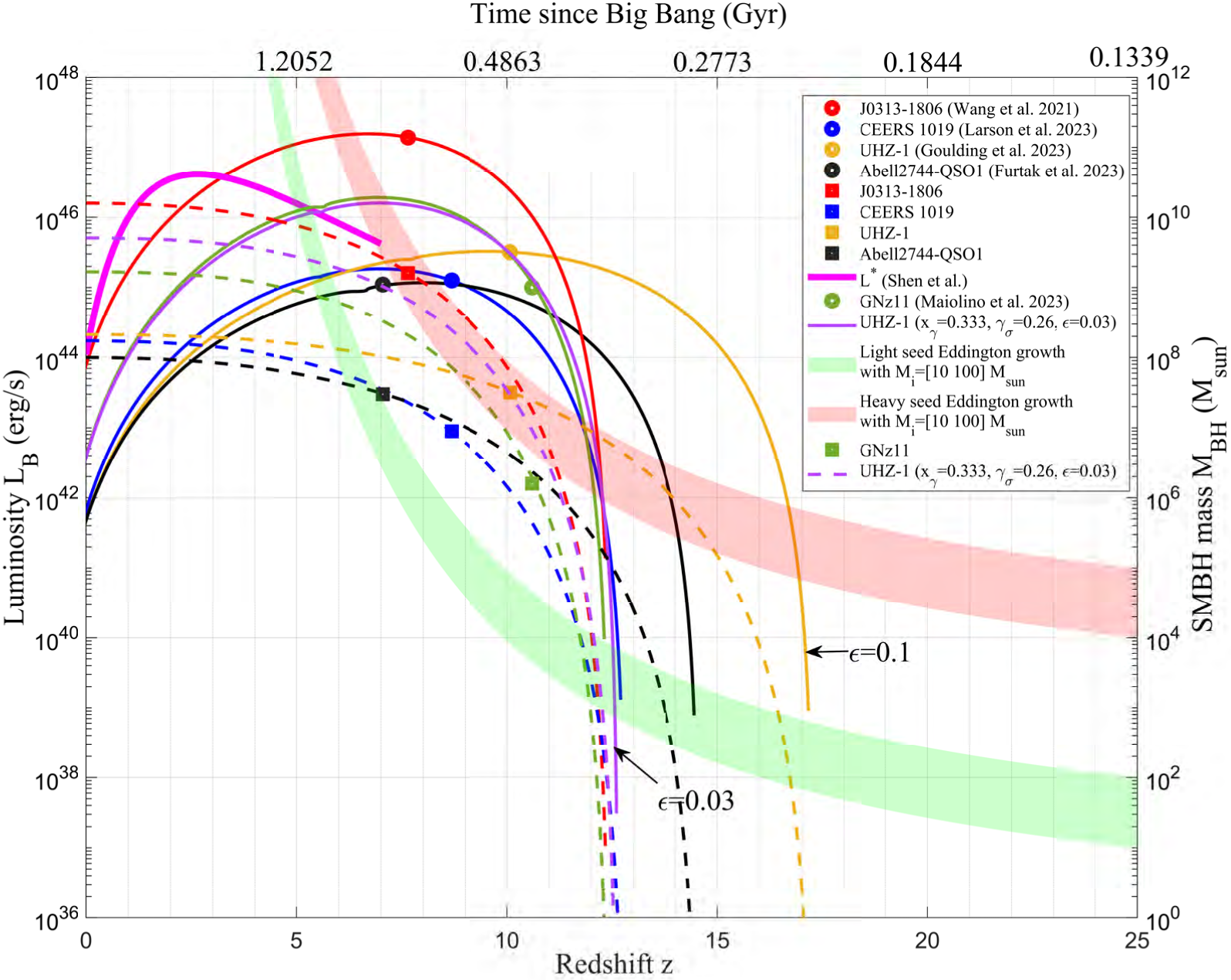}
\caption{The redshift evolution of the observed high-redshift SMBHs in Fig. \ref{fig:S34}. The circle and square symbols represent the observed luminosity $L_{ob}$ and mass $M_{ob}$. Solid and dashed lines show the calculated evolution of luminosity $L_B$ and BH mass $M_{BH}$, with $z=1/a-1$ calculated in Fig. \ref{fig:S35} by Eq. \eqref{eq:54-6}. This model does not require an initial BH seed with a certain mass (Eq. \eqref{eq:43}). For UHZ-1, two radiative efficiencies ($\epsilon=0.1$ and $\epsilon=0.03$) were used. The smaller $\epsilon$ leads to a later but faster growth of the BH mass. The fitted break luminosity $L^*$ (Fig. \ref{fig:S28} and Eq. \eqref{eq:7-3-2-3}) is also plotted. For comparison, the BH mass evolution for mass accretion at the constant Eddington limit (Eq. \eqref{eq:55}) is also provided for i) light seeds with an initial BH mass between $M_i=10M_{\odot}$ and $100M_{\odot}$ at $z_i=25$ and ii) heavy seeds with an initial BH mass between $M_i=10^4M_{\odot}$ and $10^5M_{\odot}$ (green and red shaded regions). The Eddington accretion from light seeds cannot explain the mass of J0313-1806 and UHZ-1. Either Super-Eddington growth or heavy BH seeds are required. This work demonstrates that rapid super-Eddington growth is possible in a short period of stage E1 when BHs are still small.} 
\label{fig:S33}
\end{figure}

With $a$ calculated from Eq. \eqref{eq:54-6}, Fig. \ref{fig:S33} shows the redshift evolution of both luminosity $L_B$ (solid lines) and BH mass $M_{BH}$ (dashed lines) of the observed SMBHs based on the transformation in Eq. \eqref{eq:54-4}. The observed luminosity $L_{ob}$ and the BH mass $M_{ob}$ are shown as circle and square symbols. The redshift evolution of luminosity $L_B$ and mass $M_{BH}$ passes through the observed $L_{ob}$ and $M_{ob}$. For UHZ-1, two radiative efficiencies ($\epsilon=0.1$ and $\epsilon=0.03$) were used. The smaller $\epsilon$ leads to a later but faster growth of BH. The evolution of the break luminosity $L^*$ is also plotted (solid pink line), with Abell2744-QSO1 below the break luminosity $L^*$ showing that it is in stage E2 with a declining luminosity. 

\begin{figure}
\includegraphics*[width=\columnwidth]{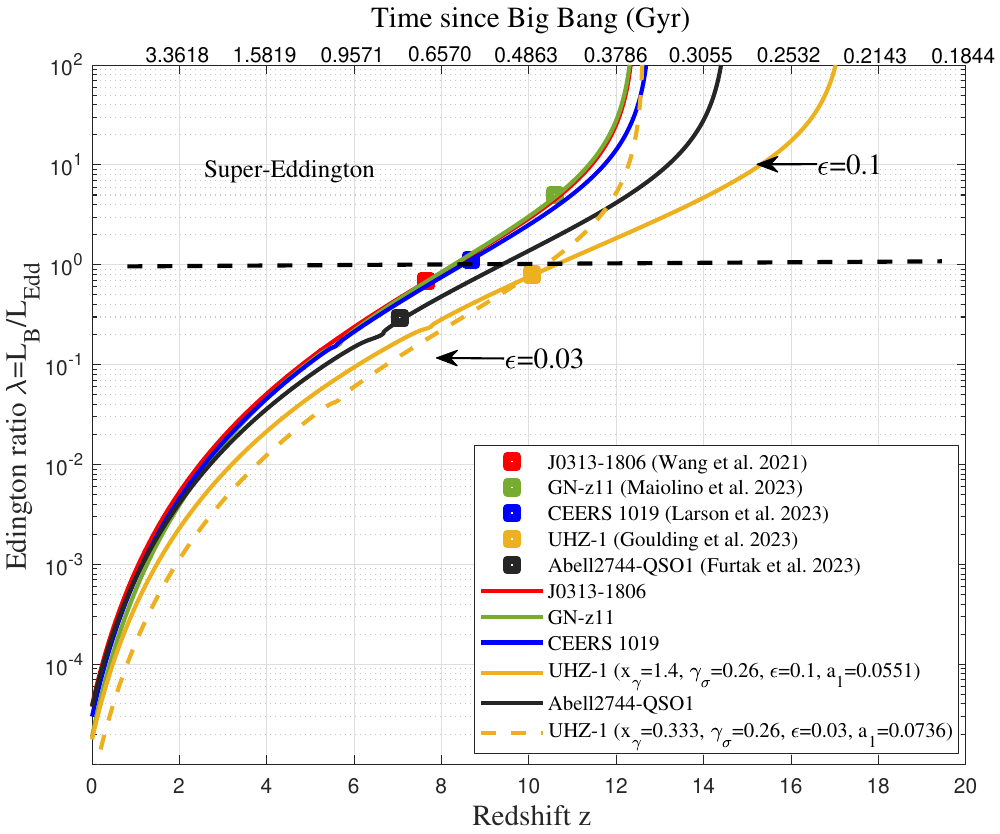}
\caption{The redshift evolution of the Eddington ratio of high-redshift SMBHs in Fig. \ref{fig:S34} to \ref{fig:S33}. Square symbols represent the observed Eddington ratio $\lambda_{ob}$, with $\lambda_{ob}\approx 5$ for GN-z11 (super-Eddington). Solid lines plot the calculated evolution of Eddington ratio $\lambda$ for every SMBH that passes exactly the observed $\lambda_{ob}$. All SMBHs have an initial super-Eddington growth with a sharply dropping Eddington ratio in a short period when they are still small ($M_{BH}<10^6M_{\odot}$). Two radiative efficiencies ($\epsilon=0.1$ and $\epsilon=0.03$) were used for UHZ-1. For $\epsilon=0.03$, the model shows a super-Eddington period of around 0.1Gyrs, compared to that of 0.26Gyrs for $\epsilon=0.1$. The smaller $\epsilon$ will significantly reduce the time of super-Eddington evolution in stage E1.} 
\label{fig:S36}
\end{figure}

For comparison, Fig. \ref{fig:S33} also presents the evolution of the BH mass that accretes at the Eddington limit with a different mass of initial BH seeds. For Eddington accretion, the BH mass grows exponentially
\begin{equation}
M_{BH} = M_i \exp \left(\frac{t-t_i}{t_{sal}}\right),
\label{eq:55}
\end{equation}
where $M_i$ is the mass of seed BH at an early epoch $z_i$. Here, $t$ and $t_i$ are the current and cosmic ages at $z_i$. The time scale $t_{sal} =4.5\times 10^8 \epsilon/(1-\epsilon)$ years is the Salpeter or e-folding time. Figure \ref{fig:S33} plots Eddington accretion with an initial mass of BH seed between $M_i=10M_{\odot}$ and $100M_{\odot}$ at $z_i=25$ (green shaded region for the light seed scenario). This is selected from the reasonable range of seed BH mass from a Pop III star remnant at $z\approx 20-30$. For that range of seed BH mass, it is challenging for the Eddington accretion to produce SMBH mass that matches the observed J0313-1806 and UHZ-1. A possible solution is the heavy seed scenario with an initial mass between $M_i=10^4M_{\odot}$ and $10^5M_{\odot}$ (red shaded region), or the formation of massive BH seeds through the direct collapse of pristine massive gas clouds \citep{Loeb:1994-Collapse-of-Primordial-Gas-Clouds,Lodato:2006-Supermassive-black-hole-formation-during}. However, at high redshift, if the average luminosity for all SMBHs of the same mass evolves at a fixed Eddington ratio (Model 3 in Eq. \eqref{eq:7-1-9-3} and Fig. \ref{fig:S22}), the resulting BH mass function will not likely match the observations and numerical solutions. We conclude that high-redshift SMBHs should not evolve at a fixed Eddington ratio. 

Figure \ref{fig:S36} presents the evolution of the Eddington ratio $\lambda$ of five observed high redshift SMBHs in Table \ref{tab:2-2}. The Eddington ratio is calculated from the evolution of the BH mass and luminosity in Fig. \ref{fig:S33}. Symbols represent the data from the observations. The solid lines show the model's prediction with $\epsilon=0.1$, while the dashed line shows the variation with $\epsilon=0.03$. Super-Eddington growth is indicated by $\lambda>1$. This is possible for a short period with larger $\varepsilon_b$ in the early Universe and radiatively inefficient accretion (small $\epsilon$) when the SMBHs are still relatively small. Due to the extremely rapid super-Eddington growth in a very short period with a relatively small BH mass, the effects of feedback on BH growth might be limited. Finally, the predicted path of evolution is independent of the mass of BH seeds. Both light and heavy seeds will follow the same path of evolution. However, this model allows for super-Eddington growth, so the light seed scenario can still be possible. 

\label{lastpage}


\section{SMBH and host galaxy data}
\onecolumn 
\begin{deluxetable}{cccccccccccccccccc}
\tablecaption{Samples of SMBHs and their host galaxies}
\tablecolumns{18}
\tablewidth{0pt}
\tabletypesize{\scriptsize}
\tablehead{
  \colhead{Galaxy} &
  \colhead{Type} &
  \colhead{$M_{BH}$} &
  \colhead{Ref.} &
  \colhead{$L_B$} &
  \colhead{Ref.} &
  \colhead{$\sigma_b$} &
  \colhead{Ref.} &
  \colhead{$\sigma_B$}  &
  \colhead{$\sigma_p$} &
  \colhead{$M_b$} &
  \colhead{Ref.} &
  \colhead{$r_b$}  &
  \colhead{$r_B$} &
  \colhead{$r_p$} &
  \colhead{$r_x$} &
  \colhead{$r_s$} &
  \colhead{$\varepsilon_b$} 
  \\
  \colhead{Name} &
  \colhead{} &
  \colhead{($M_\odot$)} &
  \colhead{} &
  \colhead{(erg/s)} &
  \colhead{} &
  \colhead{(km/s)} &
  \colhead{} &
  \colhead{(km/s)} &
  \colhead{(km/s)} &
  \colhead{($M_\odot$)} &
  \colhead{} &
  \colhead{(kpc)} &
  \colhead{(kpc)} &
  \colhead{(kpc)} &
  \colhead{(kpc)} &
  \colhead{(kpc)} &
  \colhead{($m^2/s^3$)} 
 }
 
\startdata
    Cygnus A & Seyfert & 2.7E+09 & 5     & 2.7E+45 & 2     & 270.0 & 1     & 67.1  & 38.2  & 1.6E+12 & 1     & 31.6  & 4.8E-01 & 9.0E-02 & 3.1E-03 & 2.6E-07 & 2.0E-05 \\
    A1836-BCG&         & 3.9E+09 & 1     & 3.3E+42 & 5     & 288.0 & 1     & 89.6  & 7.2   & 7.6E+11 & 1     & 13.2  & 4.0E-01 & 2.0E-04 & 3.1E-03 & 3.7E-07 & 5.9E-05 \\
    Circinus & Seyfert & 1.1E+06 & 5     & 4.8E+42 & 2     & 158.0 & 1     & 29.2  & 7.9   & 3.0E+09 & 1     & 0.2   & 1.1E-03 & 2.1E-05 & 3.7E-06 & 1.1E-10 & 7.4E-04 \\
    IC 1262  &         &         &       & 3.6E+43 & 63    & 232.5 & 63    &       & 13.0  & 9.3E+11 & 63    & 24.7  &       & 4.4E-03 &       &       & 1.6E-05 \\
    IC 1459  &         & 2.5E+09 & 5     & 1.3E+42 & 3a    & 340.0 & 1     & 99.4  & 5.6   & 6.6E+11 & 1     & 8.2   & 2.1E-01 & 3.8E-05 & 1.8E-03 & 2.4E-07 & 1.5E-04 \\
    IC 1633  &         &         &       & 8.3E+42 & 63    & 356.6 & 63    &       & 9.0   & 2.4E+12 & 63    & 27.0  &       & 4.4E-04 &       &       & 5.4E-05 \\
    IC 2560  & Seyfert & 5.0E+06 & 5     & 1.2E+42 & 5     & 137.0 & 1     & 22.7  & 5.6   & 2.3E+10 & 1     & 1.8   & 8.0E-03 & 1.2E-04 & 2.3E-05 & 4.8E-10 & 4.7E-05 \\
    IC 4296  &         & 1.3E+09 & 5     & 1.6E+42 & 3a    & 322.0 & 1     & 69.3  & 6.0   & 1.6E+12 & 1     & 22.2  & 2.2E-01 & 1.4E-04 & 1.4E-03 & 1.2E-07 & 4.9E-05 \\
    IC 5267  &         &         &       & 6.2E+40 & 63    & 167.7 & 63    &       & 2.6   & 1.5E+11 & 63    & 7.6   &       & 3.0E-05 &       &       & 2.0E-05 \\
    IC 5358  &         &         &       & 1.1E+44 & 63    & 214.2 & 63    &       & 17.2  & 1.6E+12 & 63    & 50.2  &       & 2.6E-02 &       &       & 6.3E-06 \\
    Sgr A*   &         & 4.1E+06 & 1     & 1.9E+36 & 3a    & 105.0 & 1     & 19.3  & 0.2   & 1.1E+10 & 1     & 1.4   & 9.0E-03 & 9.6E-09 & 2.2E-05 & 3.9E-10 & 2.6E-05 \\
    NGC193   &         & 2.5E+08 & 59    & 1.6E+41 & 59    & 187.0 & 59    & 70.5  & 3.4   & 1.9E+10 & 59    & 0.8   & 4.1E-02 & 4.4E-06 & 2.7E-04 & 2.4E-08 & 2.7E-04 \\
    NGC 205  &         & 3.8E+04 & 5     & 4.8E+35 & 58    & 35.0  & 13    & 5.1   & 0.1   & 3.3E+08 & 13    & 0.4   & 1.2E-03 & 2.5E-08 & 1.1E-06 & 3.7E-12 & 3.6E-06 \\
    NGC 221  &         & 2.5E+06 & 5     & 1.5E+37 & 3a    & 75.0  & 1     & 21.0  & 0.3   & 8.0E+08 & 1     & 0.2   & 4.5E-03 & 1.7E-08 & 1.2E-05 & 2.4E-10 & 6.7E-05 \\
    NGC 224  &         & 1.4E+08 & 5     & 1.4E+37 & 3a    & 160.0 & 1     & 45.4  & 0.3   & 4.4E+10 & 1     & 2.5   & 5.7E-02 & 2.1E-08 & 2.7E-04 & 1.4E-08 & 5.4E-05 \\
    NGC 315  & BCG     & 1.7E+09 & 3     & 7.6E+42 & 3a    & 341.0 & 11    & 81.6  & 8.8   & 1.2E+12 & 11    & 14.9  & 2.0E-01 & 2.6E-04 & 1.5E-03 & 1.6E-07 & 8.6E-05 \\
    NGC 326  &         &         &       & 1.3E+42 & 63    & 231.9 & 63    &       & 5.7   & 1.4E+12 & 63    & 38.3  &       & 5.6E-04 &       &       & 1.1E-05 \\
    NGC 383  &         & 5.8E+08 & 59    & 9.5E+41 & 59    & 240.0 & 59    & 55.4  & 5.2   & 5.0E+11 & 59    & 12.5  & 1.5E-01 & 1.3E-04 & 8.5E-04 & 5.5E-08 & 3.6E-05 \\
    NGC 499  &         &         &       & 8.9E+42 & 63    & 253.3 & 63    &       & 9.2   & 5.1E+11 & 63    & 11.5  &       & 5.4E-04 &       &       & 4.6E-05 \\
    NGC 507  & BCG     & 1.6E+09 & 3     & 7.3E+41 & 3a    & 331.0 & 12    & 78.1  & 4.9   & 1.3E+12 & 12    & 16.6  & 2.2E-01 & 5.4E-05 & 1.6E-03 & 1.6E-07 & 7.1E-05 \\
    NGC 524  &         & 8.7E+08 & 5     & 1.8E+40 & 5     & 235.0 & 1     & 67.1  & 1.9   & 2.6E+11 & 1     & 6.8   & 1.6E-01 & 3.8E-06 & 1.0E-03 & 8.3E-08 & 6.2E-05 \\
    NGC 533  &         &         &       & 1.3E+43 & 63    & 271.2 & 63    &       & 10.1  & 1.1E+12 & 63    & 22.4  &       & 1.2E-03 &       &       & 2.9E-05 \\
    NGC 541  &         & 3.9E+08 & 59    & 4.3E+41 & 59    & 191.0 & 59    & 48.5  & 4.3   & 2.1E+11 & 59    & 8.3   & 1.4E-01 & 9.4E-05 & 6.8E-04 & 3.7E-08 & 2.7E-05 \\
    NGC 708  &         &         &       & 3.0E+43 & 63    & 222.2 & 63    &       & 12.5  & 7.6E+11 & 63    & 22.0  &       & 3.9E-03 &       &       & 1.6E-05 \\
    NGC 720  &         &         &       & 6.5E+41 & 63    & 235.6 & 63    &       & 4.8   & 2.5E+11 & 63    & 6.4   &       & 5.3E-05 &       &       & 6.6E-05 \\
    NGC 741  &         &         &       & 5.2E+42 & 63    & 286.0 & 63    &       & 8.0   & 1.0E+12 & 63    & 17.6  &       & 3.9E-04 &       &       & 4.3E-05 \\
    NGC 821  &         & 1.7E+08 & 5     & 4.4E+39 & 2     & 209.0 & 1     & 49.2  & 1.4   & 1.3E+11 & 1     & 4.3   & 5.6E-02 & 1.2E-06 & 2.8E-04 & 1.6E-08 & 6.9E-05 \\
    NGC 1023 &         & 4.1E+07 & 5     & 1.0E+40 & 2     & 205.0 & 1     & 41.5  & 1.7   & 6.9E+10 & 1     & 2.4   & 2.0E-02 & 1.3E-06 & 8.7E-05 & 4.0E-09 & 1.2E-04 \\
    NGC 1052 & BCG     & 1.7E+08 & 59    & 3.5E+40 & 59    & 191.0 & 59    & 53.8  & 2.3   & 5.6E+10 & 59    & 2.2   & 4.9E-02 & 3.8E-06 & 2.7E-04 & 1.7E-08 & 1.0E-04 \\
    NGC 1068 & Seyfert & 8.4E+06 & 5     & 2.5E+44 & 19a   & 151.0 & 1     & 30.2  & 21.2  & 1.5E+10 & 1     & 0.9   & 7.6E-03 & 2.6E-03 & 2.6E-05 & 8.1E-10 & 1.2E-04 \\
    NGC 1194 & Seyfert & 7.1E+07 & 5     & 5.5E+44 & 19a   & 148.0 & 1     & 42.8  & 25.7  & 2.0E+10 & 1     & 1.3   & 3.2E-02 & 6.9E-03 & 1.4E-04 & 6.8E-09 & 8.0E-05 \\
    NGC 1266 &         &         &       & 1.1E+41 & 63    & 94.4  & 63    &       & 3.0   & 1.6E+10 & 63    & 2.6   &         & 8.6E-05 &         &         & 1.1E-05 \\
    NGC 1277 &         & 1.7E+10 & 5     & 2.1E+41 & 55    & 403.0 & 14    & 224.5 & 3.6   & 1.8E+11 & 14    & 1.6   & 2.8E-01 & 1.1E-06 & 4.4E-03 & 1.6E-06 & 1.3E-03 \\
    NGC 1300 &         & 7.6E+07 & 5     & 8.2E+40 & 6     & 218.0 & 1     & 63.2  & 2.8   & 2.1E+10 & 1     & 0.6   & 1.5E-02 & 1.4E-06 & 9.4E-05 & 7.3E-09 & 5.3E-04 \\
    NGC 1320 & Seyfert & 6.0E+06 & 21    & 1.4E+44 & 19a   & 250.0 & 21    & 45.2  & 18.4  & 1.8E+10 & 21    & 0.4   & 2.4E-03 & 1.6E-04 & 1.1E-05 & 5.8E-10 & 1.2E-03 \\
    NGC 1316 &         & 1.7E+08 & 5     & 1.8E+40 & 3a    & 226.0 & 1     & 57.1  & 1.9   & 9.3E+10 & 1     & 2.6   & 4.2E-02 & 1.7E-06 & 2.4E-04 & 1.6E-08 & 1.4E-04 \\
    NGC 1332 & BCG     & 1.5E+09 & 5     & 1.2E+40 & 3a    & 327.7 & 15    & 135.7 & 1.8   & 6.9E+10 & 15    & 0.9   & 6.5E-02 & 1.4E-07 & 7.1E-04 & 1.4E-07 & 1.2E-03 \\
    NGC 1374 &         & 5.9E+08 & 59    & 6.3E+39 & 59    & 167.0 & 59    & 66.7  & 1.5   & 3.3E+10 & 59    & 1.7   & 1.1E-01 & 1.2E-06 & 6.9E-04 & 5.7E-08 & 8.9E-05 \\
    NGC 1386 & Seyfert & 1.2E+06 & 19    & 3.2E+42 & 19a   & 95.0  & 22    & 16.0  & 7.1   & 5.0E+09 & 22    & 0.8   & 3.8E-03 & 3.3E-04 & 8.3E-06 & 1.2E-10 & 3.5E-05 \\
    NGC 1399 & BCG     & 5.0E+08 & 5     & 9.4E+39 & 3a    & 337.0 & 1     & 88.4  & 1.7   & 2.3E+11 & 1     & 2.9   & 5.3E-02 & 3.5E-07 & 4.1E-04 & 4.8E-08 & 4.2E-04 \\
    NGC 1400 &         & 4.7E+09 & 5     & 7.4E+40 & 3a    & 279.0 & 16    & 124.9 & 2.8   & 1.5E+11 & 16    & 2.7   & 2.4E-01 & 2.7E-06 & 2.5E-03 & 4.5E-07 & 2.6E-04 \\
    NGC 1404 &         &         &       & 4.2E+42 & 63    & 228.1 & 63    &       & 7.6   & 3.3E+11 & 63    & 9.0   &         & 3.4E-04 &         &         & 4.3E-05 \\
    NGC 1407 & BCG     & 4.7E+09 & 5     & 7.4E+40 & 3a    & 305.0 & 16    & 108.2 & 2.8   & 4.7E+11 & 16    & 7.3   & 3.3E-01 & 5.5E-06 & 3.0E-03 & 4.5E-07 & 1.3E-04 \\
    NGC 1482 &         &         &       & 3.3E+41 & 63    & 108.5 & 63    &       & 4.0   & 1.1E+11 & 63    & 13.8  &         & 7.0E-04 &         &         & 3.0E-06 \\
    NGC 1550 & BCG     & 3.9E+09 & 5     & 4.5E+39 & 5     & 270.0 & 17    & 108.1 & 1.4   & 2.1E+11 & 17    & 4.2   & 2.7E-01 & 5.6E-07 & 2.5E-03 & 3.7E-07 & 1.5E-04 \\
    NGC 1600 &         &         &       & 2.7E+42 & 63    & 331.4 & 63    &       & 6.8   & 1.5E+12 & 63    & 19.4  &         & 1.7E-04 &         &         & 6.1E-05 \\
    NGC 1700 &         &         &       & 1.2E+42 & 63    & 233.1 & 63    &       & 5.6   & 3.4E+11 & 63    & 9.0   &         & 1.3E-04 &         &         & 4.6E-05 \\
    NGC 2273 & Seyfert & 7.5E+06 & 19    & 2.0E+44 & 19    & 170.0 & 21    & 36.3  & 20.1  & 9.5E+09 & 21    & 0.5   & 4.7E-03 & 7.8E-04 & 1.9E-05 & 7.2E-10 & 3.3E-04 \\
    NGC 2434 &         &         &       & 1.2E+41 & 63    & 183.7 & 63    &       & 3.2   & 1.2E+11 & 63    & 5.0   &         & 2.5E-05 &         &         & 4.0E-05 \\
    NGC 2549 &         & 1.5E+07 & 5     & 4.3E+40 & 51    & 145.0 & 1     & 31.2  & 2.4   & 1.8E+10 & 1     & 1.2   & 1.2E-02 & 5.7E-06 & 4.4E-05 & 1.4E-09 & 8.0E-05 \\
    NGC 2748 &         & 4.4E+07 & 5     & 1.1E+39 & 6     & 115.0 & 1     & 31.3  & 1.0   & 1.7E+10 & 1     & 1.9   & 3.7E-02 & 1.1E-06 & 1.3E-04 & 4.3E-09 & 2.7E-05 \\
    NGC 2768 &         &         &       & 1.2E+41 & 63    & 184.2 & 63    &       & 3.1   & 1.5E+11 & 63    & 6.2   &         & 3.0E-05 &         &         & 3.3E-05 \\
    NGC 2778 &         & 1.5E+07 & 5     & 2.2E+38 & 6     & 175.0 & 1     & 41.5  & 0.6   & 1.1E+10 & 1     & 0.5   & 6.9E-03 & 2.6E-08 & 3.1E-05 & 1.4E-09 & 3.4E-04 \\
    NGC 2787 & LINER   & 4.1E+07 & 5     & 7.9E+39 & 2     & 189.0 & 1     & 45.4  & 1.6   & 2.9E+10 & 1     & 1.2   & 1.6E-02 & 6.9E-07 & 7.7E-05 & 3.9E-09 & 1.9E-04 \\
    NGC 2892 &         & 2.7E+08 & 59    & 1.1E+42 & 59    & 295.0 & 59    & 60.8  & 5.5   & 4.1E+11 & 59    & 6.8   & 6.0E-02 & 4.4E-05 & 3.5E-04 & 2.6E-08 & 1.2E-04 \\
    NGC 2960 & Sayfert & 1.1E+07 & 5     & 3.2E+40 & 52a   & 166.0 & 1     & 34.4  & 2.2   & 1.6E+10 & 1     & 0.8   & 7.5E-03 & 2.1E-06 & 2.9E-05 & 1.0E-09 & 1.8E-04 \\
    NGC 2974 & Seyfert & 1.7E+08 & 7     & 2.0E+42 & 48    & 233.0 & 1     & 55.2  & 6.3   & 1.3E+11 & 1     & 3.5   & 4.6E-02 & 6.9E-05 & 2.5E-04 & 1.6E-08 & 1.2E-04 \\
    NGC 3031 & Seyfert & 6.5E+07 & 5     & 1.1E+42 & 2     & 143.0 & 1     & 46.6  & 5.4   & 1.0E+10 & 1     & 0.7   & 2.4E-02 & 3.9E-05 & 1.2E-04 & 6.2E-09 & 1.3E-04 \\
    NGC 3079 & Seyfert & 2.4E+06 & 19    & 4.1E+43 & 19a   & 146.0 & 1     & 22.1  & 13.4  & 1.7E+10 & 1     & 1.1   & 4.0E-03 & 8.9E-04 & 1.1E-05 & 2.3E-10 & 8.8E-05 \\
    NGC 3091 & BCG     & 3.7E+09 & 5     & 2.6E+42 & 59    & 297.0 & 17    & 103.7 & 6.8   & 4.1E+11 & 17    & 6.7   & 2.8E-01 & 7.8E-05 & 2.5E-03 & 3.6E-07 & 1.3E-04 \\
    NGC 3115 &         & 9.0E+08 & 5     & 8.2E+39 & 3a    & 230.0 & 1     & 77.2  & 1.6   & 1.2E+11 & 1     & 3.3   & 1.2E-01 & 1.1E-06 & 8.7E-04 & 8.6E-08 & 1.2E-04 \\
    NGC 3227 & Seyfert & 2.1E+07 & 5     & 5.6E+42 & 2     & 133.0 & 1     & 44.0  & 8.2   & 3.0E+09 & 1     & 0.2   & 8.9E-03 & 5.7E-05 & 4.1E-05 & 2.0E-09 & 3.1E-04 \\
    NGC 3245 &         & 2.4E+08 & 5     & 3.0E+40 & 2     & 205.0 & 1     & 59.1  & 2.2   & 6.8E+10 & 1     & 2.3   & 5.6E-02 & 2.9E-06 & 3.2E-04 & 2.3E-08 & 1.2E-04 \\
    NGC 3310 &         & 4.2E+07 & 5     & 2.1E+41 & 5     & 84.0  & 46    & 41.6  & 3.6   & 8.0E+08 & 45    & 0.2   & 2.0E-02 & 1.3E-05 & 8.8E-05 & 4.0E-09 & 1.2E-04 \\
    NGC 3351 &         & 8.6E+06 & 5     & 5.5E+39 & 57    & 57.0  & 18    & 17.5  & 1.4   & 1.8E+09 & 18    & 0.8   & 2.3E-02 & 1.3E-05 & 5.4E-05 & 8.3E-10 & 7.5E-06 \\
    NGC 3368 &         & 7.7E+06 & 5     & 1.2E+38 & 5     & 193.0 & 21    & 44.8  & 0.6   & 6.5E+09 & 21    & 0.2   & 3.1E-03 & 6.1E-09 & 1.5E-05 & 7.4E-10 & 9.3E-04 \\
    NGC 3377 &         & 1.8E+08 & 5     & 1.6E+39 & 2     & 145.0 & 1     & 46.1  & 1.1   & 3.1E+10 & 1     & 2.1   & 6.8E-02 & 8.3E-07 & 3.3E-04 & 1.7E-08 & 4.6E-05 \\
    NGC 3379 & LINER   & 1.8E+08 & 5     & 2.3E+39 & 2     & 206.0 & 1     & 56.0  & 1.2   & 6.8E+10 & 1     & 2.3   & 4.6E-02 & 4.2E-07 & 2.6E-04 & 1.7E-08 & 1.2E-04 \\
    NGC 3384 &         & 1.1E+07 & 5     & 4.4E+39 & 6     & 143.0 & 1     & 28.4  & 1.4   & 2.0E+10 & 1     & 1.4   & 1.1E-02 & 1.2E-06 & 3.7E-05 & 1.0E-09 & 6.7E-05 \\
    NGC 3393 & Seyfert & 1.6E+07 & 5     & 2.7E+42 & 49    & 184.0 & 1     & 28.5  & 6.8   & 1.0E+11 & 1     & 4.3   & 1.6E-02 & 2.2E-04 & 5.3E-05 & 1.5E-09 & 4.7E-05 \\
    NGC 3414 &         & 2.5E+08 & 7     & 1.3E+41 & 50    & 205.0 & 1     & 63.5  & 3.2   & 5.0E+10 & 1     & 1.7   & 5.1E-02 & 6.3E-06 & 3.1E-04 & 2.4E-08 & 1.6E-04 \\
    NGC 3489 & Seyfert & 5.9E+06 & 5     & 1.0E+40 & 5     & 129.0 & 37    & 20.0  & 1.7   & 3.8E+10 & 37    & 3.3   & 1.2E-02 & 7.3E-06 & 3.1E-05 & 5.7E-10 & 2.1E-05 \\
    NGC 3557 &         &         &       & 7.8E+41 & 63    & 264.1 & 63    &       & 5.0   & 5.8E+11 & 63    & 12.0  &         & 8.1E-05 &         &         & 5.0E-05 \\
    NGC 3585 &         & 3.3E+08 & 5     & 1.5E+40 & 2     & 213.0 & 1     & 53.9  & 1.9   & 1.8E+11 & 1     & 5.7   & 9.3E-02 & 3.8E-06 & 5.0E-04 & 3.2E-08 & 5.5E-05 \\
    NGC 3607 &         & 1.4E+08 & 5     & 6.3E+39 & 2     & 229.0 & 1     & 49.8  & 1.5   & 1.6E+11 & 1     & 4.4   & 4.5E-02 & 1.2E-06 & 2.3E-04 & 1.3E-08 &8.9E-05\\
    NGC 3608 & LINER   & 4.7E+08 & 5     & 9.7E+39 & 2     & 182.0 & 1     & 55.9  & 1.7   & 9.7E+10 & 1     & 4.2   & 1.2E-01 & 3.3E-06 & 6.8E-04 & 4.5E-08 &4.6E-05\\
    NGC 3665 &         & 5.8E+08 & 59    & 1.7E+41 & 59    & 219.0 & 59    & 60.1  & 3.4   & 2.1E+11 & 59    & 6.3   & 1.3E-01 & 2.4E-05 & 7.6E-04 & 5.5E-08 &5.4E-05\\
    NGC 3842 & BCG     & 9.1E+09 & 5     & 1.3E+42 & 59    & 270.0 & 23    & 86.3  & 5.7   & 1.6E+12 & 23    & 30.6  & 1.0E+00 & 2.8E-04 & 7.7E-03 & 8.7E-07 &2.1E-05\\
    NGC 3862 &         & 2.6E+08 & 59    & 5.8E+41 & 59    & 209.0 & 59    & 43.8  & 4.6   & 3.6E+11 & 59    & 11.9  & 1.1E-01 & 1.3E-04 & 5.1E-04 & 2.5E-08 &2.5E-05\\
    NGC 3923 &         &         &       & 7.9E+41 & 63    & 246.6 & 63    &       & 5.0   & 4.2E+11 & 63    & 10.0  &         & 8.5E-05 &         &         &4.8E-05\\
    NGC 3945 &         & 8.8E+06 & 5     & 5.6E+40 & 5     & 182.0 & 24    & 39.8  & 2.6   & 1.0E+10 & 24    & 0.4   & 4.5E-03 & 1.2E-06 & 2.0E-05 & 8.5E-10 &4.5E-04\\
    NGC 3955 &         &         &       & 3.3E+40 & 63    & 94.4  & 63    &       & 2.3   & 1.4E+10 & 63    & 2.2   &         & 3.0E-05 &         &         &1.2E-05\\
    NGC 3982 & Seyfert & 8.0E+07 & 5     & 9.5E+41 & 60    & 78.0  & 44    & 26.0  & 5.2   & 1.1E+10 & 45    & 2.6   & 9.7E-02 & 7.9E-04 & 3.0E-04 & 7.7E-09 &5.9E-06\\
    NGC 3998 & Seyfert & 8.5E+08 & 5     & 4.4E+42 & 2     & 305.0 & 1     & 118.2 & 7.7   & 5.5E+10 & 1     & 0.9   & 5.0E-02 & 1.4E-05 & 4.8E-04 & 8.1E-08 &1.1E-03\\
    NGC 4026 &         & 1.8E+08 & 5     & 5.4E+39 & 2     & 180.0 & 1     & 51.8  & 1.4   & 5.2E+10 & 1     & 2.3   & 5.5E-02 & 1.2E-06 & 2.9E-04 & 1.7E-08 &8.2E-05\\
    NGC 4036 & LINER   & 7.7E+07 & 59    & 3.0E+40 & 59    & 182.0 & 59    & 42.6  & 2.2   & 6.2E+10 & 59    & 2.7   & 3.5E-02 & 4.8E-06 & 1.6E-04 & 7.4E-09 &7.2E-05\\
    NGC 4041 & Seyfert & 6.4E+06 & 5     & 1.7E+40 & 51    & 95.0  & 25    & 25.7  & 1.9   & 2.5E+09 & 25    & 0.4   & 7.9E-03 & 3.3E-06 & 2.5E-05 & 6.1E-10 &7.0E-05\\
    NGC 4073 &         &         &       & 3.6E+43 & 63    & 267.0 & 63    &       & 13.0  & 1.0E+12 & 63    & 20.8  &         & 2.4E-03 &         &         &3.0E-05\\
    NGC 4104 &         &         &       & 1.2E+43 & 63    & 291.0 & 63    &       & 9.8   & 1.6E+12 & 63    & 26.4  &         & 1.0E-03 &         &         &3.0E-05\\
    NGC 4125 &         &         &       & 2.5E+41 & 63    & 238.2 & 63    &       & 3.8   & 2.5E+11 & 63    & 6.4   &         & 2.5E-05 &         &         &6.8E-05\\
    NGC 4143 & LINER   & 1.4E+08 & 5     & 1.4E+41 & 5     & 271.0 & 27    & 66.0  & 3.3   & 9.3E+10 & 29    & 1.8   & 2.6E-02 & 3.2E-06 & 1.7E-04 & 1.3E-08 &3.5E-04\\
    NGC 4151 & Seyfert & 4.5E+07 & 2     & 7.7E+42 & 2     & 93.0  & 1     & 28.1  & 8.9   & 1.0E+10 & 1     & 1.7   & 4.6E-02 & 1.4E-03 & 1.5E-04 & 4.3E-09 &1.6E-05\\
    NGC 4203 & LINER   & 3.8E+07 & 5     & 4.0E+41 & 5     & 110.0 & 27    & 28.8  & 4.2   & 1.8E+10 & 29    & 2.1   & 3.7E-02 & 1.2E-04 & 1.3E-04 & 3.7E-09 &2.1E-05\\
    NGC 4258 & Seyfert & 3.8E+07 & 5     & 1.1E+42 & 2     & 115.0 & 1     & 33.0  & 5.4   & 1.1E+10 & 1     & 1.2   & 2.8E-02 & 1.2E-04 & 1.1E-04 & 3.6E-09 &4.1E-05\\
    NGC 4261 & LINER   & 5.3E+08 & 5     & 5.2E+41 & 3a    & 315.0 & 1     & 76.3  & 4.5   & 3.6E+11 & 1     & 5.2   & 7.4E-02 & 1.5E-05 & 5.2E-04 & 5.1E-08 &1.9E-04\\
    NGC 4278 & LINER   & 6.6E+08 & 3     & 4.7E+41 & 3a    & 232.5 & 31    & 77.7  & 4.4   & 9.0E+10 & 29    & 2.4   & 9.0E-02 & 1.6E-05 & 6.4E-04 & 6.3E-08 &1.7E-04\\
    NGC 4291 & BCG     & 9.8E+08 & 5     & 1.6E+40 & 6     & 242.0 & 1     & 81.3  & 1.9   & 1.3E+11 & 1     & 3.2   & 1.2E-01 & 1.5E-06 & 8.9E-04 & 9.4E-08 &1.4E-04\\
    NGC 4303 & Seyfert & 4.5E+06 & 2     & 1.2E+40 & 2     & 84.0  & 1     & 23.1  & 1.8   & 1.6E+09 & 1     & 0.3   & 6.8E-03 & 3.0E-06 & 2.0E-05 & 4.3E-10 &5.9E-05\\
    NGC 4321 &         & 2.7E+07 & 5     & 2.7E+39 & 5     & 69.0  & 18    & 21.1  & 1.2   & 5.8E+09 & 18    & 1.8   & 5.0E-02 & 9.4E-06 & 1.3E-04 & 2.6E-09 &6.1E-06\\
    NGC 4342 &         & 4.5E+08 & 5     & 2.1E+40 & 2     & 225.0 & 1     & 104.3 & 2.0   & 1.2E+10 & 1     & 0.3   & 3.4E-02 & 2.5E-07 & 3.0E-04 & 4.4E-08 &1.1E-03\\
    NGC 4365 &         &         &       & 6.5E+40 & 63    & 246.9 & 63    &       & 2.7   & 2.4E+11 & 63    & 5.6   &         & 7.2E-06 &         &         &8.7E-05\\
    NGC 4374 & Seyfert & 9.3E+08 & 5     & 6.1E+40 & 3a    & 296.0 & 1     & 80.2  & 2.6   & 3.6E+11 & 1     & 5.9   & 1.2E-01 & 4.2E-06 & 8.6E-04 & 8.9E-08 &1.4E-04\\
    NGC 4382 &         & 1.3E+07 & 5     & 5.7E+39 & 5     & 182.0 & 38    & 21.4  & 1.5   & 3.3E+11 & 38    & 14.4  & 2.3E-02 & 7.4E-06 & 6.3E-05 & 1.2E-09 &1.4E-05\\
    NGC 4388 & Seyfert & 7.3E+06 & 5     & 4.2E+42 & 5     & 107.0 & 1     & 24.8  & 7.6   & 6.2E+09 & 1     & 0.8   & 9.7E-03 & 2.8E-04 & 2.9E-05 & 7.0E-10 &5.1E-05\\
    NGC 4406 &         &         &       & 1.6E+42 & 63    & 230.0 & 63    &       & 6.0   & 2.6E+11 & 63    & 7.0   &         & 1.2E-04 &         &         &5.6E-05\\
    NGC 4435 &         & 8.0E+06 & 5     & 6.2E+40 & 61    & 156.0 & 26    & 30.7  & 2.6   & 1.5E+10 & 26    & 0.9   & 6.9E-03 & 4.4E-06 & 2.5E-05 & 7.7E-10 &1.4E-04\\
    NGC 4438 & LINER   & 5.6E+07 & 3     & 1.2E+40 & 3a    & 142.0 & 39    & 56.2  & 1.8   & 3.3E+09 & 39    & 0.2   & 1.5E-02 & 4.6E-07 & 8.1E-05 & 5.4E-09 &3.9E-04\\
    NGC 4457 &         &         &       & 8.5E+40 & 63    & 113.3 & 63    &       & 2.9   & 3.2E+10 & 63    & 3.6   &         & 5.9E-05 &         &         &1.3E-05\\
    NGC 4459 & HII     & 7.0E+07 & 5     & 1.3E+40 & 6     & 167.0 & 1     & 36.5  & 1.8   & 7.9E+10 & 1     & 4.1   & 4.3E-02 & 5.0E-06 & 1.7E-04 & 6.7E-09 &3.7E-05\\
    NGC 4472 & Seyfert & 2.5E+09 & 3     & 3.2E+40 & 62    & 250.0 & 40    & 83.7  & 2.2   & 3.4E+11 & 29    & 7.8   & 2.9E-01 & 5.6E-06 & 2.2E-03 & 2.4E-07 &6.5E-05\\
    NGC 4473 &         & 9.0E+07 & 5     & 5.0E+39 & 2     & 190.0 & 1     & 42.4  & 1.4   & 9.2E+10 & 1     & 3.7   & 4.1E-02 & 1.5E-06 & 1.8E-04 & 8.6E-09 &6.1E-05\\
    NGC 4477 & Seyfert & 8.4E+07 & 5     & 1.6E+41 & 63    & 144.8 & 31    & 36.8  & 3.4   & 4.5E+10 & 29    & 3.1   & 5.1E-02 & 3.8E-05 & 2.1E-04 & 8.1E-09 &3.2E-05\\
    NGC 4486 &         & 6.2E+09 & 5     & 9.3E+41 & 3a    & 375.0 & 1     & 134.0 & 5.2   & 6.0E+11 & 1     & 6.1   & 2.8E-01 & 1.7E-05 & 3.0E-03 & 5.9E-07 &2.8E-04\\
    NGC 4486A&         & 1.4E+07 & 5     & 2.2E+39 & 6     & 111.0 & 1     & 32.0  & 1.2   & 4.1E+09 & 1     & 0.5   & 1.2E-02 & 5.4E-07 & 4.2E-05 & 1.4E-09 &9.2E-05\\
    NGC 4486B&         & 6.0E+08 & 5     & 1.5E+39 & 5     & 169.8 & 47    & 97.0  & 1.0   & 5.6E+09 & 47    & 0.3   & 5.2E-02 & 6.5E-08 & 4.4E-04 & 5.8E-08 &5.7E-04\\
    NGC 4501 & Seyfert & 7.9E+07 & 5     & 2.4E+40 & 5     & 130.0 & 21    & 41.9  & 2.1   & 1.3E+10 & 21    & 1.1   & 3.7E-02 & 4.6E-06 & 1.6E-04 & 7.6E-09 &6.5E-05\\
    NGC 4526 &         & 4.5E+08 & 5     & 3.1E+40 & 5     & 195.4 & 31    & 60.0  & 2.2   & 9.4E+10 & 29    & 3.5   & 1.0E-01 & 5.3E-06 & 6.0E-04 & 4.3E-08 &6.8E-05\\
    NGC 4548 & LINER   & 3.4E+07 & 5     & 2.4E+41 & 5     & 143.7 & 31    & 42.3  & 3.7   & 8.7E+09 & 43    & 0.6   & 1.6E-02 & 1.0E-05 & 7.0E-05 & 3.3E-09 &1.6E-04\\
    NGC 4552 & LINER   & 5.0E+08 & 1     & 2.5E+40 & 3a    & 252.0 & 1     & 76.5  & 2.1   & 1.1E+11 & 1     & 2.5   & 7.0E-02 & 1.4E-06 & 4.9E-04 & 4.8E-08 &2.1E-04\\
    NGC 4555 &         &         &       & 2.7E+42 & 63    & 344.0 & 63    &       & 6.8   & 1.0E+12 & 63    & 12.6  &         & 9.7E-05 &         &         &1.0E-04\\
    NGC 4564 &         & 8.8E+07 & 5     & 9.7E+39 & 2     & 162.0 & 1     & 41.8  & 1.7   & 4.4E+10 & 1     & 2.4   & 4.1E-02 & 2.6E-06 & 1.8E-04 & 8.5E-09 &5.7E-05\\
    NGC 4594 & LINER   & 6.7E+08 & 5     & 1.9E+41 & 3a    & 240.0 & 1     & 64.5  & 3.5   & 2.7E+11 & 1     & 6.8   & 1.3E-01 & 2.1E-05 & 8.1E-04 & 6.4E-08 &6.6E-05\\
    NGC 4596 &         & 7.7E+07 & 5     & 2.4E+39 & 6     & 136.0 & 1     & 37.9  & 1.2   & 2.6E+10 & 1     & 2.0   & 4.4E-02 & 1.3E-06 & 1.8E-04 & 7.4E-09 &4.0E-05\\
    NGC 4621 &         & 4.0E+08 & 1     & 1.0E+39 & 53    & 211.0 & 1     & 73.6  & 1.0   & 4.4E+10 & 1     & 1.4   & 6.0E-02 & 1.3E-07 & 4.1E-04 & 3.8E-08 &2.1E-04\\
    NGC 4636 & LINER   & 3.1E+08 & 3     & 2.9E+40 & 3a    & 174.3 & 31    & 73.7  & 2.2   & 1.3E+10 & 29    & 0.6   & 4.7E-02 & 1.2E-06 & 3.2E-04 & 3.0E-08 &2.8E-04\\
    NGC 4649 &         & 4.7E+09 & 5     & 7.0E+39 & 3a    & 385.0 & 1     & 135.9 & 1.5   & 4.9E+11 & 1     & 4.8   & 2.1E-01 & 3.0E-07 & 2.3E-03 & 4.5E-07 &3.9E-04\\
    NGC 4696 & BCG     & 7.2E+08 & 3     & 2.4E+39 & 3a    & 251.2 & 37    & 57.1  & 1.2   & 6.8E+11 & 37    & 15.5  & 1.8E-01 & 1.6E-06 & 1.0E-03 & 7.0E-08 &3.3E-05\\
    NGC 4697 & BCG     & 2.0E+08 & 5     & 7.2E+39 & 3a    & 177.0 & 1     & 44.8  & 1.5   & 1.1E+11 & 1     & 5.1   & 8.2E-02 & 3.4E-06 & 3.9E-04 & 1.9E-08 &3.6E-05\\
    NGC 4710 &         &         &       & 1.6E+40 & 63    & 116.5 & 63    &       & 1.9   & 3.2E+10 & 63    & 3.4   &         & 1.4E-05 &         &         &1.5E-05\\
    NGC 4742 &         & 1.4E+07 & 5     & 1.1E+40 & 6     & 90.0  & 1     & 23.8  & 1.7   & 6.2E+09 & 1     & 1.1   & 2.0E-02 & 7.9E-06 & 5.9E-05 & 1.3E-09 &2.1E-05\\
    NGC 4751 &         & 2.4E+09 & 5     & 2.6E+40 & 5     & 355.0 & 42    & 123.2 & 2.1   & 2.8E+11 & 42    & 3.1   & 1.3E-01 & 6.9E-07 & 1.3E-03 & 2.3E-07 &4.6E-04\\
    NGC 4782 &         &         &       & 1.9E+42 & 63    & 308.5 & 63    &       & 6.2   & 8.5E+11 & 63    & 12.8  & 0.0E+00 & 1.0E-04 &         &         &7.4E-05\\
    NGC 4826 & Seyfert & 1.6E+06 & 5     & 7.8E+38 & 5     & 126.0 & 21    & 24.0  & 0.9   & 3.5E+09 & 21    & 0.3   & 2.2E-03 & 1.1E-07 & 6.5E-06 & 1.5E-10 &2.0E-04\\
    NGC 4889 &         & 2.1E+10 & 5     & 5.2E+41 & 5     & 347.0 & 17    & 137.0 & 4.5   & 1.2E+12 & 17    & 14.7  & 9.1E-01 & 3.2E-05 & 9.9E-03 & 2.0E-06 &9.2E-05\\
    NGC 4936 &         &         &       & 1.9E+42 & 63    & 278.2 & 63    &       & 6.2   & 8.0E+11 & 63    & 14.8  &         & 1.6E-04 &         &         &4.7E-05\\
    NGC 4945 & Seyfert & 1.4E+06 & 5     & 1.0E+39 & 2     & 134.0 & 1     & 25.6  & 0.9   & 3.0E+09 & 1     & 0.2   & 1.7E-03 & 8.4E-08 & 5.2E-06 & 1.3E-10 &3.2E-04\\
    NGC 5005 &         & 1.6E+08 & 3     & 2.5E+41 & 3a    & 139.0 & 18    & 45.3  & 3.8   & 2.5E+10 & 18    & 1.8   & 6.3E-02 & 3.6E-05 & 3.0E-04 & 1.5E-08 &4.7E-05\\
    NGC 5018 &         &         &       & 2.4E+41 & 63    & 206.5 & 63    &       & 3.7   & 2.3E+11 & 63    & 7.8   &         & 4.6E-05 &         &         &3.7E-05\\
    NGC 5044 & BCG     & 5.1E+08 & 3     & 2.6E+40 & 3a    & 239.9 & 37    & 52.9  & 2.1   & 5.6E+11 & 37    & 14.0  & 1.5E-01 & 9.8E-06 & 8.0E-04 & 4.9E-08 &3.2E-05\\
    NGC 5077 & BCG     & 8.6E+08 & 5     & 1.1E+41 & 6     & 222.0 & 1     & 65.9  & 3.0   & 2.1E+11 & 1     & 6.1   & 1.6E-01 & 1.6E-05 & 1.0E-03 & 8.2E-08 &5.8E-05\\
    NGC 5128 & Seyfert & 5.7E+07 & 5     & 2.6E+41 & 2     & 150.0 & 1     & 36.9  & 3.8   & 3.6E+10 & 1     & 2.3   & 3.4E-02 & 3.8E-05 & 1.4E-04 & 5.5E-09 &4.7E-05\\
    NGC 5252 & Seyfert & 1.0E+09 & 2     & 2.5E+44 & 2     & 190.0 & 1     & 56.7  & 21.1  & 2.4E+11 & 1     & 9.6   & 2.5E-01 & 1.3E-02 & 1.4E-03 & 9.6E-08 &2.3E-05\\
    NGC 5328 & BCG     & 4.7E+09 & 59    & 2.7E+42 & 59    & 331.0 & 59    & 109.9 & 6.8   & 6.6E+11 & 59    & 8.7   & 3.2E-01 & 7.5E-05 & 2.9E-03 & 4.5E-07 &1.4E-04\\
    NGC 5353 &         &         &       & 6.9E+41 & 63    & 283.5 & 63    &       & 4.8   & 3.6E+11 & 63    & 6.4   &         & 3.2E-05 &         &         &1.2E-04\\
    NGC 5419 & BCG     & 7.2E+09 & 59    & 5.0E+42 & 59    & 367.0 & 59    & 110.0 & 7.9   & 1.7E+12 & 59    & 18.2  & 4.9E-01 & 1.8E-04 & 4.5E-03 & 7.0E-07 &8.8E-05\\
    NGC 5490 & BCG     & 5.4E+08 & 59    & 1.3E+42 & 59    & 257.0 & 59    & 68.3  & 5.7   & 2.3E+11 & 59    & 5.0   & 9.4E-02 & 5.5E-05 & 6.1E-04 & 5.2E-08 &1.1E-04\\
    NGC 5516 &         & 3.7E+09 & 5     &         &       & 328.2 & 35    & 111.7 &       & 4.6E+11 & 35    & 6.2   & 2.4E-01 &       & 2.3E-03 & 3.5E-07   &1.9E-04\\
    NGC 5532 &         &         &       & 2.8E+42 & 63    & 277.8 & 63    &       & 6.9   & 7.8E+11 & 63    & 14.6  &         & 2.2E-04 &         &         &4.8E-05\\
    NGC 5576 &         & 2.7E+08 & 5     & 4.1E+39 & 6     & 183.0 & 1     & 46.3  & 1.3   & 1.5E+11 & 1     & 6.5   & 1.0E-01 & 2.6E-06 & 5.0E-04 & 2.6E-08 &3.1E-05\\
    NGC 5643 & Seyfert & 2.8E+06 & 34    & 1.0E+43 & 19    & 130.0 & 34    & 25.4  & 9.5   & 5.5E+09 & 34    & 0.5   & 3.5E-03 & 1.8E-04 & 1.1E-05 & 2.6E-10 &1.5E-04\\
    NGC 5813 & LINER   & 7.1E+08 & 1     & 7.3E+39 & 3a    & 230.0 & 1     & 74.9  & 1.6   & 1.1E+11 & 1     & 3.0   & 1.0E-01 & 9.2E-07 & 7.2E-04 & 6.8E-08 &1.3E-04\\
    NGC 5845 &         & 4.9E+08 & 5     & 1.9E+40 & 2     & 234.0 & 1     & 87.9  & 2.0   & 3.7E+10 & 1     & 1.0   & 5.2E-02 & 5.7E-07 & 4.0E-04 & 4.7E-08 &4.3E-04\\
    NGC 5846 & BCG     & 1.1E+09 & 7     & 9.2E+40 & 3a    & 238.0 & 1     & 67.1  & 2.9   & 3.5E+11 & 1     & 8.9   & 2.0E-01 & 1.6E-05 & 1.3E-03 & 1.1E-07 &4.9E-05\\
    NGC 5866 &         &         &       & 5.2E+40 & 63    & 161.6 & 63    &       & 2.5   & 9.4E+10 & 63    & 5.2   &         & 2.0E-05 &         &         &2.6E-05\\
    NGC 6086 & BCG          & 3.7E+09 & 5     &         &       & 318.0 & 32    & 86.8  &       & 1.4E+12 & 32    & 19.9  & 4.1E-01 &         & 3.1E-03 & 3.6E-07 & 5.2E-05 \\
    NGC 6098 &              &         &       & 3.1E+42 & 63    & 275.3 & 63    &       & 7.0   & 1.1E+12 & 63    & 21.6  &         & 3.6E-04 &         &         & 3.1E-05 \\
    NGC 6107 &              &         &       & 7.3E+42 & 63    & 240.9 & 63    &       & 8.7   & 1.3E+12 & 63    & 31.2  &         & 1.5E-03 &         &         & 1.5E-05 \\
    NGC 6166 &              & 1.9E+09 & 3     & 2.7E+41 &  3a   & 302.0 & 33    & 62.4  & 3.8   & 2.8E+12 & 33    & 44.2  & 3.9E-01 & 9.0E-05 & 2.4E-03 & 1.8E-07 & 2.0E-05 \\
    NGC 6251 & Seyfert      & 6.1E+08 & 5     & 5.0E+43 & 2     & 290.0 & 1     & 66.3  & 14.1  & 5.6E+11 & 1     & 9.6   & 1.1E-01 & 1.1E-03 & 7.2E-04 & 5.9E-08 & 8.2E-05 \\
    NGC 6264 & Seyfert      & 3.1E+07 & 5     & 1.3E+42 & 65    & 158.0 & 1     & 40.4  & 5.6   & 1.6E+10 & 1     & 0.9   & 1.5E-02 & 4.2E-05 & 6.7E-05 & 3.0E-09 & 1.4E-04 \\
    NGC 6269 &              &         &       & 1.3E+43 & 63    & 317.9 & 63    &       & 10.2  & 1.3E+12 & 63    & 18.2  &         & 5.9E-04 &         &         & 5.7E-05 \\
    NGC 6278 &              &         &       & 2.1E+41 & 63    & 193.2 & 63    &       & 3.6   & 1.4E+11 & 63    & 5.4   &         & 3.5E-05 &         &         & 4.3E-05 \\
    NGC 6323 &              & 1.0E+07 & 5     & 8.7E+43 & 56a   & 158.0 & 1     & 35.5  & 16.2  & 1.0E+10 & 1     & 0.6   & 6.6E-03 & 6.3E-04 & 2.6E-05 & 9.7E-10 & 2.2E-04 \\
    NGC 6338 &              &         &       & 4.2E+43 & 63    & 348.4 & 63    &       & 13.5  & 2.1E+12 & 63    & 24.8  &         & 1.4E-03 &         &         & 5.5E-05 \\
    NGC 6482 &              &         &       & 1.0E+43 & 63    & 316.8 & 63    &       & 9.5   & 5.2E+11 & 63    & 7.4   &         & 2.0E-04 &         &         & 1.4E-04 \\
    NGC 6861 &              & 2.1E+09 & 5     & 1.8E+40 & 5     & 388.8 & 35    & 134.0 & 1.9   & 2.5E+11 & 35    & 2.3   & 9.6E-02 & 2.9E-07 & 1.0E-03 & 2.0E-07 & 8.2E-04 \\
    NGC 6868 &              &         &       & 1.3E+42 & 63    & 250.1 & 63    &       & 5.7   & 5.4E+11 & 63    & 12.4  & 0.0E+00 & 1.5E-04 &         &         & 4.1E-05 \\
    NGC 7052 &              & 4.0E+08 & 5     & 7.7E+41 & 2     & 266.0 & 1     & 63.5  & 5.0   & 2.9E+11 & 1     & 5.9   & 8.0E-02 & 3.9E-05 & 4.9E-04 & 3.8E-08 & 1.0E-04 \\
    NGC 7176 &              &         &       & 4.0E+41 & 63    & 245.9 & 63    &       & 4.2   & 5.0E+11 & 63    & 12.0  &         & 6.1E-05 &         &         & 4.0E-05 \\
    NGC 7196 &              &         &       & 7.6E+41 & 63    & 277.9 & 63    &       & 5.0   & 4.2E+11 & 63    & 7.8   &         & 4.4E-05 &         &         & 8.9E-05 \\
    NGC 7332 &              & 1.3E+07 & 8     & 1.7E+39 & 54    & 122.0 & 8     & 26.6  & 1.1   & 1.5E+10 & 8     & 1.5   & 1.5E-02 & 1.0E-06 & 4.8E-05 & 1.2E-09 & 4.1E-05 \\
    NGC 7457 & Galaxy       & 9.0E+06 & 5     & 1.6E+39 & 6     & 67.0  & 1     & 15.8  & 1.1   & 7.0E+09 & 1     & 2.2   & 2.9E-02 & 8.9E-06 & 6.4E-05 & 8.6E-10 & 4.3E-06 \\
    NGC 7582 & Seyfert      & 5.5E+07 & 5     & 3.2E+43 & 19a   & 156.0 & 1     & 29.5  & 12.6  & 1.3E+11 & 1     & 7.7   & 5.2E-02 & 4.1E-03 & 1.8E-04 & 5.3E-09 & 1.6E-05 \\
    NGC 7618 &              &         &       & 9.4E+42 & 63    & 292.8 & 63    &       & 9.3   & 9.4E+11 & 63    & 15.8  &         & 5.1E-04 &         &         & 5.1E-05 \\
    NGC 7619 & BCG          & 2.3E+09 & 5     & 2.0E+42 & 59    & 292.0 & 17    & 90.9  & 6.3   & 4.5E+11 & 17    & 7.5   & 2.3E-01 & 7.5E-05 & 1.8E-03 & 2.2E-07 & 1.1E-04 \\
    NGC 7626 &              & 3.8E+08 & 59    & 6.2E+41 & 59    & 234.0 & 59    & 55.8  & 4.7   & 2.8E+11 & 59    & 7.4   & 1.0E-01 & 6.0E-05 & 5.5E-04 & 3.7E-08 & 5.6E-05 \\
    NGC 7768 & BCG          & 1.3E+09 & 5     &         &       & 257.0 & 36    & 68.4  &       & 5.7E+11 & 36    & 12.4  & 2.3E-01 &         & 1.5E-03 & 1.3E-07 & 4.4E-05 \\
    UGC 408  &              &         &       & 1.2E+42 & 63    & 197.6 & 63    &       & 5.6   & 3.1E+11 & 63    & 11.4  &         & 2.6E-04 &         &         & 2.2E-05 \\
    UGC 1841 & Seyfert      & 3.0E+08 & 59    & 6.6E+42 & 59    & 295.0 & 59    & 45.7  & 8.5   & 1.9E+12 & 59    & 31.0  & 1.2E-01 & 7.5E-04 & 5.5E-04 & 2.8E-08 & 2.7E-05 \\
    Mrk 1216 & LINER        & 4.9E+09 & 59    & 4.4E+42 & 59    & 324.1 & 59    & 172.1 & 7.7   & 6.6E+10 & 59    & 0.9   & 1.4E-01 & 1.2E-05 & 1.7E-03 & 4.7E-07 & 1.2E-03 \\
  
\enddata

\tablenotetext{a}{BH bolometric luminosity $L_B$ directly provided by the reference in column (6). Otherwise (without 'a'), $L_B$ was estimated using the X-ray luminosity in the 2-10 kev band, that is, $L_B= L_{2-10} \times 15.8$ \citep{Ho:2009-Radiatively-Inefficient-A}.}

\tablerefs{
 (1) \citealt{Benedetto:2013-New-calibration-and-some-}; 
 (2) \citealt{Gultekin:2009-The-Fundamental-Plane-of-}; 
 (3) \citealt{Inayoshi:2020-Universal-Transition-Diag}; 
 (4) \citealt{Zhang:2009-A-Census-of-X-Ray-Nuclear}; 
 (5) \citealt{Gultekin:2019-The-Fundamental-Plane-of-}; 
 (6) \citealt{Gultekin:2012-A-Chandra-Survey-of-Super}; 
 (7) \citealt{Hu:2008-The-black-hole-mass-stell}; 
 (8) \citealt{Haring:2004-On-the-black-hole-mass-bu}; 
 (9) \citealt{Marconi:2003-The-relation-between-blac}; 
 (10) \citealt{McConnell:2013-Revisiting-the-Scaling-Re}; 
 (11) \citealt{Boizelle:2021-Black-Hole-Mass-Measureme}; 
 (12) \citealt{Bettoni:2003-The-black-hole-mass-of-lo}; 
 (13) \citealt{McConnachie:2012-The-Observed-Properties-o}; 
 (14) \citealt{van_den_Bosch:2012-An-over-massive-black-hol}; 
 (15) \citealt{Rusli:2011-The-central-black-hole-ma}; 
 (16) \citealt{Spolaor:2008-The-early-type-galaxies-N}; 
 (17) \citealt{Bogdan:2018-Correlation-between-the-T}; 
 (18) \citealt{Das:2003-Central-mass-concentratio}; 
 (19  \citealt{Brightman:2017-X-Ray-Bolometric-Correcti}; 
 (20) \citealt{Marin:2016-Are-there-reliable-method}; 
 (21) \citealt{Davis:2019-Black-Hole-Mass-Scaling-R}; 
 (22) \citealt{Bennert:2006-Size-and-properties-of-th}; 
 (23) \citealt{Benedetto:2013-New-calibration-and-some-}; 
 (24) \citealt{Dullo:2016-Complex-central-structure}; 
 (25) \citealt{Marconi:2003-Is-there-really-a-black-h}; 
 (26) \citealt{Coccato:2006-NGC-4435--a-bulge-dominat}; 
 (27) \citealt{Sarzi:2002-Limits-on-the-mass-of-the}; 
 (28) \citealt{Fisher:2010-Bulges-of-Nearby-Galaxies}; 
 (29  \citealt{Cappellari:2011-The-ATLAS-3D--project---I}; 
 (30) \citealt{Cappellari:2012-The-Atlas3D-project-XX.-Mass-size}; 
 (31) \citealt{Beifiori:2012-On-the-correlations-betwe}; 
 (32) \citealt{McConnell:2011-The-Black-Hole-Mass-in-th}; 
 (33) \citealt{Bender:2015-STRUCTURE-AND-FORMATION-O}; 
 (34) \citealt{Garcia_Bernete:2021-Multiphase-feedback-proce}; 
 (35) \citealt{Rusli:2013-THE-INFLUENCE-OF-DARK-MAT}; 
 (36) \citealt{McConnell:2012-DYNAMICAL-MEASUREMENTS-OF}; 
 (37) \citealt{Samir:2016-The-fundamental-plane-of-early-type-galaxies}; 
 (38) \citealt{Gultekin:2011-Is-There-a-Black-Hole-in-}; 
 (39) \citealt{Perez:2009-Near-infrared-imaging-and}; 
 (40) \citealt{Chae:2018-Modeling-Nearly-Spherical}; 
 (41) \citealt{Sahu:2019-Revealing-Hidden-Substruc}; 
 (42) \citealt{Lakhchaura:2019-Correlations-between-supe}; 
 (43) \citealt{Weinzirl:2009-BULGE-n-AND-B-T-IN-HIGH-M}; 
 (44) \citealt{Beifiori:2009-Upper-Limits-on-the-Masse}; 
 (45) \citealt{Sanchez_Portal:2004-Structural-parameters-of-}; 
 (46) \citealt{Ho:2009-A-Search-for--Dwarf--Seyf}; 
 (47) \citealt{Samir:2016-The-fundamental-plane-of-early-type-galaxies}; 
 (48) \citealt{Burtscher:2015-Obscuration-in-active-gal}; 
 (49) \citealt{Koss:2015-Broadband-Observations-of}; 
 (50) \citealt{Gonzalez_Martin:2009-An-X-ray-view-of-82-LINER}; 
 (51) \citealt{Williams:2022-LeMMINGs---IV--The-X-ray-}; 
 (52) \citealt{Gao:2017-The-Megamaser-Cosmology-P}; 
 (53) \citealt{Wrobel:2008-Outflow-dominated-emissio}; 
 (54) \citealt{Nagar:2005-Radio-sources-in-low-lumi}; 
 (55) \citealt{Fabian:2013-X-ray-emission-from-the-u}; 
 (56) \citealt{Kuo:2020-The-Megamaser-Cosmology-P}; 
 (57) \citealt{Swartz:2006-Chandra-observations-of-c}; 
 (58) \citealt{Urquhart:2022-X-Ray-and-Radio-Observations-of-Central}; 
 (59) \citealt{Lakhchaura:2019-Correlations-between-supe}; 
 (60) \citealt{Kammoun:2020-A-Hard-Look-at-Local--Opt}; 
 (61) \citealt{Machacek:2004-Chandra-observations-of-N}; 
 (62) \citealt{Maccarone:2011-A-new-globular-cluster-bl}; 
 (63) \citealt{Babyk:2018-X-Ray-Scaling-Relations-of-Early-type}; 
 (64) \citealt{Gultekin:2009-THE-M-sigma-AND-M-L-RELAT}; 
 (65) \citealt{Castangia:2013-New-Compton-thick-AGN}; 
}

\tablecomments{This table lists the SMBH and host galaxy data used in our co-evolution analysis. The columns list the galaxy name, the galaxy type (from SIMBAD astronomical database), the BH mass $M_B$, the estimated BH bolometric luminosity $L_B$, the bulge velocity dispersion $\sigma_b$, the bulge mass $M_b$. The five length scales are computed quantities from these data, that is, the bulge size $r_b=G M_b/(3\sigma_b^2)$ \citep{Marconi:2003-The-relation-between-blac}, the BH sphere of influence $r_B$ (Eq. \eqref{eq:15}), the radiation scale $r_p$ (Eq. \eqref{eq:16}), and the dissipation scale $r_x$ (Eq. \eqref{eq:17}), and the Schwarzschild radius $r_s$ (Eq. \eqref{eq:15}). The rate of energy flow is estimated as $\varepsilon_b=\sigma_b^3/r_b$. The velocity dispersions on the scales $r_B$ and $r_p$ are also presented as $\sigma_B$ and $\sigma_p$ (Fig. \ref{fig:3}).}

\label{tab:A1}
\end{deluxetable}

\end{document}